\documentclass[preprint,11pt,number,sort&compress]{elsarticle}
\usepackage[utf8]{inputenc}
\usepackage{fullpage}
\usepackage[hidelinks]{hyperref}
\usepackage{amsmath}
\usepackage{amssymb}
\usepackage{amsthm}
\usepackage[stamp=false]{draftwatermark}
\usepackage{empheq}
\usepackage{slashed}
\usepackage{cleveref}
\usepackage{tabularx, ragged2e}
\usepackage{xcolor}
\usepackage[binary-units,range-phrase=--,range-units=single,exponent-product=\times]{siunitx}[=v2]
\usepackage[noprefixcmds,sicmds]{hepunits}
\usepackage{siunitx}
\usepackage{orcidlink}
\usepackage{url}
\usepackage{lineno}{ \linenumbers}
\newcolumntype{L}[1]{>{\raggedright\let\newline\\\arraybackslash\hspace{0pt}}m{#1}}
\newcolumntype{C}[1]{>{\centering\let\newline\\\arraybackslash\hspace{0pt}}m{#1}}
\newcolumntype{R}[1]{>{\raggedleft\let\newline\\\arraybackslash\hspace{0pt}}m{#1}}

\usepackage{braket}

\usepackage{bm}
\crefname{section}{Section}{Sections}
\crefname{appendix}{Appendix}{Appendices}
\usepackage{xspace}

\newcommand{\mee}           {\ensuremath{m_{\text{ee}}}\xspace}

\newcommand{\pt}            {\ensuremath{p_{\mathrm{T}}}\xspace}

\newcommand{\ptee}          {\ensuremath{p_{\mathrm{T,ee}}}\xspace}

\newcommand{\sNN}           {\ensuremath{\sqrt{s_{\mathrm{NN}}}}\xspace}

\newcommand{\lele}            {\ensuremath{\mathrm{l^{+}l^{-}}}\xspace}

\newcommand{\MeVc}          {\ensuremath{\mathrm{MeV}\kern-0.05em/\kern-0.02em c}\xspace}
\newcommand{\MeVcc}         {\ensuremath{\mathrm{MeV}\kern-0.05em/\kern-0.02em c^{2}}\xspace}
\newcommand{\GeVc}          {\ensuremath{\mathrm{GeV}\kern-0.05em/\kern-0.02em c}\xspace}
\newcommand{\GeVcc}         {\ensuremath{\mathrm{GeV}\kern-0.05em/\kern-0.02em c^{2}}\xspace}

\newcommand{\piZ}          {\ensuremath{\pi^{0}}\xspace}

\newcommand{\lsim}{\raisebox{-4pt}{$\,\stackrel{\textstyle <}{\sim}\,$}}

\newcommand{\eg}{\textit{e.g.}}
\newcommand{\ie}{\textit{i.e.}}

\newcommand{\bvec}[1]{\ensuremath{\boldsymbol{#1}}}

\title{Anomalous soft photons: status and perspectives}
\date{May 2024}

\journal{Physics Reports}

\begin{document}

\author[R1]{R.~Bailhache \orcidlink{0000-0001-7987-4592}}
\author[R2]{D.~Bonocore \orcidlink{0000-0002-8672-5584}}
\author[R3]{P.~Braun-Munzinger \orcidlink{0000-0003-2527-0720}}
\author[R4]{X.~Feal \orcidlink{0000-0002-3258-2020}}
\author[R5]{S.~Floerchinger \orcidlink{0000-0002-3428-4625}}
\author[R6]{J.~Klein \orcidlink{0000-0002-1301-1636}}
\author[R7]{K.~K\"{o}hler}
\author[R8]{P.~Lebiedowicz \orcidlink{0000-0003-1963-6263}}
\author[R1]{C.M.~Peter \orcidlink{0009-0008-9207-9911}}
\author[R9]{R.~Rapp \orcidlink{0000-0003-3020-4402}}
\author[R7]{K.~Reygers \orcidlink{0000-0001-9808-1811}}
\author[R8]{W.~Sch\"{a}fer \orcidlink{0000-0002-0764-0372}}
\author[R1]{H.S.~Scheid \orcidlink{0000-0003-1184-9627}}
\author[R3]{K.~Schweda \orcidlink{0000-0001-9935-6995}}
\author[R7]{J.~Stachel \orcidlink{0000-0003-0750-6664}}
\author[R10]{H.~van~Hees \orcidlink{0000-0003-0729-2117}}
\author[R7]{C.A.~van Veen \orcidlink{0000-0003-1199-4445}}
\author[R7]{M.~V\"{o}lkl \orcidlink{0000-0002-3478-4259}}

\address[R1]{Institut f\"{u}r Kernphysik, Johann Wolfgang Goethe-Universit\"{a}t Frankfurt, Frankfurt, Germany}
\address[R2]{Technical University of Munich, TUM School of Natural Sciences, Physics Department T31, James-Franck-Straße 1, D-85748, Garching, Germany}
\address[R3]{Research Division and ExtreMe Matter Institute EMMI, GSI Helmholtzzentrum f\"ur Schwerionenforschung GmbH, Darmstadt, Germany}
\address[R4]{IGFAE, Universidade de Santiago de Compostela, Campus Vida, 15705 Santiago de Compostela, Spain}
\address[R5]{Theoretisch-Physikalisches Institut, Friedrich-Schiller-Universit\"{a}t Jena, Max-Wien-Platz 1, 07743 Jena, Germany}
\address[R6]{European Organization for Nuclear Research (CERN), Geneva, Switzerland}
\address[R7]{Physikalisches Institut, Ruprecht-Karls-Universit\"{a}t Heidelberg, Heidelberg, Germany}
\address[R8]{Institute of Nuclear Physics Polish Academy of Sciences, Radzikowskiego 152, PL-31342 Krak{\'o}w, Poland}
\address[R9]{Cyclotron Institute and Department of Physics and Astronomy, Texas A\&M University, College Station, TX 77843-3366, USA}
\address[R10]{Institut f\"ur Theoretische Physik, Goethe-Universit\"at Frankfurt am Main, Max-von-Laue-Strasse 1, D-60438 Frankfurt am Main, Germany and Helmholtz Research Academy Hesse for FAIR, Campus Frankfurt, D-60438 Frankfurt, Germany}

\begin{keyword}
EMMI-RRTF-ER20-01 \sep TUM-HEP-1496-24
\end{keyword}

\begin{abstract}
This report summarizes the work of the EMMI Rapid Reaction Task Force on ``Real and Virtual Photon Production at Ultra-Low Transverse Momentum and Low Mass at the LHC''. We provide an overview of the soft-photon puzzle, i.e., of the long-standing discrepancy between experimental data and predictions based on Low's soft-photon theorem, also referred to as ``anomalous'' soft photon production, and we review the current theoretical understanding of soft radiation and soft theorems. We also focus on low-mass dileptons as a tool for determining the electrical conductivity of the medium produced in high-energy nucleus-nucleus collisions. We discuss how both topics can be addressed with the planned ALICE~3 detector at the LHC.
\end{abstract}

\maketitle

\tableofcontents

\section{Introduction}
The measurement of low-energy real and virtual photons in high-energy collisions of hadrons and nuclei provides access to fundamental physics questions. In this report we focus on soft-photon production and tests of the range of validity of Low's soft photon theorem as well as on the extraction of the electrical conductivity of the medium created in nucleus-nucleus collisions with the aid of soft virtual photons. In experimental tests of Low's theorem, striking discrepancies between predictions and data were found. No agreement exists on their possible origin, despite more than 40 years of research. On the theory side, there is a considerable interest in soft theorems as they are related to symmetries reflecting the infrared structure of gravity and gauge theory. The electrical conductivity is a key property that characterizes the produced quark-gluon plasma and the hadron gas. The production rate of low-energy virtual photons, measurable as $e^+e^-$ pairs in the experiment, is sensitive to the electrical conductivity. However, extracting this electrical conductivity is challenging and so far few experimental constraints exist. The planned ALICE~3 detector at the LHC provides the opportunity to quantitatively address both questions. We summarize the current theoretical understanding of the two topics and discuss experimental opportunities and challenges.

\section{The soft-photon puzzle}

\subsection{Introduction}
Bremsstrahlung produced in scattering processes can be attributed to accelerated charges between the incoming initial-state and outgoing final-state charged particles, as well as to accelerations happening in the short-lived intermediate state. However, bremsstrahlung photons with a sufficiently long wavelength cannot resolve the intermediate state and consequently their production is solely determined by the overall change of currents from the initial to the final state. Photons which satisfy this long-wavelength criterion are called soft photons. For inelastic hadronic interactions, the order of magnitude for the spatial extent of the intermediate state and its lifetime are usually estimated to be $\Delta x \approx 1\,\mathrm{fm}$ and $\Delta t \approx 1\,\mathrm{fm}/c$, respectively. The wavelength of a 100\,MeV photon is about $12\,\mathrm{fm}$. Thus, photons with energies less than $10\text{--}100\,\mathrm{MeV}$ are expected to qualify as soft photons for inelastic hadronic interactions.

Particle production in hadronic collisions at high energies is often modeled as the fragmentation of a QCD string spanned along the beam axis. The various string fragments can be viewed as particle sources that move along the beam direction with different velocities. Following an old argument by Gribov \cite{Gribov:1966hs}, for such systems the transverse momentum with respect to the beam axis has been traditionally used to characterize the softness of a produced photon. As the mean transverse momentum of pions is in the order of $p_T^\pi \approx 300\text{--}500\,\mathrm{MeV}/c$, soft photons are often defined as photons whose transverse momentum $k_{\text{T}}$ satisfies $k_{\text{T}} \ll p_T^\pi$.

In 1958 Francis Low wrote a seminal paper on how to relate the production of soft photons with the production of charged particles involved in a scattering process \cite{Low:1958sn}. Low considered 2-to-2 scattering processes plus an additional photon in the final state and based on a quantum mechanical treatment, he provided expressions for $\sigma_0$ and $\sigma_1$ in the power expansion
\begin{equation}
\label{eq:multipole_expansion_cross_sections}
\sigma = \frac{\sigma_0}{\omega_k} + \sigma_1 + \omega_k \sigma_2 + \ldots
\end{equation}
of the cross section where $\omega_k$ is the photon energy corresponding to a four-momentum $k$. A generalized
expression for the leading-power term for $n$-to-$m$ scattering
processes is tested in various hadronic scattering
experiments. Surprisingly, some of the experiments found a significant
excess of low-energy photons compared to the expectation from the leading term $\sigma_0$ in Low's
theorem. The ratio of measured-over-expected photons is in the range
between 4--8, see Tab.~\ref{tab:soft_photon_measurements}. Moreover, an
excess by a factor of four is also found by the DELPHI experiment in
$\text{e}^+ \text{e}^- \to n\, \mathrm{jets}$ collisions. This is now
referred to as the soft-photon puzzle, and one refers to the excess as
anomalous soft-photon production. The experimental situation, however,
is unsatisfactory, as some experiments did not observe an excess above
the expectation from Low's theorem. Previous reviews of this subject
and the existing measurements can be found in \cite{Lichard:1994qt,Balek:1989rx,Balek:1991md,Belogianni:2002ic,
DELPHI:2005yew,Perepelitsa:2009ata,Wong:2014ila}.

\subsubsection{Overview of soft-photon measurements}

Here we give a brief overview of data on soft-photon production in
high-energy collisions, and we highlight selected results. Important
results are summarized in Tab.~\ref{tab:soft_photon_measurements}. To
study the soft-photon signal, all experiments subtract the background
photons coming from the decay of hadrons, which is dominated by the decay
\mbox{$\pi^0 \to \gamma \gamma$}. A remaining serious experimental background
for the inner-bremsstrahlung signal as calculated with Low's formula
then is external bremsstrahlung created by electrons and positrons in
the detector material.

Early bubble-chamber experiments gave conflicting results on soft-photon
production. While no excess above the expected inner-bremsstrahlung
signal is found in $\pi^-\mathrm{p}$ collisions with a beam momentum of
10.5\,GeV/$c$ at the Stanford Linear Accelerator Center (SLAC)
\cite{Goshaw:1979kq}, a strong enhancement is found by the WA27
collaboration in $70\,\mathrm{GeV}/c$ $\mathrm{K}^+ \mathrm{p}$
collisions using the Big European Bubble Chamber (BEBC) at the CERN
Super Proton Synchrotron (SPS) \cite{Chliapnikov:1984ed}. A few years
later, anomalous soft-photon production was reported at higher
energies in $\pi^+\mathrm{p}$ and $\mathrm{K}^+\mathrm{p}$ collisions at 250\,GeV/$c$ by the EHS-NA22 collaboration using the Rapid Cycling Bubble Chamber (RCBC)
\cite{EHSNA22:1991sdp}.

A first confirmation of a photon excess from an experiment not using a
bubble chamber came from the SOPHIE/WA83 experiment at CERN's Omega
spectrometer. Employing electromagnetic calorimeters, this experiment
found a significant low-$p_T$ excess in $\pi^-\mathrm{p}$ collisions at
280\,GeV/$c$ beam momentum \cite{SOPHIEWA83:1992czx}. However, a few years
later, Antos et al.\ did
not find a significant excess above the expected hadronic
inner bremsstrahlung in p--Be collisions at 450\,GeV/$c$ in two independent
measurements, one based on the photon conversion method and the other
based on a barium-fluoride calorimeter \cite{Antos:1993wv}. Using also
barium-fluoride calorimeters, the same conclusion was reached by
Tincknell et al.\ for p--Be and p--W collisions at 18\,GeV/$c$
\cite{Tincknell:1996ks}.

\begin{figure}[tbh]
  \centering
  \includegraphics[width=\textwidth]{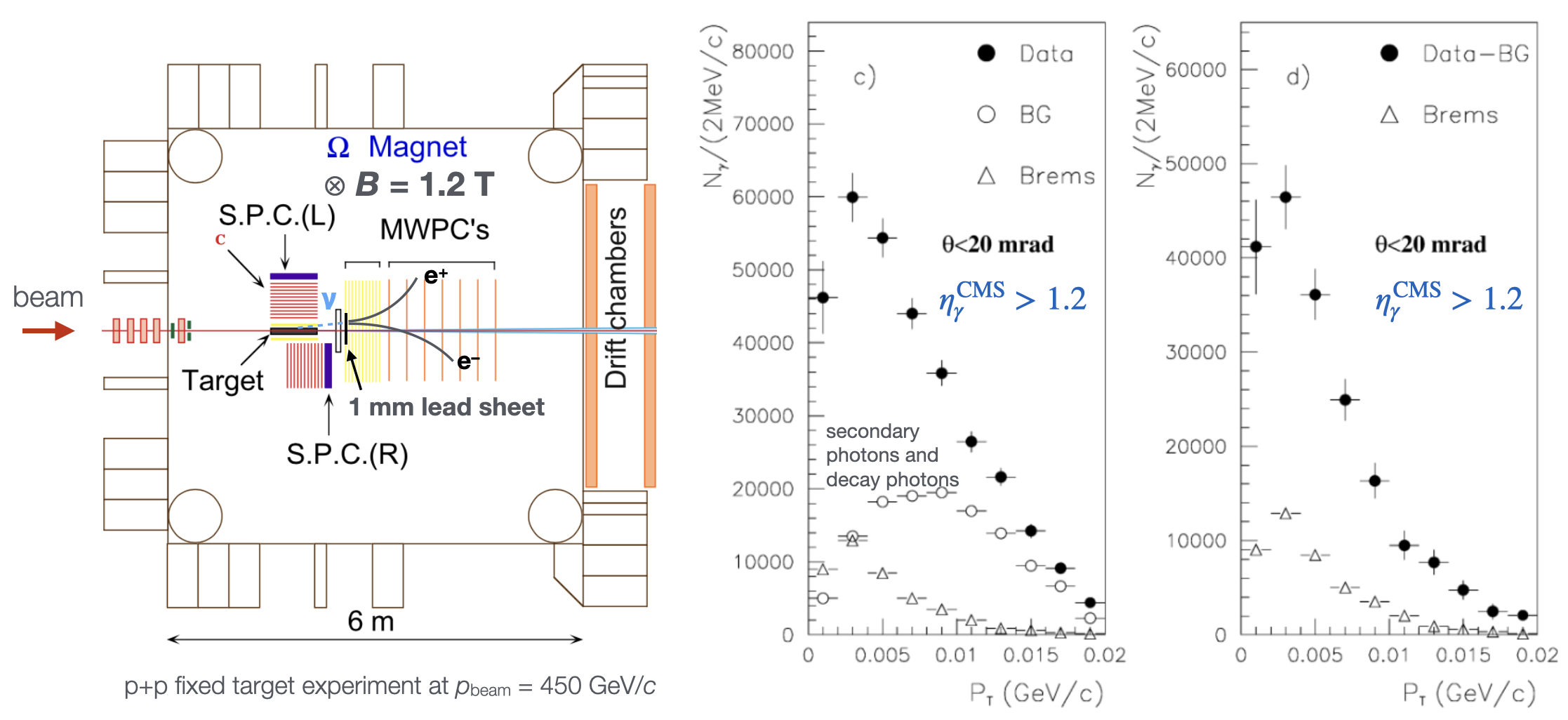}
  \caption{Left panel: Setup of the WA102 experiment, located at CERN's Omega spectrometer. Photons from p--p collisions in a liquid hydrogen target are measured by reconstructing $\text{e}^+ \text{e}^-$ tracks resulting from photon conversions in a 1\,mm thick lead converter \cite{Belogianni:2002ic}. Right panel: The left-hand figure shows the measured photon $p_T$ spectrum along with the expected background photons (BG, open circles) and inner-bremsstrahlung signal (Brems, open triangles). The background photons comprise the photons from hadron decays and secondary photons. Secondary photons are external bremsstrahlung photons, resulting from electrons and positrons produced in a photon conversion. The right-hand figure shows the measured photon yield after the subtraction of the background. The excess above the inner bremsstrahlung signal has approximately the same shape as the inner bremsstrahlung spectrum.}
  \label{fig:wa102_soft_photons}
\end{figure}
In a series of measurements at CERN's Omega spectrometer, again, a
significant photon excess was observed
\cite{WA91:1997cnv,Belogianni:2002ib,Belogianni:2002ic}. These
measurements were based on the photon-conversion method, allowing the
reconstruction of the photons' line of flight. This made it possible to
suppress background of photons not coming from the target
region. An excess above the expected inner-bremsstrahlung signal of a
factor of about 4--5 was observed in these experiments in
$\pi^-\mathrm{p}$ collisions at 280\,GeV/$c$ and pp collisions at 450\,GeV/$c$. A first publication quoted an even larger excess, however, later a problem in the calculation of the bremsstrahlung signal was identified \cite{Belogianni:2002ib}.

\begin{figure}[tbh]
  \centering
  \includegraphics[width=\textwidth]{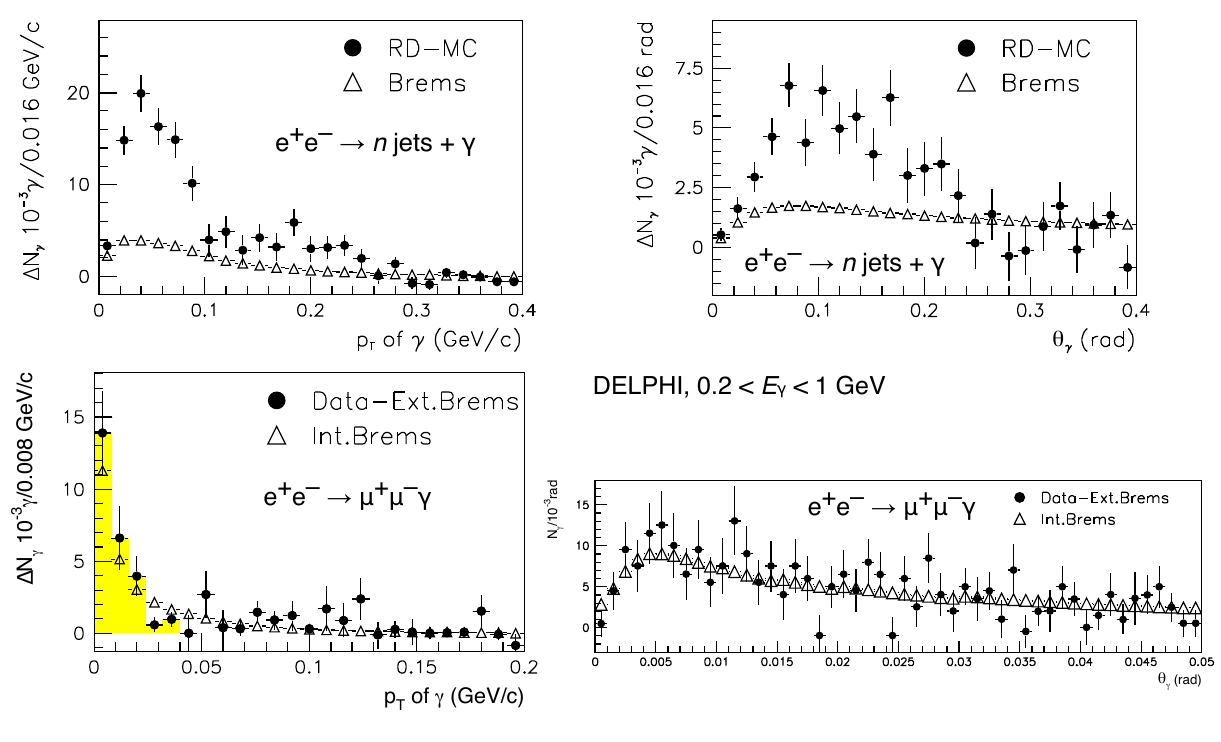}
  \caption{DELPHI results on soft-photon production in $\text{e}^+ \text{e}^- \to n\,\mathrm{jets}$ (upper panels) and $\text{e}^+ \text{e}^- \to \mu^+\mu^-$ (lower panels) at $\sqrt{s} = 91\,\mathrm{GeV}$ as a function of the transverse momentum $p_T$ and the angle $\theta$ relative to the reconstructed jet axis or muon direction, respectively \cite{DELPHI:2005yew,DELPHI:2007nmh}. The measured photon yield shows an excess above the inner-bremsstrahlung signal, as calculated based on Low's theorem in $\text{e}^+ \text{e}^- \to n\,\mathrm{jets}$. In $\text{e}^+ \text{e}^- \to \mu^+\mu^-$, a suppression of photons with angles $\theta \lesssim m_\mu/E_\mu \approx 0.0023$ due to the dead-cone effect is visible.}
  \label{fig:delphi_soft_photons}
\end{figure}
\begin{figure}[th]
  \centering
  \includegraphics[width=\textwidth]{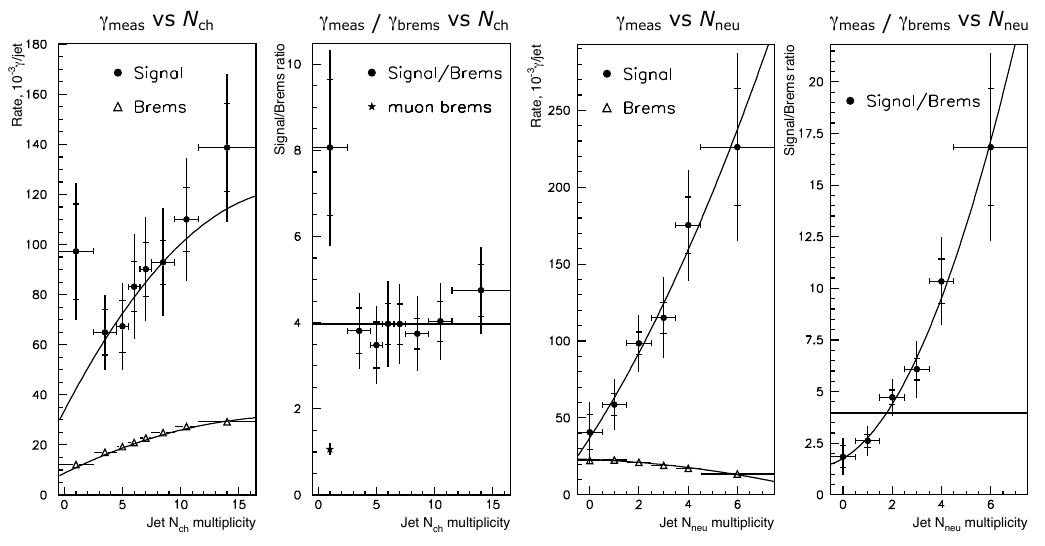}
  \caption{Photon yield per jet and the ratio $\gamma_\mathrm{meas}/\gamma_\mathrm{brems}$ as a function of the charged ($N_\mathrm{ch}$) and neutral ($N_\mathrm{neu}$) multiplicity of the jets in $e^+e^- \to n\,\mathrm{jets}$ events, as measured by DELPHI \cite{DELPHI:2010cit}. $N_\mathrm{neu}$ is calculated based on the number of showers in the electromagnetic calorimeter and the number of converted photons, and is defined in a way that it approximates the number of $\pi^0$'s in the jet.}
  \label{fig:delphi_ch_neu_dep}
\end{figure}

An important addition to the set of soft-photon measurements was made by the DELPHI experiment in $\text{e}^+ \text{e}^- \to \mathrm{jets}$ reactions at the Z pole ($\sqrt{s} = 91\,\mathrm{GeV}$) \cite{DELPHI:2005yew}. At relatively small transverse momenta with respect to the jet axis ($p_T < 80\,\mathrm{MeV}/c$), an excess of photons above the inner-bremsstrahlung signal of about a factor 4 was observed. In $\text{e}^+ \text{e}^- \to \mu^+\mu^-$ collisions, however, the measured photon signal agreed with the calculated inner-bremsstrahlung signal \cite{DELPHI:2005yew}, see Fig.~\ref{fig:delphi_soft_photons}. In particular, the dependence of the photon yield on the angle $\theta$ between the radiating muon and the emitted photon exhibited an expected depletion at small angles $\theta \lesssim m_\mu / E_\mu$ known as dead-cone effect, see Sec.~\ref{sec:leading_soft_theorem_classical_radiation}. These findings suggest that the anomalous soft-photon production is related to the production of hadrons. Studying the characteristics of the photon excess in $\text{e}^+ \text{e}^- \to \mathrm{jets}$, DELPHI found that the ratio $\gamma_\mathrm{meas}/\gamma_\mathrm{brems}$ of the measured signal over the expected inner-bremsstrahlungs signal was constant as a function of the charged-particle multiplicity within a jet. The neutral-particle multiplicity of a jet (dominated by the number $\pi^0$'s) was estimated based on the number of clusters reconstructed in the electromagnetic calorimeter and the number of reconstructed conversions of photons with energies above 1\,GeV. Somewhat surprisingly, the ratio $\gamma_\mathrm{meas}/\gamma_\mathrm{brems}$ was found to strongly increase with the multiplicity of neutral particles of a jet, reaching an excess up to about a factor 15 or larger for large neutral-particle multiplicities \cite{DELPHI:2010cit}. This is illustrated in Fig.~\ref{fig:delphi_ch_neu_dep}.

In Ref.~\ref{fig:delphi_ch_neu_dep}, the DELPHI collaboration stresses that the dependence of the photon excess on the multiplicity of neutral particles of a jet is a particular challenge for most model approaches. As explained in this paper, this feature of the data might point to a general picture, where the extra radiation results from dipole radiation of $q \bar q$ quark pairs formed in the fragmentation process. The dipole moment of a neutral $q \bar q$ pair is larger than the one of a $q \bar q$ pair with electric charge of $\pm 1$. As dipole radiation scales with the square of the dipole moment, the radiation strength of a neutral $q \bar q$ pair can be up to almost an order of magnitude higher than for a charged $q \bar q$ pair. DELPHI also points out that soft-photon production associated with $q \bar q$ production built upon the space-time structure of the Lund string fragmentation model is not strong enough to describe the data \cite{Andersson:1988nk}.

We close this section by mentioning further measurements that are interesting with regard to Low's theorem. Yu.~M.~Antipov et.\ al.\ studied pion radiative scattering ($\pi^-\mathrm{p} \to \pi^-\mathrm{p} \gamma$) at the Serpukhov accelerator \cite{SIGMA-AYaKS:1989uyb}. In this experiment, $43\,\mathrm{GeV}/c$ $\pi^-$ impinged on a liquid-hydrogen target. The scattered pion was deflected by a magnetic field, and the photon energy was measured in the forward direction with a lead glass calorimeter. Calculations based on the Low theorem were found to agree with the data up to photon energies of $40\,\mathrm{GeV}$. The results were interpreted as a confirmation of Gribov's prediction of an extended range of applicability of Low's theorem \cite{Gribov:1966hs}. The WA93 collaboration measured inclusive photons in $200\,\mathrm{GeV}$/nucleon ${}^{32}\mathrm{S}+\mathrm{Au}$ collisions with a small-acceptance BGO detector \cite{WA93:1996byj}. In the first bin of the $k_T$ spectrum ($k_T \lesssim 68\,\mathrm{MeV}/c$), an excess above the expected decay-photon yield was observed. Experimental background and hadronic bremsstrahlung were not sufficient to explain the excess \cite{kruempel_diploma_thesis}. Finally, we note that inner bremsstrahlung was also studied in radiative decays of the neutron ($\mathrm{n} \to \mathrm{p}e^- \bar \nu_e \gamma$). The sudden acceleration of the electron in the electromagnetic field of the proton gives rise to bremsstrahlung which can be calculated within quantum electrodynamics (QED). The energy spectrum of the electron has a maximum at around $200\,\mathrm{keV}$ and an endpoint energy of $782\,\mathrm{keV}$ \cite{Dubbers:2021wqv}. Bremsstrahlung photons were measured from $0.4\,\mathrm{keV}$ up to the endpoint energy \cite{RDKII:2016lpd}. No deviations from a QED calculation were found \cite{Ivanov:2017wxl}. The rather low lower-energy limit of the measurement of $0.4\,\mathrm{keV}$ appears to suggest that Low's theorem holds for radiative neutron decays.

\begin{table}[!htb]
\resizebox{\textwidth}{!}{
\setlength\extrarowheight{2pt}
\footnotesize
\begin{tabularx}{\textwidth}{L{2.3cm} X L{1.8cm} L{3.0cm} C{1.8cm} L{1.6cm} L{2.4cm}}
Exp. & year & $p_\mathrm{beam}$ or $\sqrt{s}$ & photon $k_T$ & $\gamma_\mathrm{meas}/\gamma_\mathrm{brems}$ & method & Ref. \\
\hline
$\pi^+\text{p}$ & 1979 & 10.5\,GeV/$c$ & $p_T < 20\,\mathrm{MeV}/c$ & $1.25 \pm 0.25$ & bubble chamber & Goshaw~et~al. \cite{Goshaw:1979kq} \\
$\text{K}^+\text{p}$ \linebreak WA27, CERN& 1984 & 70\,GeV/$c$ & $k_T < 60\,\mathrm{MeV}/c$ & $4.0 \pm 0.8$ & bubble chamber (BEBC) & Chliapnikov~et~al. \cite{Chliapnikov:1984ed} \\
$\pi^+\text{p}$ \linebreak CERN, EHS, NA22 & 1991 & 250\,GeV/$c$ & $k_T < 40\,\mathrm{MeV}/c$ & $6.4 \pm 1.6$ & bubble chamber (RCBC) & Botterweck~et~al. \cite{EHSNA22:1991sdp} \\
$\text{K}^+\text{p}$ \linebreak CERN, EHS, NA22 & 1991 & 250\,GeV/$c$ & $k_T < 40\,\mathrm{MeV}/c$ & $6.9 \pm 1.3$ & bubble chamber (RCBC) & Botterweck~et~al. \cite{EHSNA22:1991sdp} \\
$\pi^-\text{p}$ \linebreak CERN, WA83, OMEGA & 1993 & 280\,GeV/$c$ & $k_T < 10\,\mathrm{MeV}/c$ \newline ($0.2 < E_\gamma < 1\,\mathrm{GeV}$) & $7.9 \pm 1.4$ & calorimeter & Banerjee et al. \cite{SOPHIEWA83:1992czx} \\
p--Be & 1993 & 450\,GeV/$c$ & $k_T < 20\,\mathrm{MeV}/c$ & $< 2$ & pair conversion, calorimeter & Antos et al. \cite{Antos:1993wv} \\
p--Be, p--W & 1996 & 18\,GeV/$c$ & $k_T < 50\,\mathrm{MeV}/c$ & $< 2.65$ & calorimeter & Tincknell et al. \cite{Tincknell:1996ks} \\
$\pi^-\text{p}$ \linebreak CERN, WA91, OMEGA & 1997 & 280\,GeV/$c$ & $k_T < 20\,\mathrm{MeV}/c$ \newline ($0.2 < E_\gamma < 1\,\mathrm{GeV}$) & $7.8 \pm 1.5 $ & pair conversion & Belogianni at al. \cite{WA91:1997cnv} \\
$\pi^-\text{p}$ \linebreak CERN, WA91, OMEGA & 2002 & 280\,GeV/$c$ & $k_T < 20\,\mathrm{MeV}/c$ \newline ($0.2 < E_\gamma < 1\,\mathrm{GeV}$) & $5.3 \pm 1.0 $ & pair conversion & Belogianni at al. \cite{Belogianni:2002ib} \\
$\text{pp}$ \linebreak CERN, WA102, OMEGA & 2002 & 450\,GeV/$c$ & $k_T < 20\,\mathrm{MeV}/c$ \newline ($0.2 < E_\gamma < 1\,\mathrm{GeV}$) & $4.1 \pm 0.8 $ & pair conversion & Belogianni at al. \cite{Belogianni:2002ic} \\
$\text{e}^+ \text{e}^- \to n \, \mathrm{jets}$ \linebreak CERN, DELPHI & 2006 & 91\,GeV ($\sqrt{s}$)& $k_T < 80\,\mathrm{MeV}/c$ \newline ($0.2 < E_\gamma < 1\,\mathrm{GeV}$) & $4.0 \pm 0.3 \pm 1.0 $ & pair conversion & DELPHI \cite{DELPHI:2005yew,DELPHI:2010cit} \\
$\text{e}^+ \text{e}^- \to \mu^+ \mu^-$ \linebreak CERN, DELPHI & 2008 & 91\,GeV ($\sqrt{s}$)& $k_T < 80\,\mathrm{MeV}/c$ \newline ($0.2 < E_\gamma < 1\,\mathrm{GeV}$) & $\sim 1$ & pair conversion & DELPHI \cite{DELPHI:2007nmh}
\end{tabularx}}
\caption{\label{tab:soft_photon_measurements} Overview of soft-photon measurements. The column ``$\gamma_\mathrm{meas}/\gamma_\mathrm{brems}$'' shows the ratio of the measured photon yield over the inner-bremsstrahlung yield calculated based on Low's theorem. The column ``$p_\mathrm{beam}$ or $\sqrt{s}$'' lists the momentum of the beam for fixed-targets experiments and the center-of-mass energy for the DELPHI measurements.}
\label{tab:soft_photon_experiments}
\end{table}

\subsubsection{Attempts from theory}
Various attempts were made to explain the observed excess of
soft photons, see, e.g.,~\cite{Barshay:1989yd,SHURYAK:1989175,
  LICHARD:1990605, Lichard:1994qt, Czyz:1993ti, Botz:1994bg,
  Hatta:2010kt, Wong:PhysRevC.81.064903,
  Kharzeev:PhysRevD.89.074053}. Most models introduce an additional
source of soft-photon
production.
The authors of the latter references point out that the DELPHI results suggest an additional mechanism for soft-photon production due to non-perturbative QCD evolution and the dynamics of hadronization. However, while the proposed processes may be relevant when approaching the Low region they generally vanish as $k_{\text{T}} \rightarrow 0$.

Presently, no agreement exists on the possible origin of the observed excess in soft-photon-production. Clearly, this  calls for a new effort in measuring ultra-soft-photon production in hadronic collisions in a systematic way.

\paragraph{Paper by Barshay} In Ref.\ \cite{Barshay:1989yd}, S.\ Barshay
argues that the approach to a transient phase of coherent pion
condensation could, through the modification of the dispersion relation
for pions in such a phase, induce an enhanced production of photons in
the low-momentum regime.

\paragraph{Papers by Shuryak} In refs.\ \cite{SHURYAK:1989175,
  Shuryak:1990ie}, E.\ Shuryak discusses a model for a ``pion liquid''
that could form in hadronic collisions. Here, charged pions undergo
several random collisions inside the medium. For photon frequencies that
are small compared to the lifetime of this liquid the usual Low-theorem
result is recovered while an enhancement is observed at somewhat larger
frequencies where the in-medium collisions can be resolved.

\paragraph{Papers by Lichard and Van Hove} In Ref.\
\cite{LICHARD:1990605}, P.\ Lichard and L.\ Van Hove discuss a model
that goes back to earlier work of Van Hove \cite{VanHove:1988qt} where
photons are emitted from supposedly rather long-lived ``globs'' of very
soft partons. As they argue, such blobs of ``cold quark-gluon plasma''
could need much time to hadronize, and, during this hadronization
process, could produce photons at intermediate frequencies. A more
detailed analysis of this model is presented together with a discussion
of experimental data in a later paper \cite{Lichard:1994qt}.

\paragraph{Paper by Czy{\.z} and Florkowski} In Ref.\ \cite{Czyz:1993ti},
the authors investigate photon production from a boost-invariant model
for fragmenting color-flux tubes described as classical color strings
between quarks that are themselves created by Schwinger pair production
from the color fields. Since this fragmentation dynamics extends in
time and space, one finds an enhancement over the leading Low formula at
intermediate photon frequencies.

\paragraph{Paper by Botz, Haberl and Nachtmann} In Ref.\
\cite{Botz:1994bg}, the authors propose synchroton radiation of quarks
in hadron-hadron collisions. When quarks and antiquarks in a hadron move
through the QCD vacuum, they are under the influence of chromoelectric
and chromomagnetic vacuum fields. The chromomagnetic Lorentz force
causes deflections of the electrically charged quarks and antiquarks
leading to the emission of synchroton radiation with photon energies
below 300 MeV in the center-of-mass system of the collision in addition
of hadron bremsstrahlung.

\paragraph{Paper by Hatta and Ueda} In Ref.\ \cite{Hatta:2010kt}, the
authors determine an analog of the inclusive cross section for photon
production in electron-positron annihilation in $\mathcal{N}=4$
supersymmetric Yang-Mills theory. The calculation is done both
perturbatively and, at leading order in the large-$N$ limit, from
holography. They find a contribution with similar kinematic dependence
as bremsstrahlung photons, and speculate that this might help to resolve
the soft-photon puzzle. Since the particle content of $\mathcal{N}=4$
supersymmetric Yang-Mills theory differs substantially from the real world (it has many massless charged particles and is non-confining), it is difficult to judge whether the found mechanism is universal enough or not.

\paragraph{Paper by Wong} In Ref.\ \cite{Wong:PhysRevC.81.064903}, Wong
constructs a two-dimensional model with the gauge group
$\text{SU}(3) \times \text{U}(1)$, and finds a way to map it to a scalar
quantum field theory which contains mesons, but also a massive
photon. He then argues that the non-equilibrium dynamics of this model,
which features hadronization through meson production in its strong
sector, can describe anomalous soft-photon-production.

\paragraph{Paper by Kharzeev and Loshaj} In Ref.\
\cite{Kharzeev:PhysRevD.89.074053}, D.\ E.\ Kharzeev and F.\ Loshaj
describe a mechanism for photon production from the oscillating
electromagnetic currents that arise during the process of string
fragmentation. When a pair of highly energetic quarks separates, the QCD
string between them pulls additional quark-antiquark pairs from the
vacuum which screen the original color fields but also generate
additional ones, leading to a multistaged process of particle
production. This results in oscillating electromagnetic currents, which
in turn act as sources for electromagnetic fields or
photons.
The constructed model leads to an enhanced production of photons at
intermediate frequencies with the energy scale set by the QCD string
tension, and the width of a neutral isoscalar resonance. The latter is
taken as a fit parameter by Kharzeev and Loshaj, and they arrive thus at
a model which describes the photon yields found by the DELPHI
Collaboration in hadronic events as a function of the jet momentum,
reasonably well.

\paragraph{Attempts from perturbative QCD} Useful information in the calculation of soft-photon spectra has been brought by analyses in perturbative QCD with massless quarks. Specifically, Ma, Sterman, and Venkata \cite{Ma:2023gir} computed a three-loop correction of order $\alpha_s^3/\omega$ to the leading power form of Low's theorem. Similarly, other authors \cite{Laddha:2018myi, Bonocore:2021cbv} pointed out that the next-to-leading power term in the soft expansion receives a one-loop correction which is logarithmic in the soft momentum (i.e., schematically of order $\alpha_s\log(\omega)$), and thus might be enhanced for small momenta.
However, it still remains an open question how these perturbative results can be linked to the non-perturbative regime.

\paragraph{Conclusion from different theory attempts}
Overall, none of the proposed approaches can convincingly explain the experimentally observed excess of soft photon production.
However,
this soft-photon puzzle might hint at non-perturbative QCD physics, specifically hadronization. When the duration
$\Delta\tau$ of the collision, i.e., the time between the moment of
collision and the formation of final-state momenta is prolonged by the
hadronization dynamics, the effective expansion parameter of the soft
expansion $\omega \Delta\tau$ increases which implies that the strict
regime of validity of the soft expansion is shifted to much smaller photon
momenta. Accordingly, if the considered momentum range is kept fixed,
subleading terms in the expansion might become sizeable, which would lead to an
apparent deviation from the soft theorem.
Although the contribution of these
subleading terms near the Low divergence should be negligible, it would be interesting to measure them. Such an estimate would give access to effects that so far have not been quantified in soft photon spectra, such as the spin and the recoil of the charge emitter, the role of collinear effects and the dependence of the soft spectrum on the strong coupling constant.

\subsection{Soft radiation and soft theorems}

\subsubsection{\label{sec:leading_soft_theorem_classical_radiation}Low's leading soft theorem: classical bremsstrahlung}

The phenomenology of soft-photon radiation is closely related to the
problem of infrared divergences in QED and to the long-range nature of
the electromagnetic interaction, \ie, the vanishing mass of the photon
as dictated by the underlying Abelian gauge symmetry. In
contradistinction to the problem of ultraviolet (UV) divergences in the
perturbative evaluation of higher-order corrections, the IR problem has
been addressed as early as in 1937 by Bloch and Nordsieck
\cite{Bloch:1937pw} on the level of cross sections for transitions including the emission of arbitrary numbers of low-energy photons. They showed that in the soft photon limit the total probability for an electron transition is unaffected by the emission of an infinite number of low-energy photons while the total energy emitted remains finite, thus recovering the well-known result from classical electrodynamics. The modern interpretation and solution to this IR problem can be attributed to Yennie, Frautschi and Suura \cite{Yennie:1961ad} and to Weinberg \cite{Weinberg:1965nx} who showed that by considering the finite energy resolution of detectors --- where an arbitrary number of ``soft photons'' with energies below this resolution can escape undetected --- the cross sections for \eg, elastic scattering, in addition to radiative corrections to all orders of perturbation theory, are IR finite.

From a more physical point of view, the infrared (IR) problems are due to the long-range nature of the electromagnetic interaction and thus to the
inadequacy of naive Fock charged-particle states as the asymptotic
states of scattering theory. Rather, the true asymptotic state is a
charged particle surrounded by its own electromagnetic field. Formally,
these asymptotic states are given by so-called ``infra-particle
states'', i.e., states describing a charged particle together with a
photon-coherent state. As shown in
\cite{Bloch:1937pw,Kibble1968c,Kibble1968a,Kibble1968b,Kibble1968},
indeed when using the correct infra-particle states with the usual QED
scattering operator, all IR divergences cancel, and the corresponding
cross sections coincide with those obtained in the more conventional
Bloch-Nordsieck approach. For a recent review, including the relation to
asymptotic or large gauge symmetries, see \cite{Gaharia:2019xlh}.

Low's soft theorem \cite{Low:1958sn} states
that the amplitude of emission or absorption of very low-energy quanta
in any given transition with charged particles can always be expressed
as a \textit{soft} factor times the amplitude of the transition without
the emission or absorption of the very low-energy quanta. Low obtained
this result for spin-0 particles using the Ward identity, giving
expressions for the \textit{leading} and \textit{next-to-leading} terms
in the multipole or soft expansion of cross sections in
Eq.~\eqref{eq:multipole_expansion_cross_sections}. The extension for
particles of spin 1/2 is due to Burnett and Kroll
\cite{Burnett:1967km}. The generalization for particles of arbitrary
spin has been given by Bell and Van Royen \cite{Bell:1969yw}. The modern
treatment of Low's theorem is due to Weinberg \cite{Weinberg:1965nx},
who demonstrated the universal behavior of this soft factorization for
all Abelian theories as well as gravity, and showed the exact
cancellation of IR divergences of cross sections between unobserved real
and virtual low-energy bosons in these theories to all orders in
perturbation theory. This section presents a comprehensive introduction
to Low's theorem and the general features of classical soft-photon
radiation.

\begin{figure}[ht]
\centering
\includegraphics[scale=0.6]{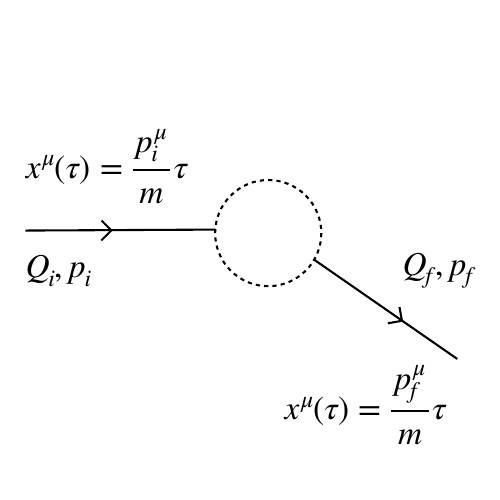}
\caption{A classical depiction of a single-particle transition
  experiment where a charge $Q_i$ is prepared with initial momentum
  $p_i$ and goes out with charge $Q_f=Q_i$ and final momentum $p_f$
  after undergoing an unspecified interaction in a localized region of
  space-time, illustrated by a blob. \label{fig:fig_3_xf}}
\end{figure}

Before introducing Low's theorem, let us consider the classic radiation
scenario of a single charged-particle transition like the one depicted
in Fig.~\ref{fig:fig_3_xf}. There, a particle carrying charge $Q_i$ in units of the electron charge $e$ is
prepared with initial momentum $p_i$. After an interaction, which does
not need to be further specified, it goes out with final momentum $p_f$
and charge $Q_f=Q_i=Q$. The classical photon field induced by this charge
during the process is easily obtained from the solution to the Maxwell
equations, namely
\begin{equation}
\label{eq:a_mu_x_def}
A_\mu(x)= \int d^4y D_{\mu\nu}(x-y) J^\nu(y) = \int\frac{d^4k}{(2\pi)^4} e^{-ik\cdot x} \bigg\{-\frac{g_{\mu\nu}}{k^2+i0^+}+(1-\xi)\frac{k_\mu k_\nu}{(k^2+i0^+)^2} \bigg\}\tilde{J}^\nu(k)
\end{equation}
where $D_{\mu\nu}(x-y)$ is the photon propagator and
$J^\mu(x)\equiv \big(\rho(x),\bm{j}(x)\big)$ the classical current
density induced by the charged particle along its path $x_\mu(\tau)$ in
the transition
\begin{equation}
J^\mu(x)= Qe \int^{\infty}_{-\infty} d\tau \dot{x}^\mu(\tau)\delta^4(x-x(\tau))\,.
\end{equation}
Here $\tau$ is the particle's proper time, and $Q$ its electric charge in units of $e$. In the last step in
Eq.~\eqref{eq:a_mu_x_def} we Fourier transformed to momentum space,
where $k^\mu$ is the momentum of the
  photon, $\mu=0,1,2,3$, $k^2=k_\mu k^\mu$, $\xi$ the gauge fixing
  parameter, and the small imaginary part $i0^+$ ensures that positive
  and negative energy modes of the gauge field propagate forward and
  backward in time, respectively. We are interested in computing the
form of this photon field at low momentum $k$ or, equivalently, at large
distances $x$. At large $x$, the particle follows the classical
straight-line trajectory of Fig.~\ref{fig:fig_3_xf} given by
\begin{equation}
\label{eq:straigth_line_trajectories}
x^\mu(\tau)=\frac{p^\mu_i}{m}\tau \,\,\,\,\, \text{for}\,\,\,\, \tau< 0\,,\,\,\,\, \,\,\,\, x^\mu(\tau)=\frac{p^\mu_f}{m}\tau\,\,\,\,\, \text{for}\,\,\,\, \tau>0\,,
\end{equation}
where $m$ is the mass of the charged particle. Using
Eq.~\eqref{eq:straigth_line_trajectories}, the current at large $x$ can
now be easily obtained as
\begin{equation}
J^\mu(x) \simeq \int^{\infty}_{0} d\tau \bigg[Q_fe\frac{p^\mu_f}{m} \bigg] \delta^4 \Big(x-\frac{p_f}{m}\tau\Big) +\int^{0}_{-\infty} d\tau \bigg[Q_ie\frac{p^\mu_i}{m} \bigg] \delta^4 \Big(x-\frac{p_i}{m}\tau\Big)\,,
\end{equation}
and its Fourier transform then reads
\begin{equation}
\label{eq:j_mu_k}
\begin{split}
\tilde{J}^\mu(k)
&= \int d^4x e^{ik\cdot x} J^\mu(x) = \int^\infty_{0} d\tau \bigg[Q_fe\frac{p^\mu_f}{m}\bigg] e^{i\frac{k\cdot p_f}{m}\tau}+\int^0_{-\infty}d\tau \bigg[Q_ie\frac{p^\mu_i}{m}\bigg] e^{i\frac{k\cdot p_i}{m}\tau} \\
&= ie \bigg\{Q_f\frac{p^\mu_f}{k\cdot p_f}-Q_i\frac{p_i^\mu}{k\cdot p_i}\bigg\}\,,
\end{split}
\end{equation}
where in the last step we integrated in $\tau$ and neglected the
boundary terms at plus and minus infinity. When substituting now
Eq.~\eqref{eq:j_mu_k} into Eq.~\eqref{eq:a_mu_x_def} we see that the
gauge-dependent terms --- those proportional to $k_\mu k_\nu$ within the
photon propagator in Eq.~\eqref{eq:a_mu_x_def} --- will vanish due to
charge conservation $\partial_\mu J^\mu(x) = 0$, namely they will give
rise to terms
\begin{equation}
\label{eq:k_mu_j_mu_null}
k_\mu \tilde{J}^\mu(k) =ie( Q_f-Q_i)=0\,.
\end{equation}
Thus, plugging Eq.~\eqref{eq:j_mu_k} into Eq.~\eqref{eq:a_mu_x_def} and
setting $Q_f=Q_i=Q$ yields
\begin{equation}
A_\mu(x)=iQe\int\frac{d^4k}{(2\pi)^4} e^{-ik\cdot x} \frac{-g_{\mu\nu}} {k^2+i0^+} \bigg\{\frac{p^\nu_f}{k\cdot p_f}-\frac{p_i^\nu}{k\cdot p_i}\bigg\}\,.
\end{equation}
The integral in $k_0$ in this expression can easily be performed by using complex analysis.
Defining the $k_0$ integration contour for $(t>0)$ in the lower half
complex plane, enclosing the $k_0=\omega_{{k}} -i0^+$ pole, where
$\omega_{{k}}=|\bm{k}|$ is the frequency of an on-shell photon, simply
gives\footnote{We neglected the absorptive mode of the photon field for
  $t<0$ corresponding to the $k_0=-\omega_{{k}}+i\varepsilon$ pole, and
  the two Lienard-Wiechert fields at large distances, corresponding to
  the Coulomb field always present in the presence of the charge, as we
  are interested in computing the probability of emitting an on-shell
  photon.}
\begin{equation}
\label{eq:a_mu_x_result1}
A_\mu(x)= Qe\int\frac{d^3\bm{k}}{(2\pi)^3} e^{-i\omega_kt+i\bm{k}\cdot \bm{x}} \frac{-g_{\mu\nu}} {2\omega_k} \bigg\{\frac{p_f^\nu}{k\cdot p_f}-\frac{p_i^\nu}{k\cdot p_i}\bigg\}\,,\,\,\,\, \text{for}\,\,\,\, t>0\,.
\end{equation}
It remains to determine the amplitude of finding within this field an
on-shell photon mode of momentum $k=(\omega_{k},\bm{k})$ and
polarization $\lambda$. This can easily be done by noticing that the sum
over physical photon polarizations can be expressed as
\begin{equation}
\label{eq:pol_sum}
\sum_{\lambda=1,2} \varepsilon_\mu(\bm{k},\lambda)\varepsilon_\nu^*(\bm{k},\lambda)=-g_{\mu\nu}-\frac{k_\mu k_\nu}{(k\cdot n)^2}+\frac{k_\mu n_\nu+k_\nu n_\mu}{(k\cdot n)}\,,
\end{equation}
where $n_\mu=(1,\bm{0})$ is a normalized time-like vector. Substituting
then $-g_{\mu\nu}$ in Eq.~\eqref{eq:a_mu_x_result1} with
Eq.~\eqref{eq:pol_sum} and noticing once again that the $k_\mu k_\nu$,
$k_\mu\eta_\nu$ and $k_\nu\eta_\mu$ terms can be neglected
because they vanish due to charge conservation, one can rewrite
Eq.~\eqref{eq:a_mu_x_result1} as
\begin{equation}
A_\mu(x)=\int \frac{d^3\bm{k}}{(2\pi)^3}\frac{1}{\sqrt{2\omega_k}}\epsilon_\mu(\bm{k},\lambda)e^{-i\omega_kt+i\bm{k}\cdot\bm{x}} M(p_f;k,\lambda\,| \,p_i)\,,
\end{equation}
which is an infinite superposition of normalized free plane waves of a photon
with on-shell momentum $k=(\omega_k,\bm{k})$ and polarization $\lambda$,
\begin{equation}
A_\mu^{\text{free}}(x)=\frac{1}{\sqrt{2\omega_k}} \epsilon_\mu(\bm{k},\lambda)e^{-i\omega_kt+i\bm{k}\cdot\bm{x}}\,,
\end{equation}
with amplitude denoted $M(p_f;k,\lambda\,|\, p_i)$
given by
\begin{equation}
\label{eq:M_fi_1_classical}
M(p_f;k,\lambda\,|\, p_i)=\frac{Qe}{\sqrt{2\omega_k}} \epsilon_\mu^*(\bm{k},\lambda) \bigg\{\frac{ p_f^\mu}{k\cdot p_f}-\frac{p_i^\mu}{k\cdot p_i}\bigg\}\,.
\end{equation}
This is the well-known result from classical electrodynamics for the
amplitude of emitting a soft photon with momentum
$k=(\omega_{k},\bm{k})$ and polarization $\lambda$ from an accelerated
charge that goes from initial momentum $p_i$ to final momentum $p_f$ due
to some interaction.

\begin{figure}[ht]
\centering
\includegraphics[scale=0.6]{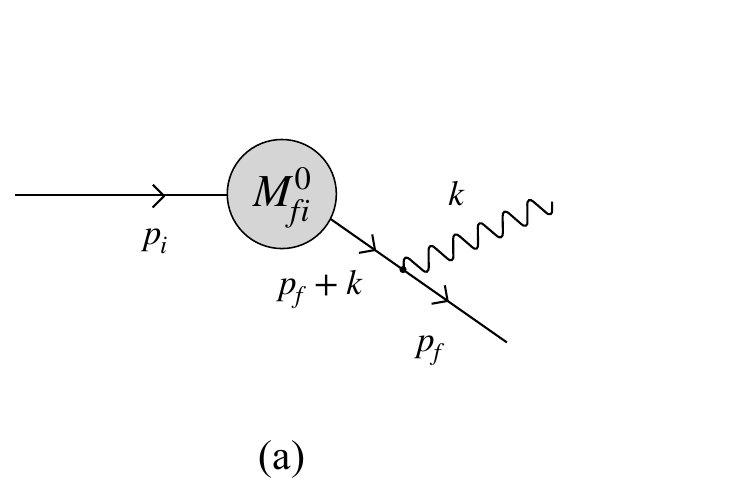}
\includegraphics[scale=0.57]{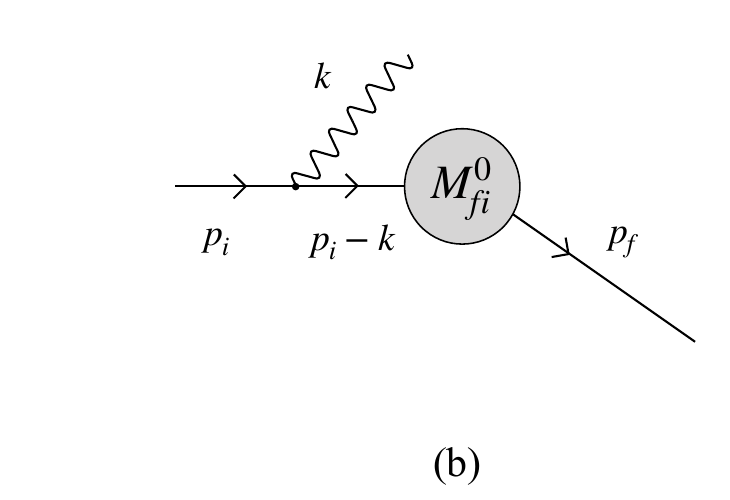}
\caption{Feynman diagrams for the emission of a photon from a single
  charged-particle transition. The initial state is prepared with a
  particle of charge $Q_i$ with momentum $p_i$ and spin $s_i$. An
  arbitrary interaction is denoted with a blob with amplitude $M_{fi}^0$
  in the absence of emission of a soft-photon. The experiment measures
  the final momentum $p_f$, spin $s_f$ and charge $Q_f$ of the particle
  after the interaction, and the momentum and polarization of the
  soft-photon emitted.}
\label{fig:fig_54_xf}
\end{figure}

Let us compare now this classical result with the full prediction from
quantum electrodynamics, and introduce to this end Low's theorem. In the
following, we summarize the treatment by Weinberg \cite{Weinberg1995}. The Feynman
diagrams for the soft-photon emission in a single charged-particle
transition are depicted in Fig.~\ref{fig:fig_54_xf}. For concreteness, we
analyze the first diagram (a) in Fig. \ref{fig:fig_54_xf}, where the
photon is emitted after the interaction, assuming the charged particle
to be a spin-1/2 particle with final charge $Q_f$, momentum $p_f$, spin
$s_f$ and mass $m$.

According to the usual Feynman rules, the diagram (a) of
Fig.~\ref{fig:fig_54_xf} requires changing the spinor $\bar{u}(p_f,s_f)$ for the outgoing charged fermion to
\begin{equation}
\label{eq:hvh.1}
Q_fe\bar{u}(p_f,s_f)  (-i) \gamma^{\mu} \frac{i
  (\slashed{p}_f+\slashed{k}+m)}{(p_f+k)^2-m^2+i 0^+}
\frac{\epsilon_{\mu}^*(k,\lambda)}{\sqrt{2\omega_k}} .
\end{equation}
For
$k=(\omega_k,\bm{k})$ with $\omega_k=|\bm{k}| \rightarrow 0$ we can
neglect $\slashed{k}$ in the numerator and use
\begin{equation}
\bar{u}(p_f,s_f)\gamma^{\mu}
(\slashed{p}_f+m)=\bar{u}(p_f,s_f) \big[\{\gamma^{\mu},\slashed{p}_f \}-(\slashed{p}_f-m)\gamma^{\mu}\big]
=\bar{u}(p_f,s_f) 2 p_f^\mu.
\end{equation}
Further, using the on-shell
conditions $p_f^2=m^2$ and $k^2=0$, we finally get
\begin{equation}
\label{eq:hvh.2}
M_{a}(p_f,s_f;k,\lambda|p_i,s_i)=\frac{Q_fe}{\sqrt{2\omega_k}} \frac{\epsilon^*(\bm{k},\lambda) \cdot p_f}{p_f \cdot k} M_{fi}^0(p_f,s_f;p_i,s_i)+\mathcal{O}(\omega_k^0).
\end{equation}
This shows that in the limit $|\bm{k}| \rightarrow 0$ for the soft
photon emitted from the \emph{outgoing} charged-particle line, the
amplitude is given by the product of the amplitude for the process
without the emission of this soft photon multiplied with a factor which
is of order $\mathcal{O}(1/\omega_k)$. If the soft photon is emitted
from the \emph{incoming} charged particle line instead like in (b) of
Fig.~\ref{fig:fig_54_xf}, in the corresponding fermion propagator we
will have $p_i-k$ and thus the sign changes compared to
Eq.~\eqref{eq:hvh.2} which results in
\begin{equation}
\label{eq:hvh.3}
M_{b}(p_f,s_f;k,\lambda|p_i,s_i)=-\frac{Q_ie}{\sqrt{2\omega_k}} \frac{\epsilon^*(\bm{k},\lambda) \cdot p_i}{p_i \cdot k} M_{fi}^0(p_f,s_f;p_i,s_i)+\mathcal{O}(\omega_k^0).
\end{equation}
If the soft photon is radiated from an inner charged-particle line of a
diagram, i.e., from any internal line within the blob in
Fig.~\ref{fig:fig_54_xf}, there is no divergence for
$\omega_k \rightarrow 0$, and thus for the \emph{leading-order
  soft-photon amplitude} $\mathcal{O}(1/\omega_k)$ the amplitude is
given by the sum of diagrams (a) and (b) of Fig.~\ref{fig:fig_54_xf} in
Eqs.~\eqref{eq:hvh.2} and \eqref{eq:hvh.3},
\begin{equation}
\label{eq:M_fi_1_quantum}
M(p_f,s_f;k,\lambda\,|\,p_i,s_i)=M_{a}+M_{b}=\frac{Qe}{\sqrt{2\omega_k}} \epsilon_\mu^*(\bm{k},\lambda)\bigg\{\frac{p_f^\mu}{k\cdot p_f}-\frac{p_i^\mu}{k\cdot p_i}\bigg\}M_{fi}^0(p_f,s_f\,|\,p_i,s_i)+\mathcal{O}(\omega_k^0),
\end{equation}
where we used $Q_f=Q_i=Q$.

Equation~\eqref{eq:M_fi_1_quantum} is Low's theorem for the special case of the
emission of a single soft photon in a $1\to 1$ transition of a spin-1/2
particle. We see that the soft factor does neither depend on the
particular features of the transition nor on the spin dimensions of the
charged particle, and it completely agrees with the classical amplitude
of soft-photon emission we obtained in Eq.~\eqref{eq:M_fi_1_classical}.
The only difference that appears is the presence of the non-radiative
transition amplitude $M_{fi}^0$, i.e., without the emission of any
photon. Indeed, in a classical calculation there is not such an
amplitude for a charged particle to go from $p_i$ to $p_f$, the final
momentum $p_f$ is completely determined by the knowledge of
the initial condition $p_i$ and the classical equations of motion.

\begin{figure}[ht]
    \centering
    \includegraphics[scale=0.45]{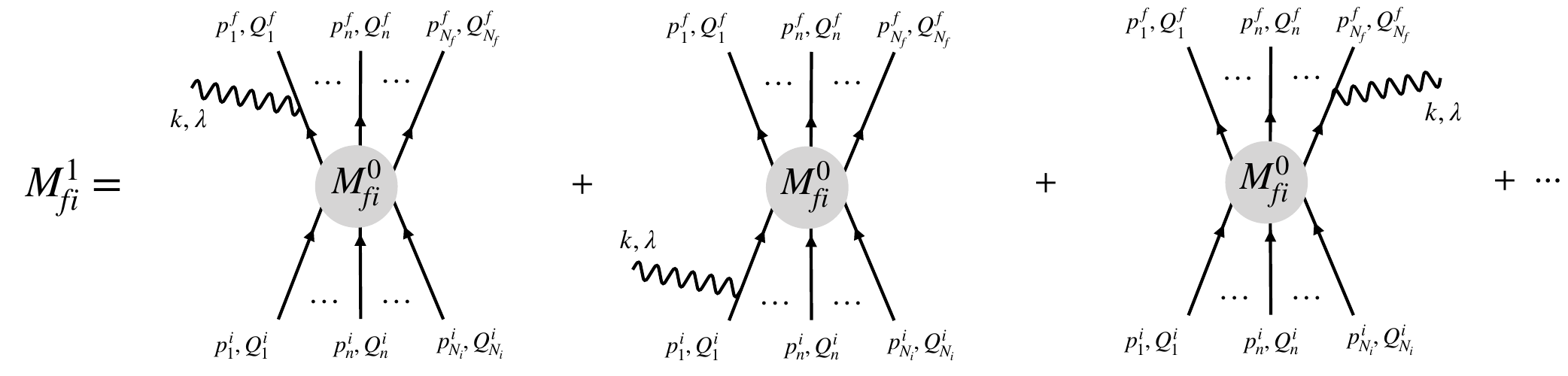}
    \caption{Set of real soft-photon diagrams - one for each external charged particle line of the transition - contributing to the total amplitude $M_{fi}^1$ of emitting a soft real photon of momentum $k$ and polarization $\lambda$ in any given $i\to f$ transition.}
    \label{fig:fig_1_xf}
\end{figure}

The previous calculation can easily be extended to the case in which the
transition involves not one but many charged particles, like the one
depicted in Fig.~\ref{fig:fig_1_xf}. Using the shorthand notation
\begin{equation}
\begin{split}
& M^1_{fi} \equiv M(p_{1}^f,s_1^f,\ldots,p_{N_f}^f,s_{N_f}^f;k,\lambda\,|\,p_1^i,s_1^i,\ldots,p_{N_i}^i,s_{N_i}^i) \,,\\
& M^0_{fi} \equiv M(p_{1}^f,s_1^f,\ldots,p_{N_f}^f,s_{N_f}^f\,|\,p_1^i,s_1^i,\ldots,p_{N_i}^i,s_{N_i}^i)\,,
\end{split}
\end{equation}
for the amplitude of the $i\to f$ transition with and without the
emission of the real soft photon, Low's theorem for arbitrary
transitions reads
\begin{equation}
\label{eq:s_fi_1_k_lp}
 M_{fi}^1 = \frac{e}{\sqrt{2\omega_k}}\sum_{n=1}^{N_i+N_f} \eta_n Q_n
  \frac{\epsilon^*(\bm{k},\lambda) \cdot p_n}{k\cdot p_n}M_{fi}^0 +\mathcal{O}(\omega_k^0)\, ,
\end{equation}
where the sum over $n$ runs over initial and final particles and
$\eta_n=+1$ or $\eta_n=-1$ if the charged particle is outgoing or
incoming, respectively.

The amplitude $M_{fi}^1$ for the production of a soft photon then
factorizes into $M_{fi}^0$, the Feynman amplitude for the production of
charged particles only (non-radiative) and the soft-photon production factor,
\begin{equation}
\label{eq:s_fi_1_k_lp_part}
\frac{e}{\sqrt{2\omega_k}}\sum_{n=1}^{N_i+N_f} \eta_nQ_n
  \frac{\epsilon^*(\bm{k},\lambda) \cdot p_n}{k\cdot p_n}.
\end{equation}
It should be noted that the process can of course be accompanied by the production of neutral particles which are in general not detected.

We remark that, from Eq.\ (\ref{eq:s_fi_1_k_lp}), the soft-photon
production factor arises from the interference of photon production from
incoming and outgoing charged particles. Importantly, one should recognize that in this soft-photon
limit all Feynman diagrams where the soft-photon line is connected to
an internal line corresponding to a virtual charged particle yield a
non-diverging and, therefore, negligible contribution to the soft-photon
production cross section. It is then not necessary to evaluate the
contribution of all possible internal loops to the cross section: Low’s
leading term is ``tree-level correct''.

The photon production amplitude in Eq.~\eqref{eq:s_fi_1_k_lp_part} has a pole whenever $p \cdot k$ vanishes (in any reference frame). In the rest frame of the photon emitting charged particle, this implies that the photon energy must be much smaller than the mass of the charged particle to approach the Low limit.

Generally we have
\begin{equation}
k \cdot p = \omega (E - |\bm{p}| \cos \theta),
\end{equation}
where $E$ is the energy of the charged particle, $|\bm{p}|$ the magnitude of its momentum, and $\theta$ the opening angle between the photon and the charged particle. Here, all quantities are evaluated  either in the frame of the radiating particle (particle frame) or in a frame with z-direction along the direction of the colliding beams (beam frame).

To start with, we consider in the particle frame ultra-relativistic particles ($E \approx |\bm{p}|$) and not too large angles $\theta$. This results in
\begin{equation}
k \cdot p \approx \omega E (1 - \cos \theta) \approx \omega E \frac{\theta^2}{2} = k_T E \frac{\theta}{2},
\end{equation}
explicitly exhibiting the connection between $k \cdot p$ and  the photon transverse momentum $k_T$. More generally, the photon energy $\omega$ can, in the particle frame  or in the beam frame, always be expressed as $\omega = k_T \cosh y_k $ with photon rapidity $y_k$ and transverse momentum $k_T$, both in the respective frame.
Clearly, $k \cdot p$ vanishes for $\omega \rightarrow 0$, leading to the ``Low'' pole. Note that strictly speaking the expression $(E - |\bm{p}| \cos \theta) $ is always positive, since there exist no massless charged particles.
The above considerations demonstrate that the divergence in the photon energy $\omega$ manifests itself as a divergence in the transverse momentum $k_T$.  This was first considered in \cite{Gribov:1966hs} with the conclusion that the applicability of the Low theorem might reach well beyond the region originally investigated by Low.

In collider experiments, the beam frame is used. Then the  Low divergence can be measured as a divergence in the $k_T$ of the photons with respect to the beam axis. This is the approach that will be taken in the ALICE 3 project. The first steps in ALICE 3 will be to test for inclusive and exclusive reactions the leading order predictions based on Low's original result with precision measurements of the associated charged particle distributions. The obvious next steps will be to extend with ALICE 3 the measurements into higher photon $k_T$ regions until $\pi^0$ decay background becomes prohibitive.

The generalization of Low's theorem in Eq.~\eqref{eq:s_fi_1_k_lp} to the
emission of many very low-energy photons of momentum $\{k_r\}$ and
polarization $\{\lambda_r\}$, with $r=1,\ldots,N_\gamma$, is given by
multiplying the amplitude of the transition $i\to f$ by $N_\gamma$ soft
factors like the one in Eq.~\eqref{eq:s_fi_1_k_lp}, one for each
low-energy photon accompanying the transition,
\begin{equation}
\label{eq:s_fi_n_k_lp}
M^{N_\gamma}_{fi} =  \prod_{r=1}^{N_\gamma}\bigg\{\frac{\epsilon_\mu^*(\bm{k}_r,\lambda_r)}{\sqrt{2\omega_k}}e\sum_{n=1}^{N_i+N_f}\eta_nQ_n\frac{ p_n^\mu}{k_r\cdot p_n}\bigg\}M^0_{fi}+\mathcal{O}(\omega_k^0)\, ,
\end{equation}
where we analogously abbreviated the radiative amplitude as
\begin{equation}
M^{N_\gamma}_{fi} \equiv M(p_{1}^f,s_1^f,\ldots,p_{N_f}^f,s_{N_f}^f,k_{1},\lambda_1,\ldots,k_{N_\gamma},\lambda_{N_\gamma}\,|\,p_1^i,s_1^i,\ldots,p_{N_i}^i,s_{N_i}^i)\,.
\end{equation}

Equation~\eqref{eq:s_fi_n_k_lp} describes the emission of $N_\gamma$
very low-energy photons as mutually independent processes
and also insensitive to the particular details of the
interactions described by the non-radiative transition amplitude
$M^0_{fi}$. The \textit{soft} factor in Eq.~\eqref{eq:s_fi_1_k_lp}
continues to exhibit the same features as the classical radiation
induced by a system of $N_i$ charges coming in from the far past in
uniform motion with momenta $\{p^i_n\}$, interacting then in some
complicated way that does not need to be specified to go into a system
of $N_f$ charges going out to the far future in uniform motion as well
with shifted momenta $\{p^f_n\}$. Note that charge is conserved, but
$N_f$ can be different from $N_i$. The \textit{leading} term in Low's
soft theorem then just reflects the fact that photons with very low
momenta $k \to 0$ radiated during the transition are sensitive only to
very large regions of the space-time $x^\mu$, $k_\mu x^\mu\sim 1$, such
that the transition from the initial to the final state becomes for them
effectively instantaneous.

The relevant parameter for the soft approximation is the dimensionless
quantity $\omega_T/\Lambda$ where $\omega_T$ is the total energy
radiated in the form of very low-energy photons during the transition
and $\Lambda$ the infrared (IR) region of soft photons
in which the leading term of the soft expansion in
Eq.~\eqref{eq:s_fi_n_k_lp} is strictly valid. To estimate $\Lambda$ we
note that in deriving Eq.~\eqref{eq:s_fi_n_k_lp} we assumed that the
total energy $\omega_T$ emitted through the $N_\gamma$ soft photons
shall be negligible compared with the energy of the charged
particles in the initial and final states,
\begin{equation}
\omega_T=\omega_1+\ldots+\omega_{N_\gamma} \ll E_n^{f,i}\,.
\end{equation}
According to Weinberg \cite{Weinberg:1965nx}, a good strategy is to set
$\Lambda$ equal to $m$, the typical mass of the external charged
particles and define the IR region of soft photons as those satisfying
\begin{equation}
\label{eq:omega_Lambda_m}
\frac{\omega_T}{\Lambda}< 1\,,\,\,\,\,\, \text{with}\,\,\,\,\,\Lambda \simeq m\,.
\end{equation}
However, while the condition in Eq.~\eqref{eq:omega_Lambda_m} is
necessary to take the soft-photon limit in any of the external
charged-particle lines of the transition it might not be sufficient as
the previous calculation ignored soft photons emitted from the inner
charged particle lines. Eventually if the typical time extent $\Delta \tau$ of the
interactions is sufficiently large --- like in processes involving
high-multiplicity events long-lived virtual charged particles and/or
interactions with extended matter\footnote{The limit of applicability of
  Low's theorem in the Bethe-Heitler cross section in the presence of
  matter is the well-known Landau-Pomeranchuk-Migdal (LPM) effect
  \cite{Landau:1953um,Migdal:1956tc}. The IR region $\Lambda\simeq 1/\Delta \tau$ in this
  case can be of the order of a few MeV, as measured by the SLAC E-146
  collaboration \cite{SLAC-E-146:1997hnd}. The QCD analog of the LPM
  effect plays a fundamental role in describing the physics of jets in
  ultra-relativistic heavy-ion collisions.} --- the corresponding IR scale
$\Lambda\simeq 1/\Delta \tau$ for which internal line emissions can be safely ignored can be
in fact much smaller than the condition in
Eq.~\eqref{eq:omega_Lambda_m}. As a consequence, the choice of $\Lambda$
will generally be process dependent. For instance, if the transition involves the production of an intermediate 10 GeV $\mathrm{J}/\psi$ --- with typical decay width of $90~\mathrm{keV}$ and mean life-time of $7 \, \mathrm{pm}/c$ in the center of mass frame --- the IR region of soft photons emitted by the charged pair decays will be of the order $\omega \lesssim \Lambda = 1/\Delta \tau \sim 30$ keV.

We turn now to analyze the general features of the observed soft-photon
radiation, considering for simplicity the emission of a single soft
photon in Eq.~\eqref{eq:s_fi_1_k_lp}. The generalization to many soft
photons is straightforward. The probability density $dI$ of finding one
very-low energy photon of frequency $\omega<\Lambda$ and momentum $k$ in
a volume $d^3\bvec{k}$ around $\bvec{k}$ accompanying the transition in
Fig.~\ref{fig:fig_1_xf} is given by taking the absolute square of the
leading term in the soft-photon approximation in
Eq.~\eqref{eq:s_fi_1_k_lp} and summing over the two possible photon
polarizations
\begin{equation}
\label{eq:dI_definition}
dI=\frac{d^3\bvec{k}}{(2\pi)^3}\sum_{\lambda=\pm 1}|M_{fi}^{1}|^2 =\frac{d^3\bvec{k}}{(2\pi)^3}\frac{e^2}{2\omega_k}  \sum_{\lambda=\pm 1} \epsilon_\mu^*(\bm{k},\lambda)\epsilon_\nu(\bm{k},\lambda) \sum_{n,m=1}^{N_i+N_f} \eta_n Q_n \eta_m Q_m \frac{p_n^\mu p_m^\nu}{k\cdot p_n k\cdot p_m}|M_{fi}^0|^2\,,
\end{equation}
where $|M_{fi}^0|^2$ is the differential cross section for the
transition without the emission of the very low-energy photon. To discuss the angular distribution of soft radiation we write
$k=(\omega_k,\bm{k})=\omega_k(1,\hat{\bm{k}})$ and
$p_n=(E_n,\bm{p}_n)=E_n(1,\bm{\beta}_n)$ and therefore
\begin{equation}
k\cdot p_n = \omega_k E_n(1-\bm{\beta}_n\cdot \hat{\bm{k}})\,,\,\,\,\, k\cdot p_m = \omega_k E_m(1-\bm{\beta}_m\cdot\hat{\bm{k}})\,,
\end{equation}
where $\beta_{n}=(1,\boldsymbol{\beta}_{n})$ and $\beta_m=(1,\boldsymbol{\beta}_m)$.
We can also re-write the polarization sum in Eq.~\eqref{eq:pol_sum_2} more conveniently as
\begin{equation}
\label{eq:pol_sum_2}
\sum_{\lambda=\pm 1}\epsilon^*_\mu(\boldsymbol{k},\lambda) \epsilon_\nu(\boldsymbol{k},\lambda)p^\mu_n p^\nu_m = \boldsymbol{p}_n\cdot \boldsymbol{p}_m -\frac{(\boldsymbol{k}\cdot\boldsymbol{p}_n)(\boldsymbol{k}\cdot \boldsymbol{p}_m)}{\omega^2}=E_nE_m\big(\boldsymbol{\beta}_n \times \hat{\boldsymbol{k}}\big)\cdot \big(\boldsymbol{\beta}_m \times \hat{\boldsymbol{k}}\big)\,.
\end{equation}
Noticing
$d^3\bm{k}=\omega_k^2d\omega_kd\Omega_{\bvec{k}}$, where $\Omega_{\bvec{k}}$ is
the soft-photon solid angle, this leads to
\begin{equation}
\label{eq:dI_definition_2}
dI=\alpha \frac{d\omega_k}{\omega_k}\frac{d\Omega_{\bvec{k}}}{(2\pi)^2} \sum_{n,m=1}^{N_i+N_f} \eta_n Q_n \eta_m Q_m \frac{\big(\boldsymbol{\beta}_n \times \hat{\boldsymbol{k}}\big)\cdot \big(\boldsymbol{\beta}_m \times \hat{\boldsymbol{k}}\big)}{(1-\bm{\beta}_n\cdot \hat{\bm{k}})(1-\bm{\beta}_m\cdot \hat{\bm{k}})}|M_{fi}^0|^2\,,
\end{equation}
where $\alpha=e^2/4\pi$. Due to the $d\omega_k/\omega_k$ term, the differential spectrum is
logarithmically divergent in the infrared limit when $\omega_k\to 0$. In
addition, the angular distribution of radiation has peaks orthogonal to
the direction of the charged particles in the far past and far future if
these are non-relativistic $|\bvec{\beta}_n|\to 0$, or mostly forward
along a small cone of angle $\theta_n\sim 1/\gamma_n$, where $\theta_n$ is the angle between $\bm{k}$ and $\bm{\beta}_n$, if these are
relativistic $|\bvec{\beta}_n|\to 1$. A direct consequence of the particular form of the angular distribution
in Eq.~\eqref{eq:dI_definition_2} is the emergence of collinear divergences
when the emitting particle is ultrarelativistic and the \textit{dead cone} of
soft-photon radiation when the emitting particle is massive. In order to
see these two related phenomena, we assume for simplicity that all charged particles in the transition go in different directions, and therefore their radiation is resolved angularly for each particle. We focus on the angular distribution of soft photon radiation around a single
charged particle in Eq.~\eqref{eq:dI_definition_2},
with charge $Q$, mass $m$ and energy $E$ and then
\begin{equation}
\label{eq:dI_angular_onecharge}
dI = \alpha Q^2 \frac{d\omega_k}{\omega_k}  \frac{d\Omega_{k}}{(2\pi)^2}  \frac{|\bm{\beta}|^2\sin^2\theta}{(1-|\boldsymbol{\beta}|\cos\theta)^2}|M_{fi}^0|^2\,,
\end{equation}
where $\theta$ is the angle between the charged particle and the soft photon. If a charged particle would be propagating along the light cone, i.e.,
with $m=0$ and $|\bvec{\beta}|=1$, in addition to the IR divergence
$1/\omega_k$, another divergence from the collinear emission at
$\theta=0$ would occur, in which soft photons are emitted in the same
exact direction as the massless charged particle. In contrast, if the
charged particle is massive, there is instead a suppression of
soft-photon radiation around $\theta=0$, commonly called the ``dead
cone''. The width of this dead cone can be obtained from the denominator
in Eq.~\eqref{eq:dI_angular_onecharge} that can be rewritten as
\begin{equation}
\label{eq:1-beta}
1-|\boldsymbol{\beta}|\cos\theta=(1-\cos\theta)+(1-|\boldsymbol{\beta}|)\cos\theta\,,
\end{equation}
with the first term corresponding to the massless charged-particle
case, with $|\bvec{\beta}|= 1$, and the second to a mass dependent
correction. Assuming the massive charged particle is relativistic, then
$\gamma\gg 1$ and $\theta\ll 1$, and we can expand
\begin{equation}
|\bm{\beta}|=\sqrt{1-\frac{1}{\gamma^2}}=1-\frac{1}{2\gamma^2}+\cdots\,,\,\,\, \cos\theta=1-\theta^2/2+\cdots\,,\,\,\, \sin\theta=\theta+\cdots,
\end{equation}
so that Eq.~\eqref{eq:dI_angular_onecharge} becomes
\begin{equation}
dI =  \alpha Q^2 \frac{d\omega_k}{\omega_k}   \frac{d\theta}{2\pi} \frac{4\theta^3}{\big[\theta^2+\theta_0^2\big]^2}|M_{fi}^0|^2,
\end{equation}
with $\theta_0=1/\gamma$. This well-known suppression of small-angle
soft radiation in the \textit{dead cone}, $\theta \lesssim \theta_0$, has
a number of phenomenological implications in heavy-ion collisions and
finds immediate application in the perturbative calculations of
heavy-quark fragmentation functions \cite{Dokshitzer:2001zm}. From
the previous equation, small-angle soft radiation reaches a maximum when
\begin{equation}
\label{eq:dI_angular-maximum}
\frac{d}{d\theta} \frac{\theta^3}{(\theta^2+\theta_0^2)^2}\bigg|_{\theta=\theta_\text{max}}=  \frac{\theta^2_\text{max}(3\theta_0^2-\theta^2_\text{max})}{(\theta^2_\text{max}+\theta_0^2)^3}=0\,,
\end{equation}
this is, at an angle $\theta_\text{max}=\sqrt{3}\theta_0=\sqrt{3}/\gamma$
with respect to the emitting ultra-relativistic charged particle. The heavy-to-light ratio of soft radiation off an ultra-relativistic heavy
particle (HP) and off a massless particle (LP) is often expressed as
\begin{equation}
\label{eq:deadcone_ratio_def}
\frac{dI_\text{HP}}{dI_\text{LP}}=\frac{\bm{\beta}^2\sin^2\theta}{\big[\big(1-\cos\theta) + (1-|\bm{\beta}|)\cos\theta \big]^2}\frac{(1-\cos\theta)^2}{\sin^2\theta}\simeq  \bigg\{1+\frac{1}{\gamma^2\theta^2}\bigg\}^{-2}=\bigg\{1+\frac{\theta_0^2}{\theta^2}\bigg\}^{-2}.
\end{equation}

\begin{figure}[htb]
  \centering
  \includegraphics[width = 0.5\textwidth]{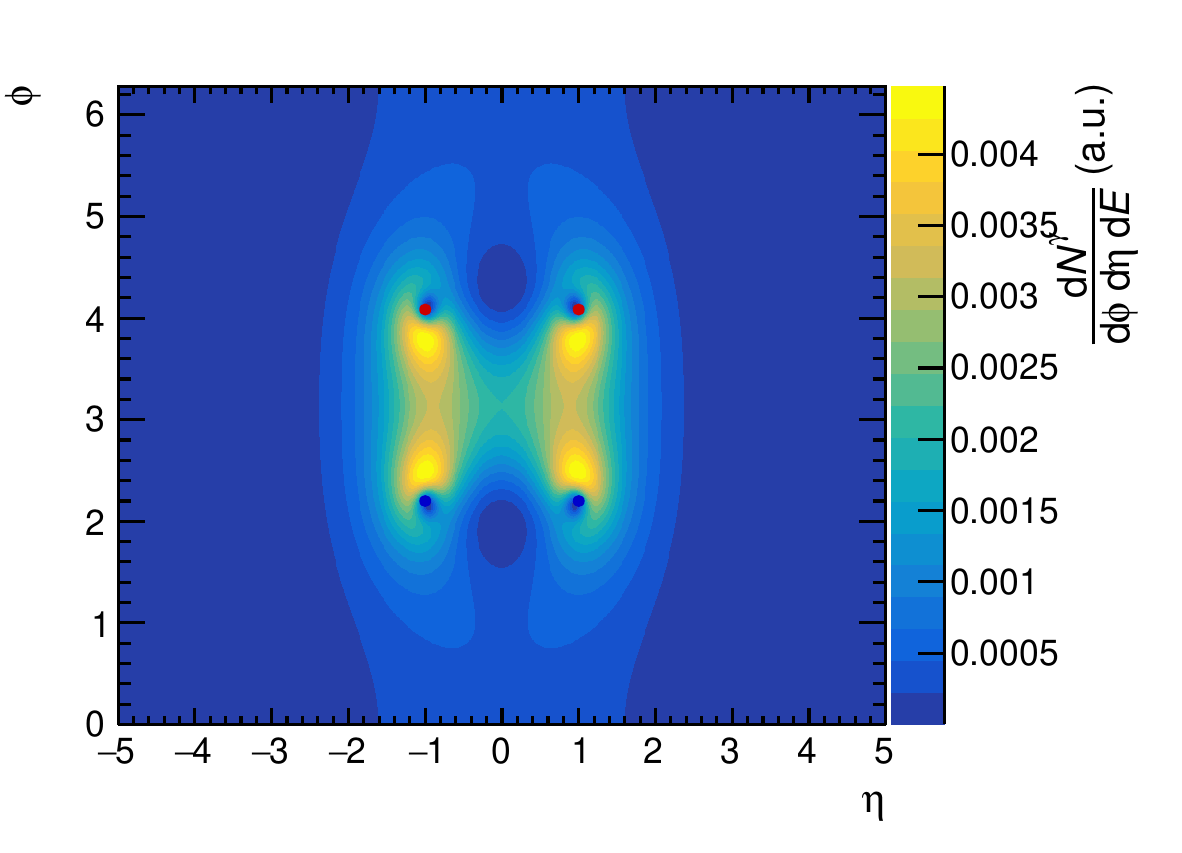}
  \caption{Intensity of photon emission of four charged particles in pseudorapidity $\eta$ and azimuth $\phi$ according to Eq.~\eqref{eq:dI_definition_3}. The direction of the particles is given by the dots, with blue and red signifying opposite values of $\eta_n Q_n$. This example shows an increased emission between particles of equal charge, and suppressed emission between particles of opposite charge.}
  \label{fig:interference}
\end{figure}

Coming back to the differential spectrum in Eq.~\eqref{eq:dI_definition_2},
it can be integrated over all emission angles to obtain the differential
rate of emission of a soft photon with a frequency between $\omega_k$
and $\omega_k+d\omega_k$. To perform this integration it is however convenient to go one step back and use the form in Eq.~\eqref{eq:dI_definition} and the polarization sum in Eq.~\eqref{eq:pol_sum} to find
\begin{equation}
\label{eq:dI_definition_3}
dI = \frac{\alpha}{(2\pi)^2} \frac{d^3\bm{k}}{\omega_k} \sum_{n,m=1}^{N_i+N_f} \eta_nQ_n\eta_mQ_m\frac{-p_n\cdot p_m }{(k\cdot p_n )(k\cdot p_m)}|M_{fi}^0|^2.
\end{equation}
We note that Eq.~\eqref{eq:dI_definition_3} constitutes a coherent sum over all emitting charged particles and thus interference is implicitly included. This is demonstrated in Fig.~\ref{fig:interference} that shows the intensity of photon emission of four charged particles in pseudorapidity $\eta$
and azimuth $\phi$. Suppression of photon emission along the direction of motion of the emitting particle is visible (dead cone) as well as an interference pattern bridging the region between charged particles.

Integrating Eq.~\eqref{eq:dI_definition_3} in the soft-photon solid angle one then gets
\begin{equation}
\frac{dI}{d\omega_k} = |M_{fi}^0|^2\frac{1}{\omega_k} \frac{\alpha}{2\pi}  \sum_{n,m=1}^{N_i+N_f} \eta_nQ_n\eta_mQ_m \frac{1}{v_{nm}}\log \frac{1+v_{nm}}{1-v_{nm}}\,,\,\,\,\,\,\,\,\, {\rm with} \  v_{nm}= \sqrt{1-\frac{m_n^2 m_m^2}{(p_n\cdot p_m)^2}},
\end{equation}
where $v_{nm}$ is the relative velocity of each charge $n$ in the rest frame of another charge $m$. The above
result is sometimes expressed in terms of four-dimensional pairwise cusp angles $\gamma_{nm}$ between charged particles as
\begin{equation}
\frac{dI}{d\omega_k} = |M_{fi}^0|^2\frac{1}{\omega_k} \frac{\alpha}{\pi}  \sum_{n,m=1}^{N_i+N_f} \eta_nQ_n\eta_mQ_m \gamma_{nm}\coth \gamma_{nm}\,,\,\,\,\,\,\,  {\rm with} \,\,\,\,\,\, \cosh \gamma_{nm}=\frac{p_n\cdot
p_m}{m_nm_m}.
\end{equation}
The total probability for emission
of a soft photon in the IR region $\omega_k<\Lambda$ logarithmically
diverges in the low-frequency limit as
\begin{equation}
I = |M_{fi}^0|^2 \frac{\alpha}{\pi}  \sum_{n,m=1}^{N_i+N_f} \eta_nQ_n\eta_mQ_m \gamma_{nm}\coth \gamma_{nm} \int^{\Lambda}_\lambda \frac{d\omega_k}{\omega_k}\,,
\end{equation}
where $\lambda$ is the IR cut-off and hence the number of emitted
photons in the IR region is effectively infinite. However, the total
emitted energy is still finite since the quantity
$\omega_k dI/d\omega_k$ is well-behaved in the $\lambda\to 0$ limit. As
it is well known these IR divergences always appear in cross sections
due to the emission of arbitrary numbers of soft real photons and cancel
exactly to all-orders in perturbation theory with the IR divergences
introduced by the existence also of arbitrary numbers of virtual soft
photons accompanying any transition.

\subsubsection{Low's leading soft theorem: symmetry considerations}

As shown in the previous section using either classical or quantum
mechanical arguments, invariant S-matrix elements for the amplitude of
emission of soft photons during a transition relate to leading power in
the $1/\omega$ expansion where $\omega$ is the emitted photon energy
to S-matrix elements for the transitions to take place without the
emission of the soft photon. In this section, we will show how this very
general result can be traced back to very fundamental aspects of the
gauge interactions, ultimately related to Lorentz and gauge invariance
and the massless nature of gauge bosons.

Gauge interactions are typically described on a very fundamental level
with Quantum Field Theories (QFTs). In these formulations, both
particles and forces are described on equal footing through \emph{local}
field operators in space-time. The analogous mechanism of the classical
``action at a distance'' or force between any two ``particles'' is hence
described by the propagation of a gauge field from one space-time point
to another. In order to make this propagation and the underlying theory
fully compatible with the (special) relativistic structure of
space-time, i.e., consistent with the limiting speed of light one
introduces a microcausality principle according to which any two
local observables in nature must necessarily commute for space-like
separated coordinates,
\begin{equation}
\label{symm.1}
[\hat{O}_1(x),\hat{O}_2(y)]=0 \quad \text{for} \quad (x-y)^2<0.
\end{equation}
These local observables can always be expressed in
terms of local quantized field operators. The latter are constructed by
finding irreducible representations of the proper orthochronous
Poincar{\'e} group, i.e., the group generated by space-time
translations, Lorentz boosts, and spatial rotations \cite{wigner39,Weinberg1995}. Any irreducible
representation can be constructed by first considering the momentum
eigenstates. Since $p^2=m^2$ is a Casimir operator of the Poincar{\'e}
group, it is one of the parameters characterizing any irreducible
representation. Accordingly, only representations with $m^2>0$ (leading
to the description of massive particles) or $m^2=0$ (leading to massless
particles) are consistent with the microcausality principle for
interacting fields in Eq.~\eqref{symm.1}.

Let us now discuss these two possible situations separately. For an
irreducible representation with $m^2>0$, any momentum eigenstate
$|\bm{p},\alpha \rangle$ can be obtained through a Lorentz boost of the
zero-momentum eigenstate $|\bm{p}=0,\sigma \rangle$. The possible
degeneracy of this state is then characterized by the action of the
Poincar{\'e} subgroup, which leaves $\bm{p}=0$ invariant. This is the
spatial rotation group SO(3) or, in quantum theory, its covering group
SU(2). In other words, $|\bm{p}=0,\sigma \rangle$ must span an
irreducible representation of SU(2). The corresponding Casimir operator
$\bm{s}^2$ defines the spin $s$ of the particle, with
$s \in \{0,1/2,1,3/2,\ldots \}$, corresponding to the eigenvalue
$s(s+1)$ for $\bm{s}^2$ and $s_z \in \{-s,-s+1,\ldots,s-1,s \}$. This
SU(2) is known as the \textit{little group} of the massive irreducible
representation of the Poincare group.

On the other hand, for $m=0$, one has instead
$p^2=(p^0)^2-\bm{p}^2=0$. Here one can choose an arbitrary light-like
``standard momentum'' vector $p=(\kappa,0,0,\kappa)$, and the little
group in this case is the Lorentz subgroup which leaves this light-like
four-vector invariant, i.e., either rotations around the $z$-axis or
``null rotations''. Altogether, one finds that the little group is
isomorphic to the symmetry group of the Euclidean plane, ISO(2). Now,
given that the ``translations'' in this group lead to continuous
eigenvalues of the corresponding generators, this would physically imply
the existence of a continuous spin/polarization of the massless
particles described by the theory. Since this has never been observed in
nature, this implies that the null rotations must be represented
trivially so that only the rotations around the $z$ axis can lead to
non-trivial operations on the fields. Since for the $z$-rotations the
possible spin values are $\{0,1/2,1,\ldots \}$, then the representation
of the Poincar{\'e} group constructed in this way leads to the right
representation of the spatial rotations.

The crucial observation to be made is that the previously mentioned
triviality of the null rotations can only be achieved within a local QFT
by assuming gauge invariance in our theory. For instance, for Abelian
gauge bosons with $s=1$ in quantum electrodynamics (QED), the standard
way to realize the representation of the Poincar{\'e} group is to use a
four-vector field $A_{\mu}(x)$ representing the photon. The massless
nature of this particle is then expressed as $\Box A_{\mu}=0$, and the
triviality of the null rotations is then achieved by assuming that for
any scalar field $\chi$, the transformed field
$A_{\mu}'=A_{\mu} + \partial_{\mu} \chi$ describes the same physical
scenario and therefore the theory is gauge invariant. It immediately
follows that only two polarization states for the photon are physical
corresponding to the helicity values
$h=\bm{p} \cdot \bm{s}/|\bm{p}| \in \{s,-s\}$ (for photons $s=1$, i.e.,
$h \in \{1,-1\}$).

To describe the interactions of such a massless gauge field with the
matter fields, for instance, electrons and positrons in QED, the
Lagrangian, and equations of motion must be also gauge invariant. The
local gauge transformations for the matter fields read
$\psi \rightarrow \exp(-\mathrm{i} q \chi)\psi$,
$\bar{\psi} \rightarrow \exp(+\mathrm{i} q \chi)\bar{\psi}$. In QED the
gauge invariant field-strength tensor is given by
\begin{equation}
\label{sym.2}
F_{\mu \nu}=\partial_{\mu} A_{\nu} - \partial_{\nu} A_{\mu},\quad\quad\quad \partial_{\mu} F^{\mu \nu}=J^{\nu},
\end{equation}
such that necessarily
\begin{equation}
\label{sym.4}
\partial_{\mu} J^{\mu}=\partial_t J^0 + \bvec{\nabla} \cdot \bm{J}=0,
\end{equation}
which implies the conservation of the electric charge,
\begin{equation}
\label{sym.5}
Q=\int_{\mathbb{R}^3} d^3 x j^0(x).
\end{equation}

We already found how this classical charge conservation in
Eq.~\eqref{eq:k_mu_j_mu_null} is precisely a Ward identity ensuring the
gauge invariance of all radiative S-matrix elements and constraining their exact
form in the soft-photon $k\to 0$ limit, namely Low's soft theorem.
The previous equivalence between soft theorems and Ward identities was
first noted by Weinberg in \cite{Weinberg:1964kqu,Weinberg:1964ew} who
showed that conservation of charge in QED and Einstein's equivalence
principle in gravity --- the equality of gravitational and inertial mass ---
follow directly from Lorentz invariance the existence of massless
particles of spins $s=1$ and $s=2$ called photons and gravitons and the
pole structure of the scattering amplitudes. In short, charge and
gravitational mass are soft photon and graviton coupling constants,
hence if a process violates charge conservation then the same process
with inner-bremsstrahlung of an extra soft photon (or graviton) would
violate Low's soft theorem and hence Lorentz invariance. Indeed, as
noted by Weinberg the S-matrix element for the emission of a photon or
graviton can be written as the product of a photon field polarization
``vector'' $\epsilon^\mu$ or ``tensor'' $\epsilon^{\mu}\epsilon^\nu$,
respectively, with a Lorentz covariant vector or tensor amplitude. The
``vector'' $\epsilon_\mu(\bvec{q},\lambda)$ is not a four-vector but
under a Lorentz transformation goes to
$\epsilon_\mu(q,\lambda) \rightarrow \epsilon_\mu(q',\lambda) + C
q'_\mu$, where the additional term, $\propto q'$, corresponds to an
additional gauge transformation. Since observables must be Lorentz
covariant and gauge independent, the $S$-matrix element for the emission
of a photon can only be Lorentz invariant and gauge independent if this
``gauge contribution'' $\propto q$ cancels. This means that the S-matrix
element must vanish if any $\epsilon_\mu$ gets replaced by the emitted
photon momentum $q_\mu$. Using Eq.~\eqref{eq:s_fi_1_k_lp}, in the
soft-photon limit this statement necessarily leads to Low's form of the
soft theorem and to charge conservation in any given transition,
\begin{equation}
\label{hvh.3}
\sum_{n=1}^{N_i+N_f} \eta_n Q_n=0.
\end{equation}
More generally, the
deep implications of gauge invariance of S-matrix amplitudes were first
analyzed by Dirac in \cite{Dirac:1955uv}. He showed around the same
time as Low that a manifestly invariant formulation of QED under
general gauge transformations prevents one from defining electron states as free field solutions without including their interactions with the photon field. In this treatment, fermions appear always accompanied by dynamical Coulomb fields around them with free field solutions $\Psi(x)$ replaced then by dressed states $\Psi(x)e^{C(x)}$, and each choice of $C(x)$ corresponding to a choice of the electric field around the charged fermion. Building on
previous work by Dollard \cite{Dollard:1964}, Chung
\cite{PhysRev.140.B1110} and Kibble
\cite{Kibble1968,Kibble1968a,Kibble1968b,Kibble1968c}, Faddeev and
Kulish \cite{Kulish:1970ut} realized that when these Coulomb fields --- encoded as coherent states of infinitely many soft photons dressing the incoming and outgoing vacuum states and encoding all pairwise interactions at asymptotic space-time distances in the transition --- are considered in scattering problems then
real and virtual IR divergences systematically cancel from all
QED amplitudes. Conversely, in the conventional treatment of scattering problems in
field perturbation theory it is assumed that the gauge interactions vanish at
very large space-time distances so that Fock vacuum states can be
effectively represented by free field solutions. As a result, real and
virtual IR divergences in Low's theorem cancel only at the cross section
level
\cite{Bloch:1937pw,Jauch1954,Jauch1955,Yennie:1961ad,Weinberg:1965nx}. A
comprehensive introduction to the Faddeev-Kulish S-matrix can be found
in the Supplement S4 of the classic textbook of Jauch and Rohrlich
\cite{Jauch1955}. More modern discussions of the Faddeev-Kulish S-matrix
and soft theorems in the context of Soft Collinear Effective Field
Theory (SCET) and worldline formulations can be found in
\cite{Hannesdottir:2019opa,Bonocore:2021cbv,Feal:2022iyn,Feal:2022ufw}.

Besides the intrinsic interest of the Faddeev-Kulish S-matrix
formulation in practical calculations for the treatment of the IR
divergences in Low's theorem the Faddeev-Kulish picture of dressed
vacuum coherent states is of fundamental importance for the discussion
of asymptotic symmetries and general gauge invariance. In the context of
gravity, Bondi, Van der Burg, Metzner, and Sachs
\cite{Bondi:1962px,Sachs:1962wk} showed a long time ago that degenerate
vacua having different numbers of soft gravitons are connected by
supertranslations of the infinite-dimensional BMS group of asymptotic
space-time symmetries. Armed with this insight, Strominger and
collaborators \cite{Strominger:2017zoo} showed that in QED and likely
QCD one can in principle construct the analog of the BMS group as the
group of large gauge transformations at every point on the celestial
sphere at null infinity. Large gauge transformations are those
satisfying $A_{\mu} \rightarrow A_{\mu}+\partial_{\mu} \chi$, with
$\chi$ a function which does not die down at spatial infinity.  They
showed how soft theorems in QED and gravity can be directly related to
the Ward identities for the infinite dimensional symmetry group of large
U(1) gauge transformations \cite{He:2014cra} and the BMS group
\cite{Strominger:2013jfa}, respectively, approaching angle dependent
constants defined through the corresponding cusp anomalous dimensions in
the celestial sphere \cite{Arkani-Hamed:2020gyp}.  We refer the reader
to \cite{Campiglia:2015qka} for a comprehensive introduction. The
spontaneous breaking of this asymptotic symmetry leads to an infinite
number of degenerate vacua characterized by different numbers of soft
photons and gravitons allowing thus a suggestive interpretation of soft
photons and gravitons as Goldstone modes of the theory for the
transition between such different vacua. The charges of the broken
generators satisfy a Kac-Moody algebra on the two-dimensional celestial
sphere at null infinity.

A direct phenomenological implication of the existence of the
Faddeev-Kulish soft-photon clouds can be found in the equivalence
between soft theorems and gravitational \cite{Strominger:2014pwa} and
electromagnetic \cite{Bieri:2013hqa} displacement (or memory)
effects. In gravity, the clock times of an array of detectors will
differ before and after the passage of a gravitational-wave pulse by a
BMS supertranslation \cite{Strominger:2014pwa}. An analogous experiment
in QED to measure this electromagnetic memory effect has been proposed
by Susskind in \cite{Susskind:2015hpa}. Likewise, the striking
similarities between the spontaneous breaking of Poincar\'e invariance
in highly occupied systems of soft gravitons and gluons has been
proposed as a possible correspondence between the IR sector of quantum
chromodynamics and gravity and hence of the dynamics of gluon
condensates in relativistic nucleus-nucleus collisions and that of black holes
\cite{Dvali:2021ooc}.

The discussion above focused on computing and describing the main
features of the soft radiation encoded in the leading power term
$\mathcal{O}(\omega^{-1}_k)$ of the multipole or soft expansion of
cross sections in Eq.~\eqref{eq:multipole_expansion_cross_sections}. As
previously emphasized, this term agrees with the classical power
spectrum of the transition and thus far has provided the main
theoretical reference for experimental analysis \cite{DELPHI:2005yew,
  DELPHI:2007nmh, DELPHI:2010cit}. In the next section we will address
the problem of computing the sub-leading power corrections of
$\mathcal{O}(\omega^{0}_k)$ in this soft expansion.

\subsubsection{\label{sec:expansion_scheme}Expansion scheme:
Leading order,
Derivative operators versus momentum shifts in next-to-leading order}

The extension of the factorization picture described above at subleading
power in the soft expansion has been investigated since the time of Low
in the 1950s and the 1960s \cite{Gell-Mann:1954wra, Low:1958sn,
	Burnett:1967km, Bell:1969yw} and has continued to
attract interest until recent years both in QCD phenomenology
\cite{DelDuca:1990gz, Bonocore:2015esa, Larkoski:2014bxa,
	 Moult:2018jjd, Boughezal:2018mvf,
	Bahjat-Abbas:2019fqa, vanBeekveld:2021hhv, Liu:2020tzd, Beneke:2019oqx,
	Broggio:2021fnr, Ravindran:2022aqr} and in more formal investigations on
	the structure of
scattering amplitudes \cite{Cachazo:2014fwa, Strominger:2017zoo}.  More
specifically, at Next-to-Leading-Power (NLP) in the soft expansion, Low,
Burnett and Kroll (LBK) derived a tree-level theorem which yields an
$\omega_k^0$ correction to the power spectrum. In the language used in
modern literature, these theorems at the amplitude level are dubbed
\emph{next-to-soft} theorems. Let us consider first for simplicity the
emission of a single real soft photon during a transition $i\to f$
involving two charged particles of momenta $p_1$ and $p_2$ either on the
initial or final state. The \textit{next-to-soft} theorem generalizes
Eq.~\eqref{eq:s_fi_1_k_lp} to
\begin{equation}
\label{eq:s_fi_1_k_nlp}
{M}^{1}_{fi}(p_1,p_2,k) =\left({\cal S}_{\text{LP}}+{\cal
S}_{\text{NLP}}\right){M}_{fi}^{0}(p_1,p_2),
\end{equation}
where the leading power (LP) and next-to-leading power (NLP) soft
factors are given by
\begin{equation}
{\cal S}_{\text{LP}}=e\sum_{n=1}^2 Q_n\eta_n\,\frac{p_n\cdot
\epsilon(k)}{p_n\cdot
	k}~, \qquad
{\cal S}_{\text{NLP}}=e\sum_{n=1}^2
Q_n\eta_n\,\frac{ik_{\nu}J^{\mu\nu}\epsilon_{\mu}(k)}{p_n\cdot
	k}~.
\label{NLPamplitude}
\end{equation}
Here, $J^{\mu\nu}$ represents the total angular momentum operator of the
hard emitting particle. More precisely, it is the sum of the orbital
angular momentum generator
\begin{equation}
L^{\mu\nu}_n=i\left(p_n^{\mu}\frac{\partial}{\partial
	p^n_{\nu}}
-p_n^{\nu}\frac{\partial}{\partial p^n_{\mu}}\right)
\label{eq:orbital}
\end{equation}
and the spin generator $S^{\mu\nu}_n$ of the Lorentz group. It is clear
that ${\cal S}_{\text{NLP}}$ in Eq.~\eqref{eq:s_fi_1_k_nlp} is not a
multiplicative factor but an operator acting on the non-radiative
amplitude.  For instance, in the
case of an amplitude with only two charged particles of spin $1/2$ in
the initial state, the spin generator is
$S^{\mu\nu}=\frac{i}{4}[\gamma^{\mu},\gamma^{\nu}]$ and
Eq.~\eqref{eq:s_fi_1_k_nlp} can be written explicitly as
\begin{equation}
\begin{split}
& {M}_{fi}^1(p_1,p_2,k) \\
&=e\sum_{n=1}^2 \eta_nQ_n\,\frac{p_n\cdot \epsilon(k)}{p_n \cdot k}\,\bar
v(p_2){\cal
	H}(p_1,p_2)
u(p_1)
-\epsilon_{\mu}(k)e\sum_{n=1}^2 \eta_nQ_n\,\bar v (p_2)
G^{\mu\nu}_n\,\frac{\partial {\cal H}(p_1 ,p_2)
}{\partial p_n^{\nu}}
u(p_1)\\
& - e\eta_1Q_1\,\bar v (p_2)
\frac{ik_{\nu}S^{\mu\nu}\epsilon_{\mu}(k)}{p_1\cdot k}{\cal H}(p_1
,p_2) u(p_1)
- e\eta_2Q_2\,\bar v (p_2) {\cal H}(p_1 ,p_2)
\frac{ik_{\nu}S^{\mu\nu}\epsilon_{\mu}(k)}{p_2\cdot k}
u(p_1)
~,
\label{eq:s_fi_1_k_nlp_fermions}
\end{split}
\end{equation}
where we defined
\begin{equation}
G^{\mu\nu}_n\equiv g^{\mu\nu}-\frac{(2p_n-k)^{\mu}k^{\nu}}{2p_n\cdot k}
=g^{\mu\nu}-\frac{p_n^{\mu}k^{\nu}}{p_n\cdot k}+{\cal O}(k)~.
\label{Gtensor}
\end{equation}
Note in particular that derivatives contained in Eq.~\eqref{eq:orbital} act on the
hard subdiagram
${\cal H}(p_1,p_2)$ only, i.e., the non-radiative
amplitude ${M}_{fi}^{0}(p_1,p_2)$ stripped off the external spinors. This
is a consequence of the way the soft expansion is performed. Specifically,
Eq.~\eqref{eq:s_fi_1_k_nlp_fermions} can be derived by assuming that momentum
conservation
in the form $p_1+p_2+k=0$ is imposed after expanding the full radiative
process ${M}_{fi}^1(p_1,p_2,k)$
in powers of the soft momentum $k$.
 This is in analogy with the original work of
Low \cite{Low:1958sn} for $2\to2$ scattering of scalar particles.

An alternative approach that has been recently proposed in the literature
\cite{Lebiedowicz:2021byo,
	Lebiedowicz:2022nnn, Lebiedowicz:2023mlz, Lebiedowicz:2023ell} consists of
	expanding the radiative amplitude on the momentum conservation surface by
	implicitly inserting a dependence over $k$ in the momenta $p_1$ and $p_2$
	(and consequently also in the spinors, if necessary). Even in the scalar case, the
	resulting expression for the expanded radiative amplitude is then different
	from the original result by Low (see \cite{Lebiedowicz:2023ell} for a detailed discussion on this point). However, one should note that once momentum conservation is enforced, the difference between the two
	expressions is only sub-subleading (i.e.\ NNLP) and thus beyond the accuracy of
	Low's theorem \cite{Balsach:2023ema, Fadin:2024tar}.
More generally, the freedom to choose the functional dependence of the
non-radiative amplitude leads to many forms of Low's theorem, which are all
formally equivalent after momentum conservation is imposed up to NNLP
corrections.
However, one should bear in mind that in photon spectra where the soft
momentum $k$ is necessarily non-vanishing these NNLP effects do bring
numerical differences that might be sizable for sufficiently large  values of $\omega_k$.
Therefore, some formulation might be more versatile and efficient than others.

Despite the slightly intricate
structure at the amplitude level, soft emissions at NLP acquire a
remarkably simple form for unpolarized cross sections. In fact, as
originally observed by Burnett and Kroll, by summing over the
polarizations and neglecting next-to-next-to-leading power (NNLP) terms
 the analogue of Eq.~\eqref{eq:s_fi_n_k_lp} for the bremsstrahlung cross
section of a single soft photon to NLP in a transition $i\to f$
involving $N$ particles of arbitrary spin is given by
\begin{equation}
\label{eq:s_fi_n_k_nlp}
|{M}_{fi}^1(p_1,\ldots,p_N,k)|^2
=e^2\sum_{n,m=1}^{N}(-\eta_nQ_n\eta_mQ_m)\bigg\{\frac{p_n\cdot p_m}{p_n\cdot k
\,p_m\cdot k}+\frac{(p_n)_{\mu} }{p_n\cdot k }
G_m^{\mu\nu} \frac{\partial}{\partial
p_m^{\nu}}\bigg\}|M_{fi}^{0}(p_1,\ldots,p_N)|^2 ~.
\end{equation}
Note in particular that both the spin and the orbital contributions of
\cref{NLPamplitude} are represented by the term proportional to the
derivative of the non-radiative process due to the unpolarized nature of the
cross section.

In analogy with Eq.~\eqref{eq:s_fi_1_k_nlp_fermions}, the  NLP
term in Eq.~\eqref{eq:s_fi_n_k_nlp} contains derivatives acting on a non-radiative amplitude
$M_{fi}^0(p_1,\ldots,p_N)$ that is evaluated for a configuration where
momentum is not conserved since $p_1+\ldots+p_N=-k\neq 0$.  While
this is not a problem in the strict $k \to 0$ limit, the finiteness of
$k$ at NLP makes the amplitude on the r.h.s. of
Eq.~\eqref{eq:s_fi_n_k_nlp} non-physical. Several possibilities to
overcome this issue have been recently put forward
\cite{Lebiedowicz:2021byo,
	Lebiedowicz:2022nnn, Lebiedowicz:2023mlz, Lebiedowicz:2023ell,
	Bonocore:2021cbv, Balsach:2023ema, Engel:2021ccn}. Reference
	\cite{Bonocore:2021cbv} in particular,
building on previous studies in the context of gluon resummation
\cite{DelDuca:2017twk}, proposed to rewrite the non-radiative amplitude
in terms of shifted kinematics as
\begin{equation}
|M_{fi}^1(p_1,\dots,p_N,k)|^2
=e^2\left(\sum_{n,m=1}^N-\eta_nQ_n\eta_mQ_m\frac{p_n\cdot p_m}{p_n\cdot k
\,p_m\cdot
	k}\right)
|M_{fi}^{0}(p_1+\delta p_1,\dots,p_N+\delta p_N)|^2
~,
\label{eq:s_fi_n_k_nlp_shifted}
\end{equation}
where the shift on the $\ell$-th momentum $p_\ell$ is defined at NLP as
\begin{equation}
\delta p_{\ell}^{\mu}=
\left(\sum_{n,m=1}^N\eta_nQ_n\eta_mQ_m\frac{p_n\cdot p_m}{p_n\cdot k
	\,p_m\cdot
	k}\right)^{-1}
\sum_{k=1}^{N}
\left(
\eta_kQ_k\eta_{\ell}Q_\ell\frac{(p_{k})_{\nu}G_{\ell}^{\nu\mu}}{p_k\cdot k}
\right)~.
\label{eq:shifts}
\end{equation}
For the simple case of two external charged particle legs these become
\begin{equation}
\delta p_1^{\mu}= \frac{1}{2} \left(-\frac{p_2\cdot k}{p_1\cdot p_2}p_1^{\mu}
+\frac{p_1\cdot k}{p_1\cdot p_2}p_2^{\mu}
-k^{\mu}\right)~,\quad
\delta p_2^{\mu}= \frac{1}{2} \left(\frac{p_2\cdot k}{p_1\cdot p_2}p_1^{\mu}
-\frac{p_1\cdot k}{p_1\cdot p_2}p_2^{\mu}
-k^{\mu}\right)~,
\label{delta12}
\end{equation}
from which one can readily see that
$\delta p_1^{\mu}+\delta p_2^{\mu}=-k^{\mu}$, thus restoring momentum
conservation in the non-radiative amplitude $M_{fi}^0(p_1,\ldots,p_N)$.
Note that the cross section in Eq.~\eqref{eq:s_fi_n_k_nlp_shifted}
contains the same soft factor as the LP approximation for the
cross section in Eq.~\eqref{eq:s_fi_n_k_lp}, and therefore can be
analogously implemented in the photon spectrum.

At this point it should be noted that while the shifted kinematics restore momentum
conservation to all-order in the soft expansion, the on-shell condition is
fulfilled only at NLP, i.e.\ $(p_i+\delta p_i)^2=m^2_i+{\cal O}(k^2)$, but it is
violated at NNLP. Although legitimate within the range of validity of Low's
theorem, this feature might be problematic for numerical
implementations. The solution to this problem exploits the fact that it is
possible to add spurious NNLP terms without invalidating Low's theorem.
Specifically, one can modify the
shifts $\delta p_i$ so that they fulfil the on-shell condition and momentum
conservation to all orders in the soft expansion. A specific form for these
modified shifts has been recently provided \cite{Balsach:2023ema} and reads
\begin{align}
\delta p_i^\mu
= A Q_i \sum_{j=1}^N \frac{\eta_j Q_j}{k \cdot p_j} p_{j\nu} G^{\nu\mu}_i
+ \frac{1}{2} \frac{A^2 Q^2_i {B}}{p_i \cdot k}
k^\mu~,
\label{eq:modshifts}
\end{align}
where
\begin{align}
A
= \frac{1}{\chi} \left(
\sqrt{1 - \frac{2\chi}{B}}
- 1
\right)~,
\qquad
B=-\sum_{ij=1}^{N} \eta_i \eta_j Q_i Q_j
\frac{p_i \cdot p_j}{(p_i \cdot k) (p_j \cdot k)} ~,
\qquad
\chi = \sum_{i=1}^N \frac{\eta_i Q^2_i}{p_i \cdot k}~.
\end{align}
One can easily verify that $\delta p_i$ in Eq.~\eqref{eq:modshifts} reduces to
Eq.~\eqref{eq:shifts} by expanding it in $k$ at NLP (thus fulfilling
Eq.~\eqref{eq:s_fi_n_k_nlp_shifted}) and that the momenta
$p_i+\delta p_i$ are both on-shell and fulfill momentum conservation to all
orders in $k$.

\subsubsection{Loop corrections}
The infrared structure of QED with massive particles to all orders in
perturbation theory is rather simple as can be seen by looking at the
structure of diagrams contributing to the exponent of the soft function
\cite{Yennie:1961ad}. One important consequence is that Low's  theorem at
leading power (i.e. $1/\omega$) remains unchanged at the loop level. Things are
different for the subleading term in the soft expansion and, when the massless
limit is considered also for the leading term. Let us consider the massive and
massless case separately, both in QED and QCD.

In the presence of massive particles, one expects the
corresponding correction to the NLP soft theorem to involve ratios of scales
containing the masses of the charged particles. In fact, these additional
scales
introduce an additional quantity
whose expression can be
read in Eq.~(4.41) of \cite{Engel:2023ifn} (see Appendix A of
\cite{Engel:2021ccn} for the results for the integrals). Remarkably, this
correction to the NLP soft theorem is just a multiplicative function
(unlike the factor ${\cal S}_{\text{NLP}}$
of Eq.~(\ref{NLPamplitude})) which involves harmonic polylogarithms.
Most importantly, the aforementioned soft function is one-loop exact and
therefore the modified ${\cal S}_{\text{NLP}}$ holds to all orders in the
electromagnetic coupling constant.

The massless case is rather different. It is relevant both in QED when one can
approximate the charged particles to be massless and, perhaps more importantly,
in the partonic calculations of perturbative QCD. Specifically, loop
corrections affect the LP term ${\cal S}_{\text{LP}}$, both for soft photons
and soft gluons. However, while the LP soft-gluon current receives corrections
already at one-loop order in QCD \cite{Catani:2000pi}, the LP factor for a
soft-photon emission is modified only starting from three loops
\cite{Chen:2023hmk, Herzog:2023sgb}. This effect originates from the coupling
of the soft photon to a loop of virtual massless particles, which yields a
correction to ${\cal S}_{\text{NLP}}$ of order $\alpha_s^3/\omega$. This
analysis has been recently generalized to all orders for the simple case of
quark-antiquark pair creation \cite{Ma:2023gir}.

The NLP term ${\cal S}_{\text{NLP}}$ also receives loop corrections in the
massless case \cite{Czakon:2023tld}. Specifically, at the one-loop level,
corrections that are non-analytic in the soft momentum modifies ${\cal
S}_{\text{NLP}}$ by a term of
order $\alpha_s\log(\mu/p_n\cdot k)$, with $\mu$ the dimensional regularization
scale  \cite{Bonocore:2021cbv}. The origin of these corrections can be traced
back to the presence of other small scales, in this case the collinear one
$p_n\cdot k$ that prevents a naive Taylor expansion in the soft momentum $k$.
Diagrammatically, this effect is due to the soft photon coupling to
a subdiagram of collinear lines (hence called \emph{radiative jet})
\cite{DelDuca:1990gz, Bonocore:2015esa, Moult:2019mog, Beneke:2019oqx,
Liu:2021mac}.

In conclusion, this recent body of work in perturbative calculations has
provided new information both in the infrared structure of quantum field
theories
and in the phenomenological
implications of power corrections to the strict soft limit for hadronic
physics whenever a perturbative treatment is possible. However, it is neither
clear how these results could provide a solution to the soft-photon
puzzle nor how they relate to the non-perturbative regime. In fact,
providing accurate theoretical predictions for photon spectra in particular
kinematical conditions remains an active field of research, especially for
processes with various hadronic final states at very high rapidity. This aspect
is of special relevance also for the planned future measurements discussed in
the next section.

\subsection{Experimental consequences and opportunities at the LHC}

The ALICE Collaboration intends to install a completely new silicon-based detector (``ALICE~3'') during the fourth Long Shutdown of the Large Hadron Collider that is presently scheduled for the years 2032–2033. ALICE~3 will possess a unique pointing resolution over a large pseudorapidity range $(-4 < \eta < 4)$, complemented by multiple subdetector systems for particle identification including silicon time-of-flight layers, a ring-imaging Cherenkov detector with high-resolution readout, a muon-identification system and an electromagnetic calorimeter \cite{ALICE:2022wwr}.

A dedicated soft-photon measurement at the LHC leveraging high collision rates and improvements in detector technology could be key to resolving the soft-photon puzzle. In particular, the energy dependence of the excess above the inner bremsstrahlung expectation as calculated by Low's leading-power expression could point to the origin of the puzzle. The production of hadrons in a non-diffractive proton-proton collision at high energies is sometimes described as the fragmentation of two ``beam jets'' pointing in opposite directions along the beam axis. The DELPHI experiment considered the transverse momentum of soft photons with respect to the jet axis. This is based on an argument by Gribov \cite{Gribov:1966hs} about an extended range of applicability of Low's theorem. For non-diffractive proton-proton collisions at colliders, a low transverse momentum relative to most produced particles would imply measuring in the forward direction and at transverse momenta  of $k_T \lesssim m_\pi$ relative to the beam axis. The background of photons from decays of neutral pions and $\eta$ mesons is a major experimental challenge. However, for photon transverse momenta of $k_T \lesssim 5\,\mathrm{MeV}/c$ the expected soft-photon signal becomes larger than the decay-photon background owing to the rest mass of the neutral pion. Therefore, the photon transverse momentum region $1 \lesssim k_T \lesssim 5\,\mathrm{MeV}/c$ is a very promising range for studying Low's theorem at the LHC.

The clean photon identification as well as the good momentum and pointing resolution at low momenta make the photon conversion method a natural choice  for the measurement of low $k_T$ photons. We consider this option in the following. Reconstructing photons through conversion into $e^+e^-$ pairs becomes possible for photon energies above $E_{\gamma,\,\mathrm{min}} \approx 50\,\mathrm{MeV}$. For lower energies, multiple scattering of the electrons and positrons in the detector material makes the momentum reconstruction difficult. In addition, Compton scattering becomes relevant for $E_\gamma \lesssim 50\,\mathrm{MeV}/c$. The photon transverse momentum is given by $k_{T,\gamma} = E_\gamma / \cosh \eta$. To measure low-$k_T$ photons with $k_T \lesssim 5\,\mathrm{MeV}/c$ therefore requires measuring at large pseudorapidities making use of the forward boost. The planned ALICE~3 detector has a solenoid magnet creating a magnetic field in the direction of the beam axis. To achieve reasonable momentum resolution for electrons and positrons from photon conversions at forward pseudorapidities, these particles need to be tracked in a dipole field which gives a much larger lever arm for the momentum reconstruction. These basic considerations were followed in the design of the Forward Conversion Tracker (FCT) of ALICE~3 as detailed in the ALICE~3 letter of intent \cite{ALICE:2022wwr}.

From the relation between Feynman amplitudes and cross sections and starting from Eq.~\eqref{eq:dI_definition_3}, the leading-power expression for the invariant yield of inner-bremsstrahlung photons from, e.g., colliding $e^+e^-$ (initial state radiation) and from final state $\mu^+\mu^-$ pairs can be calculated as
\begin{equation}
  \frac{dN_\gamma}{d^2 \bvec k_{\text{T}} \, d\eta\, d\phi} = \frac{\alpha}{(2 \pi)^2} \int d^3 \bvec p_{\mu^+}d^3\bvec p_{\mu^-}
  \; \sum_{n,m=1}^{4} \eta_n Q_n \eta_m Q_m  \frac{-(p_n \cdot p_m)}{(p_n \cdot k)(p_m \cdot k)} \; \frac{dN_{\mu^+\mu^-}}{d^3 \bvec p_{\mu^+}d^3 \bvec p_{\mu^-} } \,.
  \label{eq:brems_formula_as_used_by_experiments-mu}
\end{equation}
The $p_n$ and $\bvec{p}_n$ denote the four and three momenta of the incoming electron and positron, and the outgoing muons;  $\bvec p_{\mu^+}$ and $\bvec p_{\mu^-}$ are the  three momentum of the muons, $k$ and $\bvec k$ denote the four and three momentum of the photon ($E_\gamma = \omega_k=|\bvec k|$). As defined previously (cf.\ Eq.~\eqref{eq:s_fi_1_k_lp}), the factors $\eta_n$ are $\eta_n = 1$ for the outgoing particles (the muons) and $\eta_n = -1$ for the incoming particles (the $e^+$ and the $e^-$). Note that this definition of $\eta$ differs from the one used by the DELPHI collaboration, where the charge of the particles $Q_n$ in units of the elementary charge $e$ is included in the definition of $\eta_n$ \cite{DELPHI:2005yew,DELPHI:2007nmh}. The total number of charged particles is $N$, i.e., the sum of the number of charged particles in the initial state ($N_i$) and the final state ($N_f$). Here, $N_i = 2$ (electrons) and $N_f = 2$ (muons). The last factor in the sum is the differential production rate of the produced muons. Eq.~\eqref{eq:brems_formula_as_used_by_experiments-mu} is expressed in terms of the photon production rate per muon.

The invariant yield of inner-bremsstrahlung photons for a reaction with two incoming particles with fixed momenta and $N_f$ charged hadrons in the final state, i.e., $N = N_f + 2$, is given by
\begin{align}
  \frac{dN_\gamma}{d^2 \bvec k_{\text{T}} \, d\eta\, d\phi}  &= \frac{\alpha}{(2 \pi)^2}  \int d^3 \bvec p_1 \ldots d^3 \bvec p_{N_f}
  \; \sum_{n,m=1}^{N} \eta_n Q_n \eta_m Q_m  \frac{-(p_n \cdot p_m)}{(p_n \cdot k)(p_m \cdot k)} \; \frac{dN_\mathrm{hadrons}}{d^3 \bvec p_1 \ldots d^3 \bvec p_{N_f}} \label{eq:brems_formula_as_used_by_experiments-hadrons_a}\\
  &\equiv \frac{-\alpha}{(2 \pi)^2} \int d^3 \bvec p_1 \ldots d^3 \bvec p_{N_f}
  \; \left( \sum_{n=1}^{N} \eta_n Q_n \frac{p_n}{p_n \cdot k} \right)^2 \; \frac{dN_\mathrm{hadrons}}{d^3 \bvec p_1 \ldots d^3 \bvec p_{N_f}}  \,,
  \label{eq:brems_formula_as_used_by_experiments-hadrons_b}
\end{align}
where, again, $k$ and $\bvec k$ denote the four and three momentum of the photon,
$p_n$ and $\bvec{p}_n$ denote the four and three momenta of the incoming and outgoing charged particles; $d^3 \bvec p_1 \ldots d^3 \bvec p_{N_f}$ are the three momenta of the $N_f$ produced charged hadrons; as above, $\eta = -1$ for the incoming particles and $\eta = +1$ for the outgoing charged hadrons.
Eqs.~\eqref{eq:brems_formula_as_used_by_experiments-mu} and \eqref{eq:brems_formula_as_used_by_experiments-hadrons_a} were used by DELPHI~\cite{DELPHI:2007nmh} and other experiments, and, in general, are evaluated on an event-by-event basis, either from real data or event generators that were tuned to data. In terms of computing time, Eq.~\eqref{eq:brems_formula_as_used_by_experiments-hadrons_b} can be advantageous compared to Eq.~\eqref{eq:brems_formula_as_used_by_experiments-hadrons_a}.
We note that the expression of Eq.~\eqref{eq:brems_formula_as_used_by_experiments-hadrons_b} given in parentheses squared is necessarily negative. Through the factors $\eta_{n}$ and $Q_{n}$, the sum constitutes a difference of four-momentum vectors, with its square always being negative. This holds for an arbitrary number of charged particles due to the conservation of charge and four-momentum.

Even with $e^+,e^-$ beams, where the initial state radiation is much more intense than with hadron beams, the contribution from initial state radiation amounts to only about 1\% in the kinematic range covered by the experiment~\cite{DELPHI:2007nmh} and is thus negligible. In collider mode, photons from initial state radiation are emitted at small polar angles with respect to the beam axis. The angular distribution is suppressed in the regions $\theta < 1/\gamma$ (dead cone) and has a maximum around $\theta = \sqrt 3 / \gamma$, where $\gamma$ the Lorentz factor of the beam, see Eq.~\eqref{eq:dI_angular-maximum}. Thus, only a tiny fraction of the photons are in the range of the central barrel.

Factoring out the energy of the charged particles in Eq.~\eqref{eq:brems_formula_as_used_by_experiments-hadrons_b} replaces the four-momentum $p_i$ by the vector $(1, \bvec \beta_i)$, i.e., the soft-photon production depends explicitly on the velocities of the charged particles (cf.\ Eq.~\eqref{eq:dI_definition_2}). A detector for a soft-photon measurement should therefore be complemented with a detector for charged-particle identification, e.g., a Ring Imaging Cherenkov Detector (RICH) extending the FCT design described in the ALICE~3 letter of intent \cite{ALICE:2022wwr}.

Identifying all charged particles contributing to soft-photon production in a given acceptance can be challenging. Many of the previous soft-photon experiments listed in Tab.~\ref{tab:soft_photon_experiments} therefore used event generators to determine the input for Eq.~\eqref{eq:brems_formula_as_used_by_experiments-hadrons_b}. For instance, the DELPHI experiment used JETSET~7.3, and WA83, WA91 and WA102 used tuned versions of the FRITIOF event generator that were validated with data. In the measurements by Goshaw et al.\ and by WA27, on the other hand, the input for the calculation of the inner bremsstrahlung was derived from measurement. In the paper by WA27, it is noted that the calculation of the inner bremsstrahlung was found to be rather insensitive to the structure of the hadronic cross section.

\begin{figure}[htb]
  \centering
  \includegraphics[width = 0.5\textwidth]{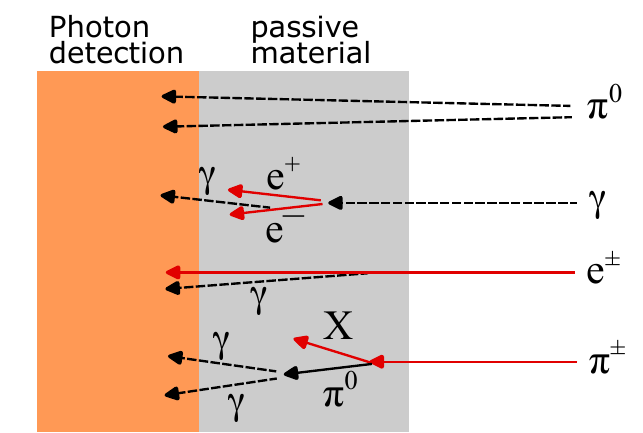}
  \caption{Illustration of background photons for a measurement of the inner-bremsstrahlung signal. Shown is the contribution from decay photons, from external bremsstrahlung created by electrons and positrons, from photon conversions, from external bremsstrahlung created by primary electrons and positrons (e.g. from $\pi^0$ Dalitz decays), and from decays of $\pi^0$'s produced in secondary interactions with the detector material.}
  \label{fig:background_photons}
\end{figure}

Experimental background of brems\-strahlung photons produced in the detector material, i.e., external bremsstrahlung, poses a formidable challenge to soft-photon
measurements. This is illustrated in Fig.~\ref{fig:background_photons}. The background stems mostly from secondary electrons which are produced through photon conversions in the detector material. In addition, background brems\-strahlung photons can also be created by electrons and positrons from $\pi^0$ Dalitz decays. The brems\-strahlung background produced in the detector material follows the same $1/k_{\text{T}}$ form as that from internal brems\-strahlung. Since particles at forward rapidities cross material such as the beam pipe at shallow angles, they are exposed to an increased effective material budget. For example, a beryllium beam pipe of \SI{500}{\um} in thickness corresponds to a radiation length of 0.14\% and 10\% at $\eta = 0$ and $\eta = 5$, respectively.

The background from external bremsstrahlung can be estimated analytically. We consider a photon detector at forward pseudorapidity with a passive material of $x/X_0$ in units of radiations lengths in front of the detector. Due to the forward boost, the electrons and positrons from photon conversions have large enough energies so that we use the approximation
\begin{equation}
\frac{dN_\gamma^\mathrm{ext.\,brems}}{d\omega_k} = \frac{x}{X_0} \left( \frac{4}{3} \frac{1}{\omega_k} - \frac{4}{3} \frac{1}{E_e} + \frac{k}{E_e^2}\right)
\approx \frac{4}{3} \frac{x}{X_0} \frac{1}{\omega_k}
\end{equation}
for the energy spectrum of the bremsstrahlung photons per electron. Here, $E_e$ is the energy of the electron or positron and $\omega_k$ is the photon energy. The mean emission angle of the radiated photon is $\theta \approx m_e / E_e = \sqrt{3}/\gamma$, i.e., the photons are emitted in a narrow cone around the electron or positron. A photon, e.g., from a $\pi^0$ decay, converts in the passive material with a probability of $7/9 \, x/X_0$. For a pseudorapidity density $dN_\gamma^\mathrm{decay}/d\eta$ of the decay photons, the transverse momentum spectrum of the external bremsstrahlung can be estimated as
\begin{equation}
\frac{dN_\gamma^\mathrm{ext.\,brems}}{dk_{\text{T}} d\eta} = \frac{28}{27} \frac{dN_\gamma^\mathrm{decay}}{d\eta} \left(\frac{x}{X_0} \right)^2 \frac{1}{k_{\text{T}}}\,.
\end{equation}
Here, it is assumed that the electrons and positrons from photon conversions traverse on average half of the passive material in front of the photon detector. For proton-proton collisions at $\sqrt{s} = 13.6\,\mathrm{TeV}$ we choose $dN_\gamma^\mathrm{decay}/d\eta \approx 4$ \cite{ALICE:2023ode}. For a passive material corresponding to 10\% of a radiation length, we then obtain a background-photon spectrum of $dN/dk_{\text{T}} d\eta \approx 0.04/k_{\text{T}}$. Evaluating Eq.~\eqref{eq:brems_formula_as_used_by_experiments-hadrons_b} for proton-proton collisions simulated with PYTHIA, we obtain for the signal of the inner bremsstrahlung $dN_\gamma^\mathrm{inner\,brems}/dk_{\text{T}} d\eta \approx 0.017/k_{\text{T}}$. This simple estimate shows that for 10\% of a radiation length of material in front of the photon detector, the signal-to-background ratio is already less than 1/2. We thus consider this an upper limit of the maximum allowable material budget.

An elegant experimental technique for measuring bremsstrahlung photons is to bend the charged particles by which the photons are emitted out of the acceptance of the photon detector. This approach was used by the SLAC E-146 collaboration for the measurement of bremsstrahlung created by electrons in various targets \cite{SLAC-E-146:1997hnd}. This experiment disentangled the soft photon from the radiating charged particle by bending the scattered electron by a dipole magnet and detecting it in a high-resolution spectrometer, while the soft photon is detected in beam direction by a BGO crystal calorimeter. A similar idea was used by the WA91 and the WA102 experiments. An isolation cut around the detected photons was applied to suppress external bremsstrahlung produced in the photon converter. This cut made use of the fact that the charged particle that emitted an inner bremsstrahlung photon was bent away by the magnet dipole field of the experiment. In the case of the ALICE~3 setup with a solenoidal field in front of the photon detector, such an isolation cut is less straightforward.

An essential aspect of the DELPHI measurement of soft photons in $e^+e^- \to n\,\mathrm{jets}$ was the rejection of jets which contained an identified electron or positron track. DELPHI systematically studied the effect of changing the electron selection procedure resulting in an increased electron identification efficiency. No effect on the signal rate was found. The hypothesis that an extra amount of external bremsstrahlung produced an apparent excess was therefore rejected. For a measurement of ultra-soft photons in proton-proton collisions at the LHC, it will also be necessary to reject events containing an electron or positron track in the acceptance of the forward photon detector. Another essential ingredient is likely to be a specially shaped beam pipe to minimize the crossing of electrons and positrons at shallow angles. These two aspects are studied in detail with the aid of GEANT simulations in the following section.

In order to address the soft-photon puzzle at the LHC, measurements in different reaction systems are required. Exclusive proton-proton reactions, e.g.\ $\mathrm{pp} \to \mathrm{pp}
+ \text{J}/\psi(\to \mu^+\mu^-\gamma)$ or $\mathrm{pp} \to \mathrm{pp} \pi^+ \pi^- \gamma$ serve as a baseline where no deviations between the data and the expected inner bremsstrahlung signal are likely to be found. In addition, ultra-peripheral Pb--Pb collisions offer a unique opportunity. Jet production in ultra-peripheral Pb--Pb collisions might be useful to study the production of ultra-soft photons in jets similar to the DELPHI measurement with little background from an underlying event. The central measurement, however, would be in non-diffractive proton-proton collisions. Studying in particular the dependence of soft-photon radiation on the multiplicity of the event can provide insights into the source of a possible excess.

\subsubsection{Measuring the forward soft-photon spectrum with the Forward Conversion Tracker of ALICE~3}
\begin{figure}
    \centering
    \includegraphics[width = 0.65\textwidth]{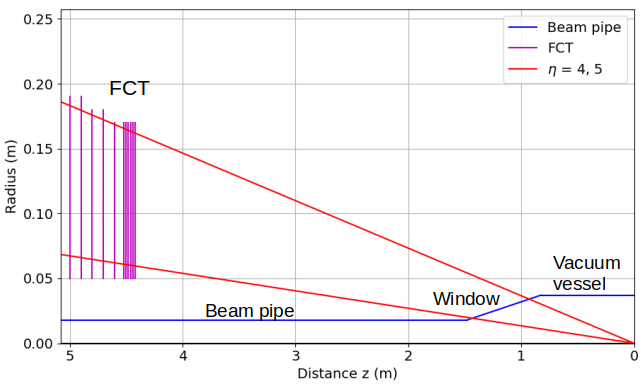}
    \caption{A schematic layout of the Forward Conversion Tracker (in magenta), a proposed soft-photon detector for ALICE 3, with a conical beam pipe (in blue) to allow for a reduced effective material path length of particles emitted in the pseudorapidity range $4 < \eta < 5$ (in red).}
    \label{fig:FCT_Window}
\end{figure}

The Forward Conversion Tracker planned for ALICE 3 is a detector that focuses on the measurement of soft-photons in the forward direction via the tracking of conversion $e^+e^-$ pairs. The vacuum vessel around the collision point of ALICE~3 will have a larger radius than the beam pipe to house its vertex locator, as described in the ALICE~3 letter of intent \cite{ALICE:2022wwr}. A slight modification of this design allows for a conical window in the beam pipe, such that particles emitted at pseudorapidity $4 < \eta < 5$ will pass through less effective material in comparison to a straight beam pipe. The design including such a window together with the Forward Conversion Tracker are shown in Fig.~\ref{fig:FCT_Window}.
The FCT consists of 11 circular silicon disks which subtend the pseudorapidity region $4 < \eta < 5$. The beam pipe and the window were constructed of beryllium with a thickness of $800$ $\mu$m, with a radius of the beam pipe of $1.8$\,cm and a radius of $3.7$ cm for the vacuum vessel. The inner radius of the disks of the FCT was kept the same at $5$\,cm, and the outer radius varied from $17$ to $19$\,cm in steps of $1$\,cm. With this design, simulation studies to investigate the magnitude of the signal photons in comparison to background photons were carried out. Two options are considered: the inclusion of a particle identification detector behind the FCT (not included in the figure) to distinguish electrons and veto the events which contain an electron/positron, and the inclusion of the conical window in the beam pipe. All combinations of including and/or excluding these two options make up the four scenarios examined.

\begin{figure}
    \includegraphics[origin=c,width=8cm]{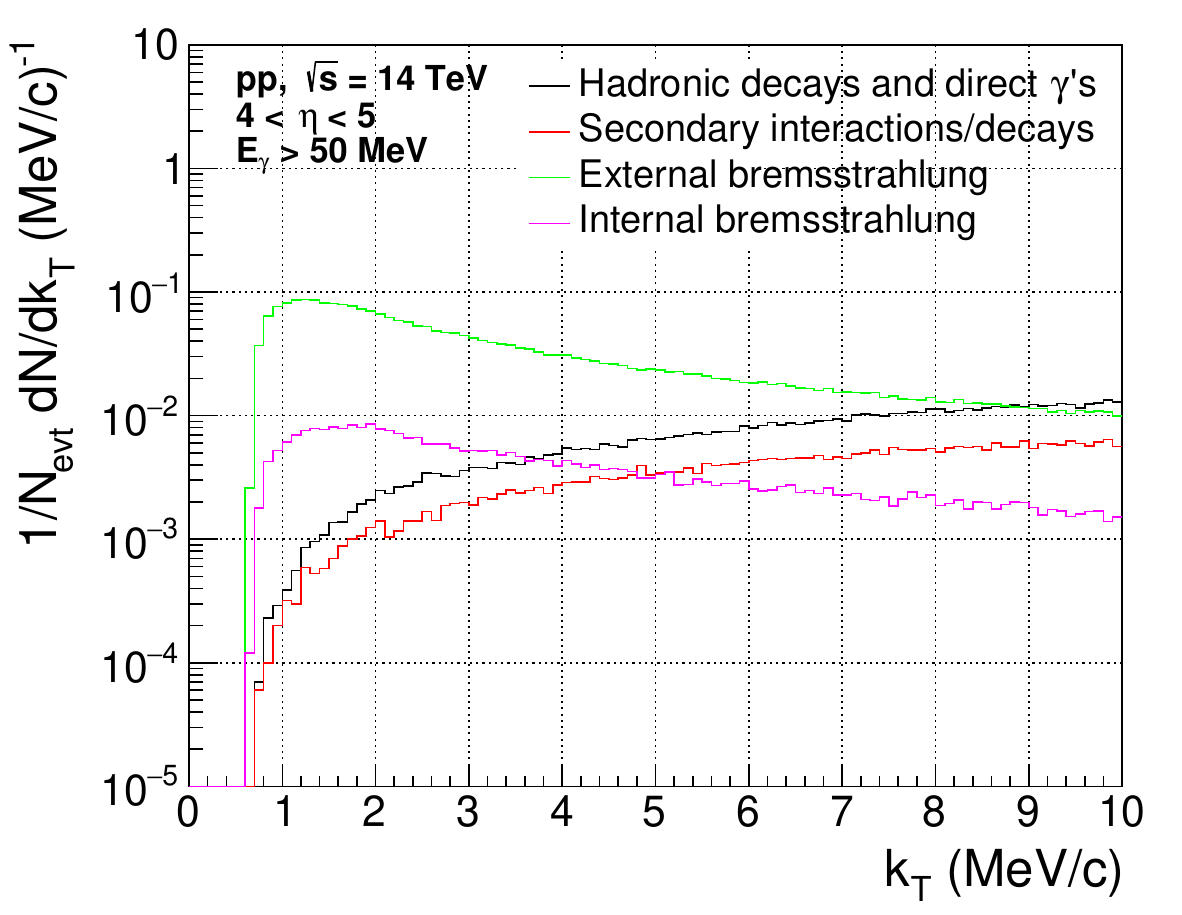}
    \includegraphics[origin=c,width=8cm]{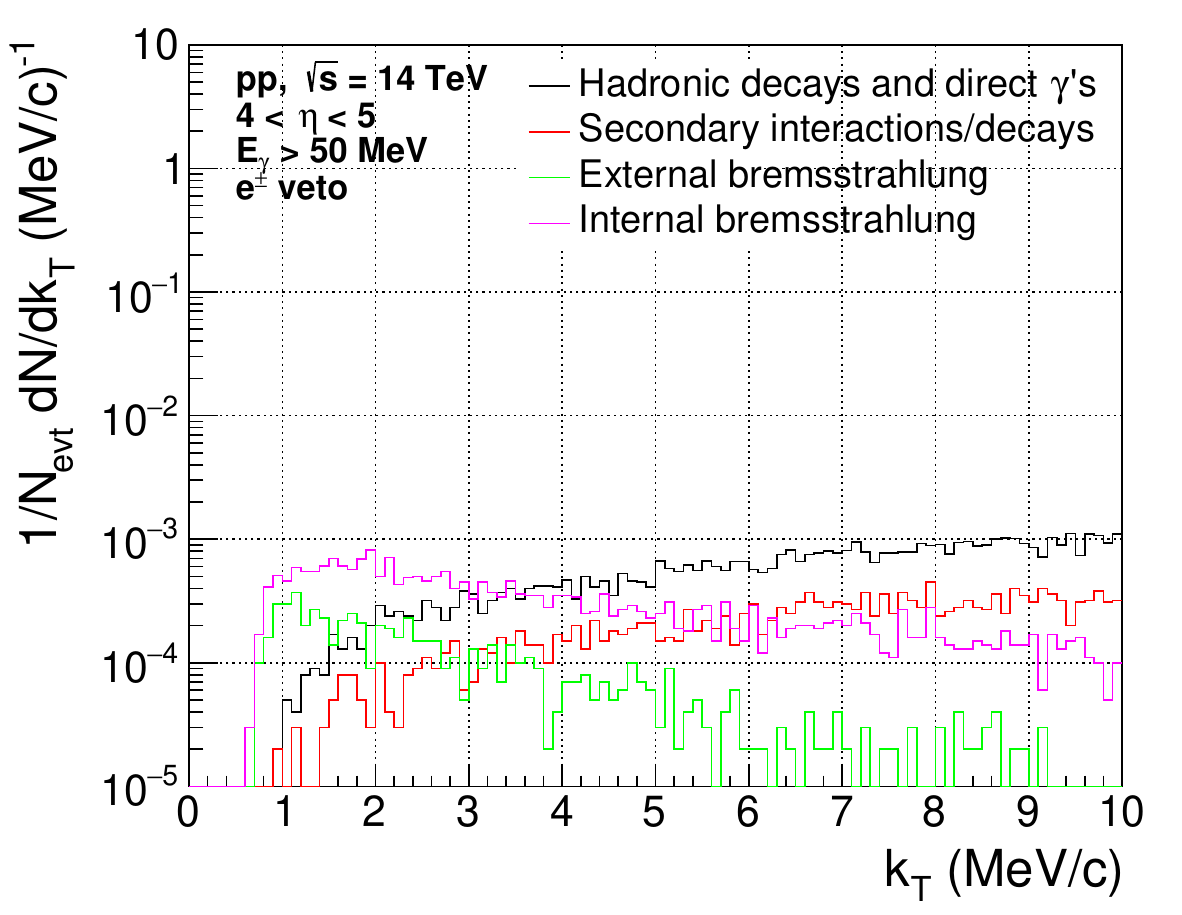}\\
    \includegraphics[origin=c,width=8cm]{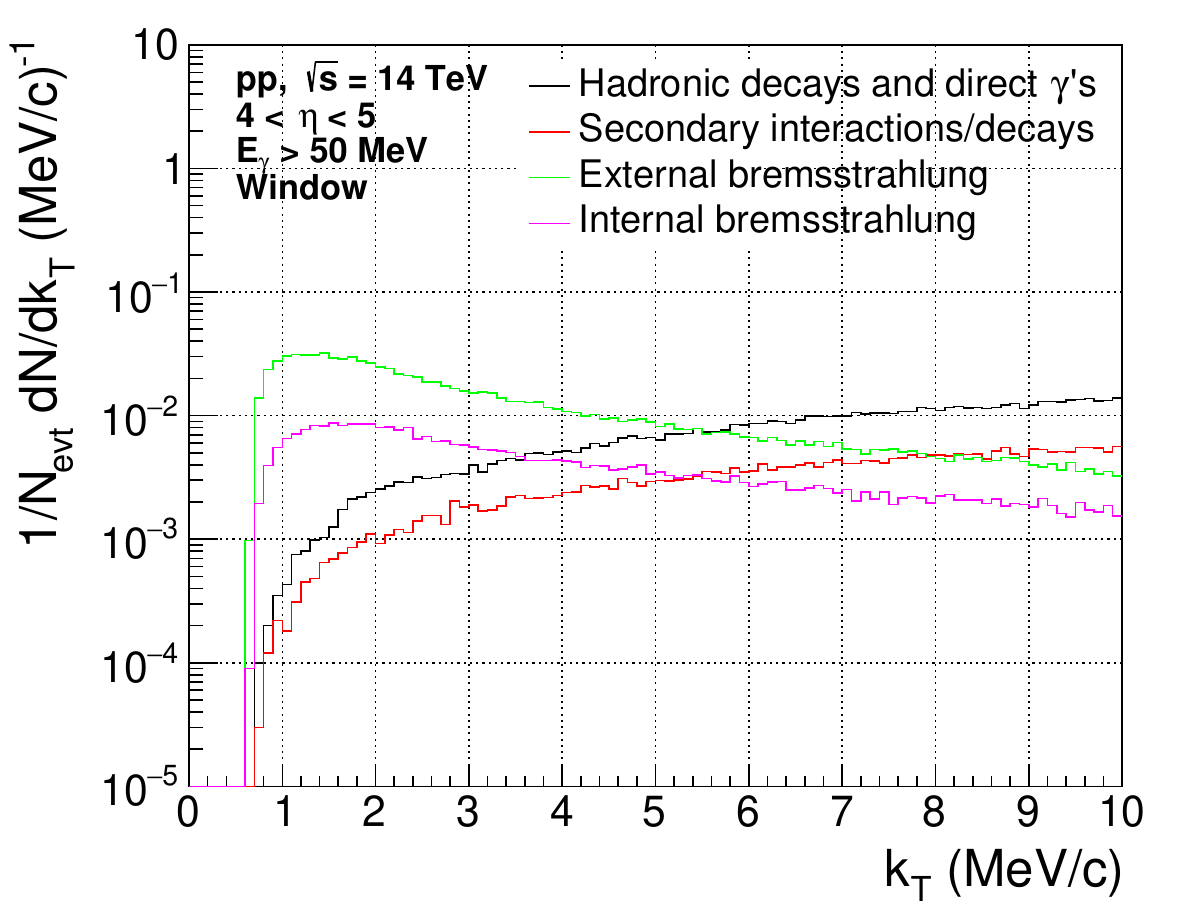}
    \includegraphics[origin=c,width=8cm]{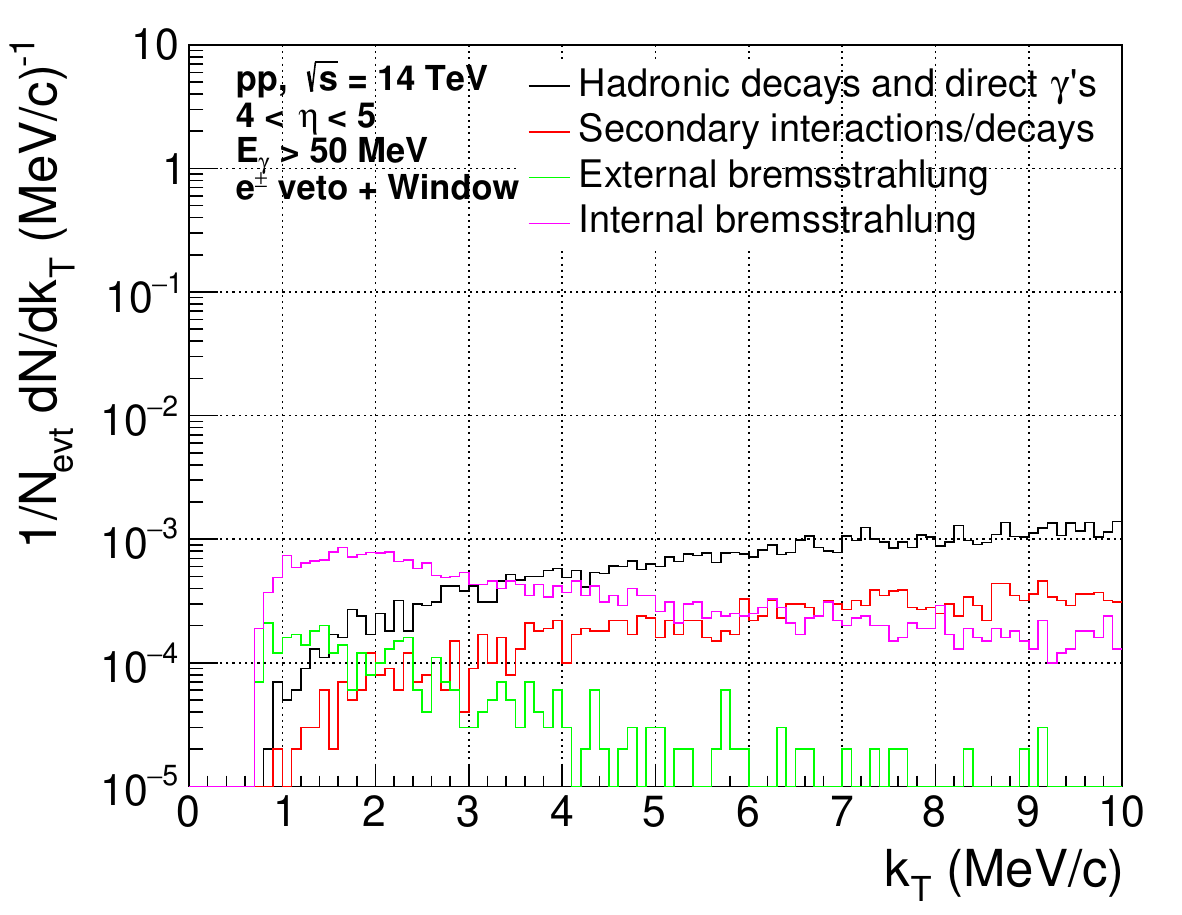}
    \caption{The photon transverse momentum spectrum for the scenarios: straight beam pipe without electron ID (top left), straight beam pipe including electron ID (top right), conical window in the beam pipe without electron ID (bottom left) and conical window in the beam pipe including electron ID (bottom right). The photon spectrum is divided in four channels: Hadronic decays and direct $\gamma$'s (black), which contains direct photons as well as photons from short-lived particle decays (e.g. $\pi^0\to\gamma\gamma$ and $\eta\to\gamma\gamma$), Secondary interactions/decays (red), which contains photons from the decay of secondary particles (e.g. $K^0_s\to\pi^0\pi^0\to 4\gamma$) and secondary interactions (e.g. $\pi^+ + \,\text{material}\to X + \pi^0\to{2\gamma}$), External bremsstrahlung (green), photons which find their origin in charged particle interactions with material radiating a bremsstrahlung photon, and Internal bremsstrahlung (magenta), which is the predicted soft-photon spectrum by Low's formula. The $k_T$ of the photons is expressed with respect to the beam axis, which is used as a proxy for the direction of the charged particles that emitted the photons.}
    \label{fig:FCT_simulation_figures}
\end{figure}

For these simulation studies, PYTHIA~8.3 was used to generate proton-proton collisions at $\sqrt{s}=14\,\text{TeV}$. The internal bremsstrahlung spectrum at $4 < \eta < 5$ was then generated according to Eq.~(\ref{eq:brems_formula_as_used_by_experiments-hadrons_b}) using a dedicated generator that takes the particle stack from PYTHIA. The generated particles (including soft photons) are then propagated through the detector setup with GEANT4 using the FTFP\_BERT\_EMV and G4OpticalPhoton physics lists. The simulated photon spectrum, separated in four different channels, measured at the first layer of the FCT, is illustrated in Fig.~\ref{fig:FCT_simulation_figures} for the four distinct scenarios.
For the ALICE~3 letter of intent \cite{ALICE:2022wwr}, a simulation was done which showed that the pointing angle resolution of the FCT for photons was good enough such that those with a pointing angle above 25\,mrad could be rejected. This rejection substantially reduced the contribution from the decay of secondary particles. The contribution from external bremsstrahlung was reduced as well but remains very significant in comparison to the signal photons as can be seen in the top left figure in Fig.~\ref{fig:FCT_simulation_figures}. From these figures it becomes clear that the inclusion of a conical window substantially reduces the external bremsstrahlung contribution, but the signal (inner bremsstrahlung) to background (all the other contributions) ratio in the region of $1 < k_T < 5$ MeV$/c$ is not yet satisfactory. The signal-to-background ratio in the region of interest increases to about unity with the inclusion of a veto on events containing electrons/positrons, and increases even further when a conical window in the beam pipe is included.
\begin{table}[!htb]
\resizebox{\textwidth}{!}{
\setlength\extrarowheight{2pt}
\footnotesize
\begin{tabularx}{\textwidth}{L{2.5cm} X L{1.8cm} L{3.1cm} C{1.8cm} L{1.8cm} L{2.0cm}}
Exp. & year & $p_\mathrm{beam}$ or $\sqrt{s}$ & photon $k_T$ & $\gamma_\mathrm{meas}/\gamma_\mathrm{brems}$ & $\gamma_\mathrm{brems}/\gamma_\mathrm{bkg}$ & Ref. \\
\hline
$\pi^+\text{p}$ & 1979 & 10.5\,GeV/$c$ & $k_T < 20\,\mathrm{MeV}/c$ & $1.25 \pm 0.25$ & 0.67 & Goshaw~et~al. \cite{Goshaw:1979kq} \\
$\pi^-\text{p}$ \linebreak CERN, WA91, OMEGA & 2002 & 280\,GeV/$c$ & $k_T < 20\,\mathrm{MeV}/c$ \newline ($0.2 < E_\gamma < 1\,\mathrm{GeV}$) & $5.3 \pm 1.0 $ & 0.47 & Belogianni at al. \cite{Belogianni:2002ib} \\
$\text{pp}$ \linebreak CERN, WA102, OMEGA & 2002 & 450\,GeV/$c$ & $k_T < 20\,\mathrm{MeV}/c$ \newline ($0.2 < E_\gamma < 1\,\mathrm{GeV}$) & $4.1 \pm 0.8 $ & 0.38 & Belogianni at al. \cite{Belogianni:2002ic} \\
$\text{e}^+ \text{e}^- \to n \, \mathrm{jets}$ \linebreak CERN, DELPHI & 2006 & 91\,GeV ($\sqrt{s}$)& $k_T < 80\,\mathrm{MeV}/c$ \newline ($0.2 < E_\gamma < 1\,\mathrm{GeV}$) & $4.0 \pm 0.3 \pm 1.0 $ & 0.036--0.013 & DELPHI \cite{DELPHI:2005yew,DELPHI:2010cit} \\
$\text{pp}$ \linebreak ALICE 3 \linebreak no w., no $\text{e}^{\pm}$ veto & $>$2034 & 14\,TeV ($\sqrt{s}$)& $1 < k_T < 5\,\mathrm{MeV}/c$ \newline ($0.1 < E_\gamma < 0.4\,\mathrm{GeV}$) & ? & 0.07--0.12 & ALICE 3 \cite{ALICE:2022wwr} \\
$\text{pp}$ \linebreak ALICE~3 \linebreak $\text{e}^{\pm}$ veto & $>$2034 & 14\,TeV ($\sqrt{s}$)& $1 < k_T < 5\,\mathrm{MeV}/c$ \newline ($0.1 < E_\gamma < 0.4\,\mathrm{GeV}$) & ? & 0.3--1.8 & ALICE 3 \cite{ALICE:2022wwr} \\
$\text{pp}$ \linebreak ALICE~3 \linebreak window & $>$2034 & 14\,TeV ($\sqrt{s}$)& $1 < k_T < 5\,\mathrm{MeV}/c$ \newline ($0.1 < E_\gamma < 0.4\,\mathrm{GeV}$) & ? & 0.2--0.3 & ALICE 3 \cite{ALICE:2022wwr} \\
$\text{pp}$ \linebreak ALICE~3 \linebreak window, $\text{e}^{\pm}$ veto & $>$2034 & 14\,TeV ($\sqrt{s}$)& $1 < k_T < 5\,\mathrm{MeV}/c$ \newline ($0.1 < E_\gamma < 0.4\,\mathrm{GeV}$) & ? & 0.33--2.5 & ALICE 3 \cite{ALICE:2022wwr}
\end{tabularx}}
\caption{\label{tab:soft_photon_experiments_s_over_b} Overview of the predicted signal ($\gamma_\mathrm{brems}$) over background ($\gamma_\mathrm{bkg}$) at various soft-photon measurements done in the past, as well as for a potential measurement at ALICE 3. In these past experiments the variable that was used to  express photon transverse momentum is $p_T$, but in this paper $k_T$ is used for consistency.}
\end{table}

It is interesting to compare ALICE~3's signal-to-background ratio with previous experiments. This is done in Tab.~\ref{tab:soft_photon_experiments_s_over_b}. It should be noted that the calculation of the signal over background of the previous experiments is based on the expected inner bremsstrahlung signal, not on the actually measured soft-photon signal. In the following, the calculation for the predicted signal over background as depicted in Table~\ref{tab:soft_photon_experiments_s_over_b} will be explained. The kinematic range in question can be inferred from the same table. For the first entry, the paper \cite{Goshaw:1979kq} mentions that the predicted number of direct photons was 671. In the same kinematic range, 1006 background photons were measured, which was inferred by deduction from the reported measured amount of direct photons which was $840\pm166$ accounting for $(45.5 \pm 8.9)\%$ of the total amount of photons measured. This then gives the signal over background of 0.67. For the second entry, the paper \cite{Belogianni:2002ib} lists the predicted yields for various soft photons in its first table, which gives a signal over background of 0.47. For the third entry, the paper \cite{Belogianni:2002ic} lists the predicted yields for various soft photons in a similar table, its first table, which gives a predicted signal over background of 0.38. For the fourth entry, it is a little more cumbersome to calculate the predicted signal over background, as no direct values for the yields were given. Instead, this paper \cite{DELPHI:2005yew} presents the Real Data (RD), Monte Carlo data (MC) and expected inner bremsstrahlung (Brems) in graphs that depict the ratio RD/MC, the difference $\text{RD} - \text{MC}$ and Brems. The MC data do not include the inner bremsstrahlung photons. To get to the expected signal over background (i.e.\ Brems/MC), the RD can be expressed as $\text{RD} = (\text{MC} + A \cdot \text{Brems})$, where $A$ is the enhancement of the expected signal which is 4.0 in this case. Combining this with $(\text{RD} - \text{MC})/\text{Brems} = A$ gives $(\text{RD}/\text{MC} - 1)/A = \text{Brems}/\text{MC}$, which in the specified kinematic region is 0.01 to 0.036.

As shown in Tab.~\ref{tab:soft_photon_experiments_s_over_b}, the signal-to-background ratio of the previous fixed-target experiments is significantly smaller than that of the DELPHI measurement in $\text{e}^+ \text{e}^- \to n \, \mathrm{jets}$. In the fixed-target experiments, the soft-photon excess is mostly within a 10--20\,mrad cone around the beam axis \cite{DELPHI:2005yew}. The background of photons from $\pi^0$ and $\eta$ decays at the correspondingly small $k_T$ with respect to the beam axis is small. This follows from the decreasing yield of $\pi^0$ and $\eta$ decay photons at low $k_T$. For a vanishing $p_T$ of a $\pi^0$ or $\eta$, the decay-photon yield actually peaks at $k_T = m/2$ (``Jacobian peak''), where $m$ is the mass of the parent particle; for a finite $p_T$, the decay-photon yield still drops rapidly to low $k_T$. For the jet production at the Z pole studied by DELPHI, the initial direction of the quark and the antiquark is determined from the reconstructed jet axis. This can be done with an accuracy of 50--60\,mrad \cite{DELPHI:2005yew}. As a result, the $k_T$ interval where the Low photons dominate is blurred and the decay-photon background becomes larger.

At the end of this section, we summarize the main results. The range $1 \lesssim k_T \lesssim 5$\,MeV/$c$ is an optimal region for testing Low's theorem, since the decay photon background is relatively small in this range. According to the relation $k_T = E_\gamma / \cosh \eta$, photons with very low transverse momentum can be measured at large pseudorapidities $\eta$ with the photon conversion method using the forward boost. The suppression of external bremsstrahlung from electrons and positrons is the main challenge for measuring the soft inner bremsstrahlung spectrum. If the ALICE~3 FCT is complemented by a detector that allows to veto events with an electron or positron in the acceptance of the FCT, a soft photon measurement at the LHC to test Low's theorem becomes possible. In addition, a window in the beam pipe to reduce the effective material in front of the FCT greatly simplifies the measurement. In our view, the ALICE~3 FCT is the best option for resolving the long-standing soft-photon puzzle in the foreseeable future.

\subsection{Specific examples: Internal bremsstrahlung in decays of $\text{J}/\psi$}

\subsubsection{Estimates of rates and spectra for $\text{J}/\psi$ production in ultraperipheral Pb--Pb collisions with STARlight}

Standard Model predictions for radiative decays of spin-one bosons such as
$Z$, $\Upsilon$, and $\mathrm{J}/\psi$ have been discussed in \cite{Fleischer:1984vt,
  Spiridonov:2004mp}. Analytical expressions for single-photon emission
that are valid down to arbitrary photon energies were derived
therein. It is photon emission from inner bremsstrahlung that determines
the line shape of these resonances when experimentally reconstructing
them from dileptons only. The first observation of the radiative decay
of $\text{J}/\psi \rightarrow e^ + e^- + \gamma$ with a photon energy
$E_\gamma > 100$ MeV was reported in \cite{FermilabE760:1996jsx}. The
relative branching ratio of
$BR(\text{J}/\psi \rightarrow e^ + e^- + \gamma) / BR(\text{J}/\psi \rightarrow e^ +
e^-) = (14.7 \pm 2.8)\%$ is in good agreement with expectations.

In ultraperipheral collisions, the two ions interact via their cloud of virtual photons.
At LHC energies, the radial
electric field lines of a lead nucleus are Lorentz contracted along the
beam direction and add coherently. As a consequence, cross sections for
photonuclear processes are enhanced quadratically by the atomic number
$Z^2 \approx 6700$ relative to the inelastic hadronic cross section.
Ultraperipheral collisions exhibit a particularly clean event topology, e.g., when a vector meson such as a $\text{J}/\psi$ meson is produced with typically rather low transverse momentum, the two decay leptons (electrons or muons) are oriented almost back-to-back and perpendicular to the beam axis with no other signal in the detector. The clean separation between electrons and muons in such collisions has already been achieved by ALICE in the central barrel~\cite{ALICE:2021gpt}.

We have used the event generator STARlight \cite{Klein:2016yzr} to calculate
$\text{J}/\psi$ meson production and their non-radiative decay to dileptons in
ultraperipheral Pb--Pb collisions as measured with the present ALICE apparatus which
covers a pseudorapidity range $(-1 < \eta < 1)$, and with the future ALICE~3
apparatus. Further, we applied Low's theorem, Eq.~\eqref{eq:brems_formula_as_used_by_experiments-mu},
to dilepton production from the non-radiative $\text{J}/\psi$ decay.
We note that any photon emission by the $\text{J}/\psi$ meson is forbidden by charge-conjugation invariance. This makes this channel unique and might be used as a benchmark.
\begin{figure}[htb]
\centering
    \includegraphics[width = 0.6\textwidth]{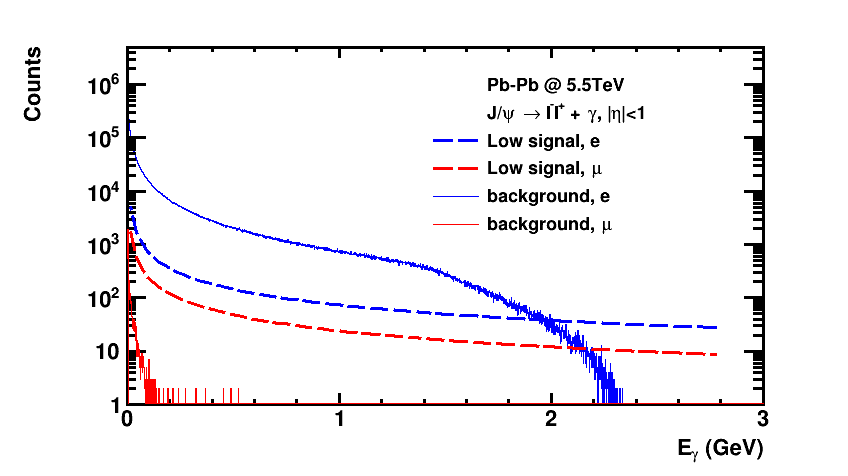}
     \caption{Soft-photon spectrum as predicted by Low's formula from
      $\text{J}/\psi$ decays into pairs of muons (red dashed curve) and electrons
      (blue dashed curve). The experimental background estimated by detailed
      Monte Carlo simulations with the present ALICE apparatus is shown
      by histograms.}
    \label{fig:jpsi-low}
\end{figure}
The soft-photon spectrum is shown in Fig.~\ref{fig:jpsi-low} for decays
to a pair of muons (red dashed curve) and electrons (blue dashed curve).
Here, the scale for the softness of photons is set by the two-body decay  and thus $m_{\text{J}/\psi}/2$.
The $1/k_{\text{T}}$
pole is clearly visible. The spectrum continues to arbitrarily large
photon energies, where Low's theorem is not applicable anymore.  Compared
to the dielectron channel, the photon yield in the dimuon channel is
only smaller by a factor of about three. This results from the velocity
$\beta$ explicitly occurring in Eq.~\eqref{eq:dI_angular_onecharge} with the muon
(electron) velocity $\beta_{\mu} = 0.9977$ ($\beta_e \approx 1$) for a
$\text{J}/\psi$ decaying at rest.

Experimental background from external bremsstrahlung of the decay
daughters in the detector material has been estimated by detailed Monte
Carlo simulations and is shown in Fig.~\ref{fig:jpsi-low} by the red
(blue) histograms for muons (electrons). While the large background for
electrons seems prohibitive for such a measurement in the central barrel,
the muon channel is essentially background free due to the much larger
mass of the muon, which suppresses the background component by a factor
$1/m$ to the power of four to six.
The lower limit in the photon energy is set by the capability of measuring the photon via conversion to electron-positron pairs in the detector material in front of the TPC.

We estimate the total number of
detected decays of $\text{J}/\psi$ mesons to $\mu^+\mu^-$ together with a
bremsstrahlung photon of $E_\gamma > 100 \,\text{MeV}$ to be about 1000, making
such a measurement possible. The estimated number of about 1000 detected
inner-bremsstrahlung photons might be reduced by a factor of up to two
in case of strong shadowing effects in the gluon distribution of the
lead nuclei, which has not been considered.

\begin{table}[htb]
    \centering
    \begin{tabular}{ccccc}
         & Pb--Pb  & $\text{p} \text{p} $ & Kr--Kr & O--O \\
         & $\sqrt{s_{NN}} = 5.52$\,TeV & $\sqrt{s_{NN}} = 14$\,TeV & $ \sqrt{s_{NN}} = 6.46$\,TeV & $\sqrt{s_{NN}} = 7$\,TeV \\
         \hline
         $\sigma$ &  41.761\,mb & 82.293\,nb & 2.974\,mb & $27.49\,\mu$b \\
         $\sigma \cdot \text{BR}(\text{e}^+ \text{e}^-) $ & 2.494\,mb & 4.914\,nb & 0.1776\,mb &  1.641$\mu$b \\
         $\sigma \cdot \text{BR}(\mu^+\mu^-) $ & 2.489\,mb & 4.905\,nb & 0.1773\,mb & 1.639\,$\mu$b\\
         L$_{\text{\rm month}}$ ($\text{nb}^{-1}$) & 5.6 & 5.1$\cdot 10^5$ & 8.4$\cdot 10^1$ & 1.6$\cdot 10^3$ \\
         \hline
         $\sigma_{\mu^\pm, \: |\eta|<4}$ & 2.044\,mb & 2.836\,nb &  0.142\,mb & 1.260\,$\mu$b\\
         $N_{\mu^\pm, \: |\eta|<4}$ & $11.5\cdot 10^6$ & $1.45\cdot 10^6$ &  $11.9\cdot 10^6$  & $2.02\cdot 10^6$  \\
         $N_{\gamma, \: E_\gamma > 100 \, {\rm MeV}}$ & $ 99 \cdot 10^3$ & $ 12 \cdot 10^3$ &  $ 102 \cdot 10^3$  & $ 17\cdot 10^3$  \\
          $N_{\gamma, \: {\rm detected} }$ & 94 & 12 & 97 & 16  \\
          \hline
         $\sigma_{\mu^\pm, \: 4< \eta <5}$ & 0.014\,mb & 0.07\,nb &  & \\
         $N_{\mu^\pm, \: 4< \eta <5}$ & $76\cdot 10^3$ & $35.7\cdot 10^3$ &   &  \\
         $N_{\gamma, \: E_\gamma > 100 \, {\rm MeV}}$ & 4.1 $ \cdot 10^3$ &  $1.9 \cdot 10^3$ &   &  \\
          $N_{\gamma, \: {\rm detected} }$ &  4  & 2
    \end{tabular}
    \caption{The total cross section calculated with STARlight, the
      number of lepton pairs decaying into the pseudorapidity range $|\eta|<4$ (central barrel) and $4 < \eta <5$ (Forward Conversion Tracker) per
      leptonic $\text{J}/\psi$ decay with the projected integrated luminosities as taken from \cite{ALICE:2022wwr}.}
    \label{alice-3-low}
\end{table}
\noindent
Estimates for the cross sections and the number of emitted dileptons and emitted and detected photons
in the pseudorapidity range $|\eta|<4$ (ALICE 3 central barrel) and $4 < \eta <5$ (ALICE 3, Forward Conversion Tracker) are summarized in
Table~\ref{alice-3-low}. The reported values for the collision energies and luminosities are taken from~\cite{ALICE:2022wwr}.
The acceptance of $\text{J}/\psi$ detection in the leptonic decay channels is
increased by a factor of 7 compared to the present ALICE setup. However, the fraction of photon conversions is reduced by about an order of magnitude due to the substantially decreased material budget when compared to the present ALICE setup. We assume a lepton tracking efficiency of 90\%, the effective material budget to be 1\% in units of radiation lengths resulting in a photon conversion probability of 0.78\%, and a reconstructing efficiency for the electron-positron pair from photon conversions of 15\%, thus an overall probability to detect a dilepton pair together with an inner-bremsstrahlung photon with energy $E_\gamma > 100 \, \text{MeV}$ to be approximately 0.095\%.

\noindent

The total number of photons generated per leptonic $\text{J}/\psi$
decay in the pseudorapidity range of $|\eta|<4$ and an energy range of
$E_\gamma \in (100\,\text{MeV}, 1.5\,\text{GeV})$ is respectively
\begin{equation}
\begin{split}
\label{Low Photonen per event alice 3}
    N_{\text{J}/\psi \rightarrow e^+ e^- \gamma} & \approx 0.2284 \: \text{per}\: \text{J}/\psi \rightarrow \text{e}^+ \text{e}^- \:\text{decay} \\
    N_{\text{J}/\psi \rightarrow \mu^+ \mu^- \gamma} & \approx 0.0861 \:
                                                \text{per}\: \text{J}/\psi
                                                \rightarrow \mu^+\mu^-
                                                \:\text{decay}.
\end{split}
\end{equation}
This number is larger than for $|\eta|<1$ since, due to the Lorentz boost at larger pseudorapidity, more photons appear above an energy threshold of $E_\gamma > 100\,\text{MeV}$. In the forward direction at  $4< \eta <5$, the boost is even larger. However, the limited angular coverage with respect to the flight direction of the emitting muon overcompensates this increase and the fraction of photon per muon pair is reduced to 5.39\%.

The background signal occurring during the interaction, where the
$\text{J}/\psi$ is being produced, has not been taken into account. The dominant
contribution is the direct-dimuon production in ultraperipheral collisions, discussed in further detail in \cite{upc} in general and in \cite{ALICE:2021gpt} with respect to the production of the $\text{J}/\psi$-vector meson. This continuum of non-resonant dimuon production has an even larger cross section than the $\text{J}/\psi$-to-dimuon-production cross section and might serve as a complementary measurement for inner bremsstrahlung.

\subsubsection{Results for the $\text{p} \text{p} \to \text{p} \text{p} (\text{J}/\psi \to \mu^{+} \mu^{-} \gamma)$ reaction in a Regge inspired toy model}

\begin{figure}
\centering
\includegraphics[width = 7cm]{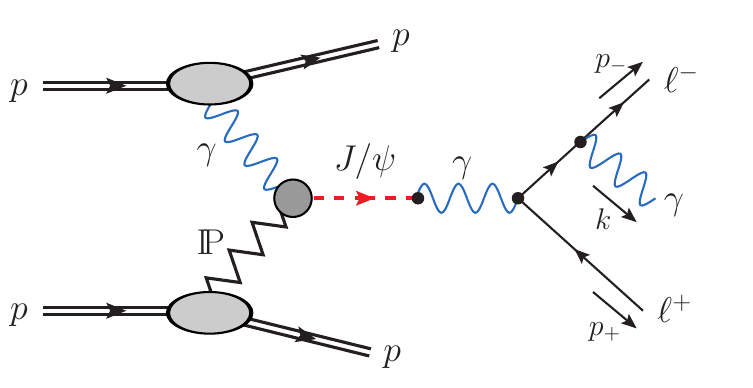}
\caption{\label{fig:diagram_pp_llgam} Photon-Pomeron
  ($\gamma$-$\mathbb{P}$) fusion mechanism for exclusive $\text{J}/\psi$
  production in $\text{p} \text{p}$ collisions with the decay
  $\text{J}/\psi \to \ell^{+}\ell^{-}\gamma$. There is also the diagram
  corresponding to the interchange of the two final leptons
  $\ell^{+}(p_{+}) \leftrightarrow \ell^{-}(p_{-})$.}
\end{figure}

An interesting process that would permit a rather clean identification
of photons is exclusive production of $\text{J}/\psi$ mesons in
$\text{p} \text{p}$ collisions,
\begin{equation}
\text{p} \text{p} \to \text{p} \text{p} (\text{J}/\psi \to \mu^{+}\mu^{-} \gamma)\,.
\label{pp_ppJpsi_mumugamma}
\end{equation}
Due to the negative charge parity of the $\text{J}/\psi$ meson the photon-Pomeron
fusion shown in Fig.~\ref{fig:diagram_pp_llgam} is the dominant production
mechanism. The full amplitude of the reaction
(Eq.~\eqref{pp_ppJpsi_mumugamma}) is a sum of the $\gamma \mathbb{P}$ and
$\mathbb{P} \gamma$ exchanges.\footnote{A similar process for the
  $\text{p} \text{p} \to \text{p} \text{p} \gamma$ reaction within the tensor-Pomeron approach is discussed in
  \cite{Lebiedowicz:2023mhe}. It is shown there that the
  photoproduction contribution wins over the diffractive bremsstrahlung
  one \cite{Lebiedowicz:2022nnn} for the photon rapidity $|y| < 4$
  and the absolute value of the transverse momentum of the photon
  $k_{\text{T}} \gtrsim 10$~MeV.
  Exclusive diffractive bremsstrahlung of one and two photons at forward rapidities in proton-proton collisions at the LHC was discussed in \cite{Lebiedowicz:2023rgc}.}

The $\text{p} \text{p} \to \text{p} \text{p} (\text{J}/\psi \to \mu^{+} \mu^{-})$
reaction can be treated in analogy to the
$\text{p} \text{p} \to \text{p} \text{p} (\phi \to \mu^{+} \mu^{-})$
reaction discussed in Sec.~III of \cite{Lebiedowicz:2019boz} within the tensor-Pomeron approach.
For simplicity, we have used the $b$-type coupling
in the $\mathbb{P} \text{J}/\psi \text{J}/\psi$ vertex and the transition
$\gamma$-$\text{J}/\psi$ is made here through the vector-meson-dominance model.
The values of the intercept and slope parameters of the Pomeron trajectory,
the coupling constants, and the form-factor parameters are
determined from a comparison of the model to the experimental data for
the $\gamma \text{p} \to \text{J}/\psi \text{p}$ reaction measured by ZEUS, H1, LHCb, and
ALICE Collaborations (see Fig.~3 of \cite{ALICE:2018oyo} for reference).
It has been checked that the model supplemented by the absorptive corrections
($\text{p} \text{p}$ rescattering) describes the LHCb data \cite{LHCb:2018rcm}
on $\text{p} \text{p} \to \text{p} \text{p} \text{J}/\psi$
at centre-of-mass energy $\sqrt{s} = 13$~TeV.
The ratio of full and Born cross sections (the gap survival factor)
is about 0.85--0.9.
In the case of Eq.~\eqref{pp_ppJpsi_mumugamma} a similar effect can be expected.

For the reaction
$\text{p} \text{p} \to \text{p} \text{p} (\text{J}/\psi \to \mu^{+} \mu^{-}
\gamma)$, we have
${\cal M}^{(2 \to 5)} = (\epsilon^{\mu}(k))^{*} {\cal M}_{\mu}$, where
$\epsilon$ is the polarization vector of the photon and
${\cal M}_{\mu} = {\cal M}^{(\gamma \mathbb{P})}_{\mu} + {\cal M}^{(\mathbb{P} \gamma)}_{\mu}$.
The amplitude ${\cal M}^{(\gamma \mathbb{P})}_{\mu}$
is obtained as in Eq.~(2.7) of \cite{Lebiedowicz:2019boz}
but with $i\Gamma^{(\phi KK)}_{\kappa}(p_{3},p_{4})$ replaced by
\begin{equation}
e g_{\text{J}/\psi \mu^{+} \mu^{-}}
\bar{u}(p_{-})
\left(
\frac{2p_{- \mu} + \gamma_{\mu} \!\! \not \!{k}}{2 p_{-} \cdot k} \gamma_{\kappa}
-
\gamma_{\kappa} \frac{2p_{+ \mu} + \not \!{k} \gamma_{\mu}}{2 p_{+} \cdot k}
\right)
v(p_{+})\,.
\label{approach_1}
\end{equation}
The coupling parameter
$g_{\text{J}/\psi \mu^{+} \mu^{-}} = 8.197 \times 10^{-3}$ is estimated from
the decay rate $\text{J}/\psi \to \mu^{+} \mu^{-}$ (neglecting radiative
corrections). For the $\mathbb{P} \gamma$-exchange we have the same
structure with the replacements Eq.~(2.23) of \cite{Lebiedowicz:2019boz}.
The $\text{p} \text{p} \to \text{p} \text{p} (\text{J}/\psi \to \mu^{+} \mu^{-} \gamma)$ reaction
is calculated within exact $2 \to 5$ kinematics using
the event generator GenEx~\cite{Kycia:2017ota}.

In the following we shall compare the full decay approach
(Eq.~\eqref{approach_1}) to the soft-photon approximation (SPA) where we keep
only the pole terms $\propto \omega^{-1}$ in the radiative amplitude.
The ${\cal M}_{\mu}$ amplitude is then expressed by the amplitude
without radiation
${\cal M}^{(\text{p} \text{p} \to \text{p} \text{p} (\text{J}/\psi \to \mu^{+} \mu^{-}))}$
times the photon emission factor
\begin{equation}
{\cal M}_{\mu}
\propto
{\cal M}^{(2 \to 4)}\;
e
\left(
\frac{p_{- \mu}}{p_{-} \cdot k} - \frac{p_{+ \mu}}{p_{+} \cdot k}
\right)
\,.
\label{approach_2}
\end{equation}

\begin{figure}
\includegraphics[origin=c,width=7cm]{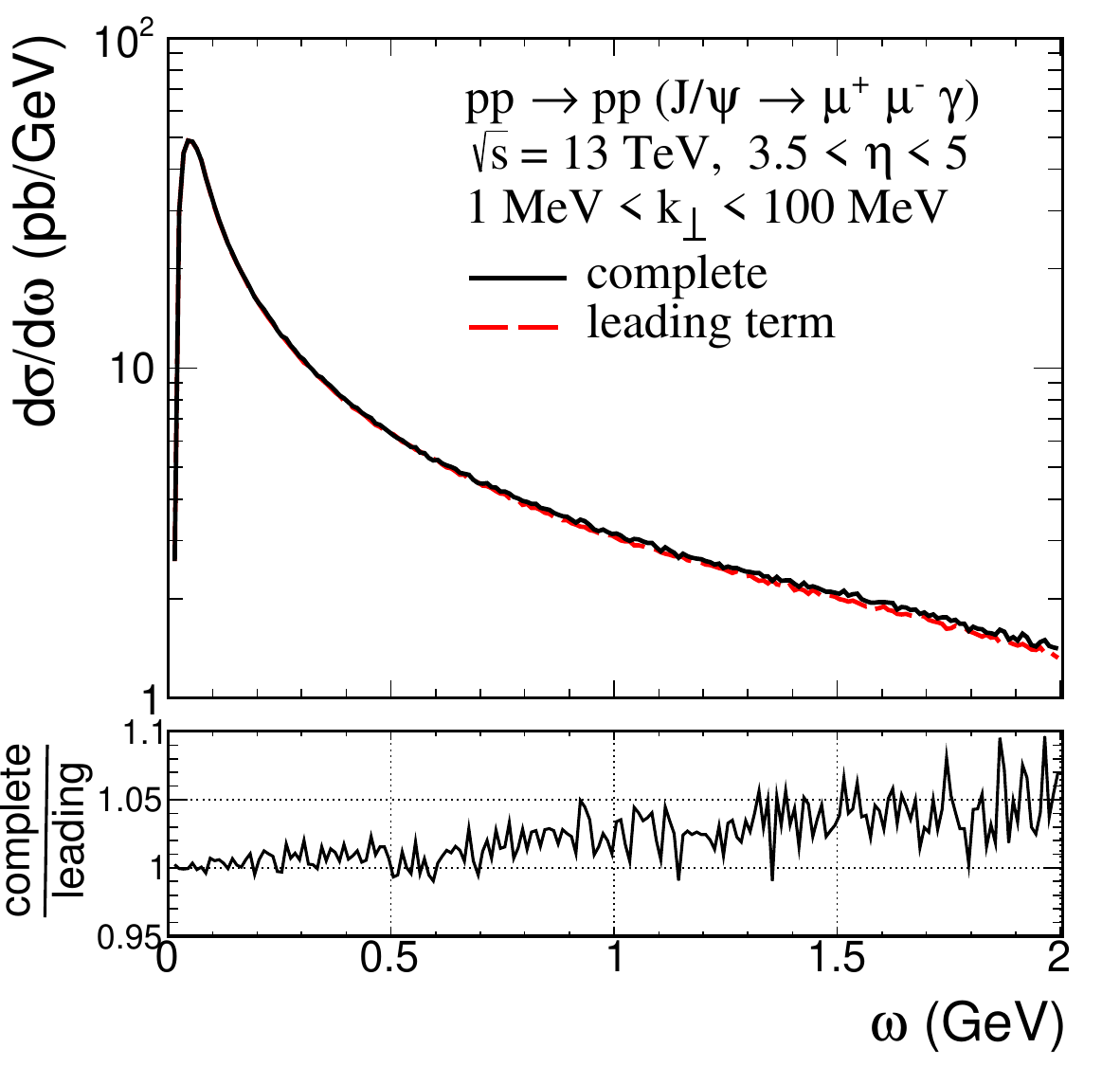}
\includegraphics[origin=c,width=7cm]{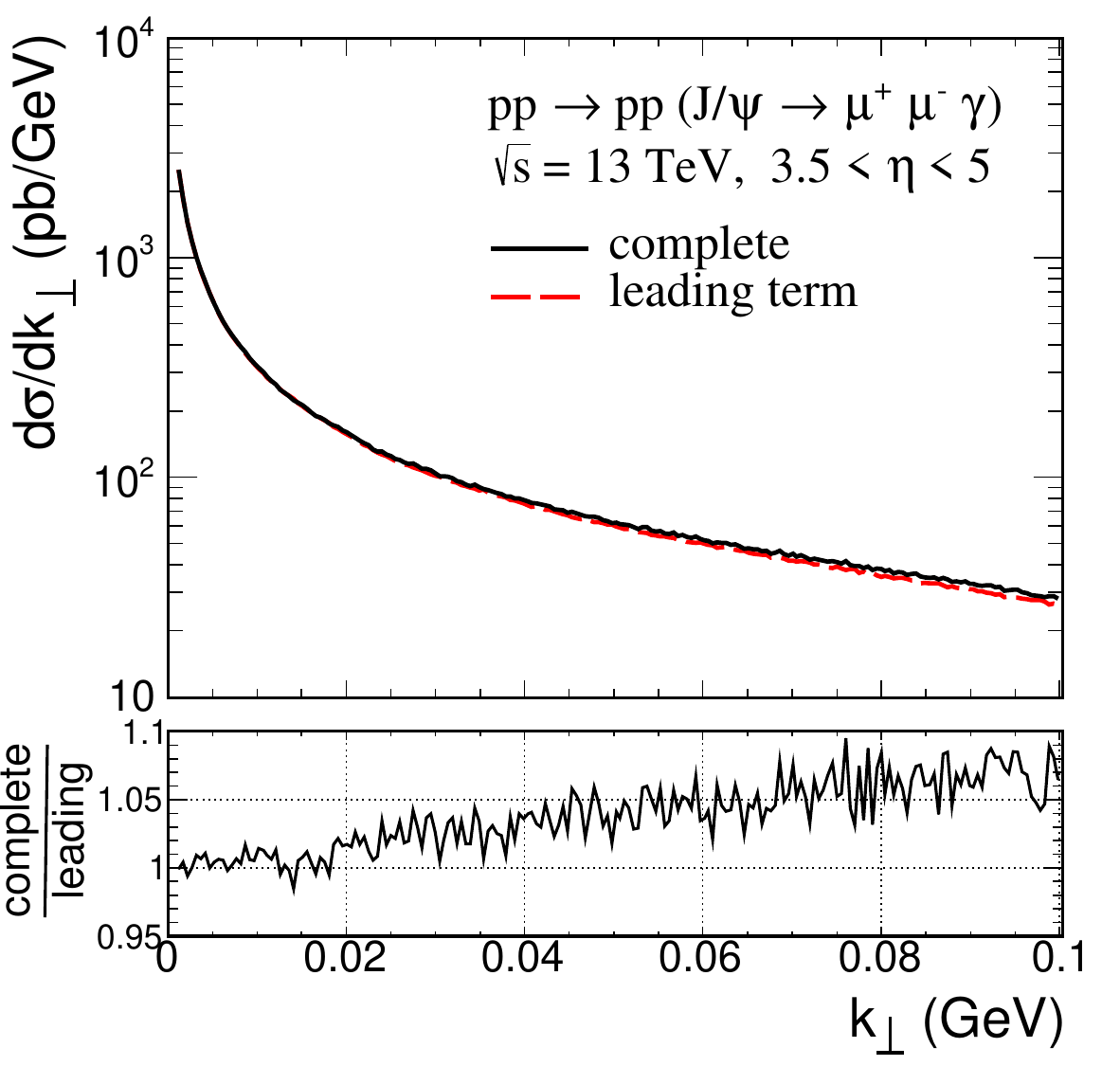}\\
\includegraphics[origin=c,width=7cm]{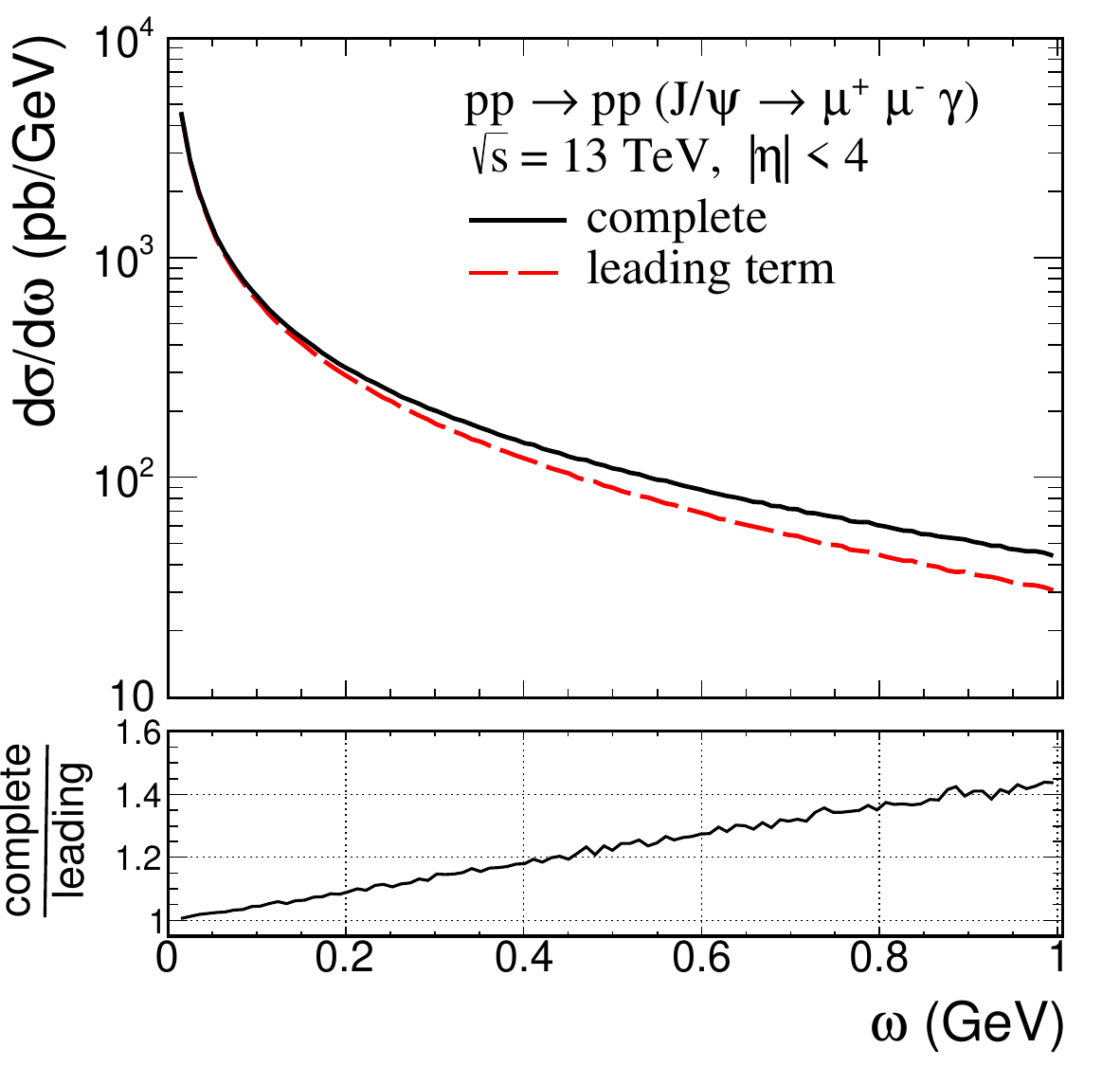}
\includegraphics[origin=c,width=7cm]{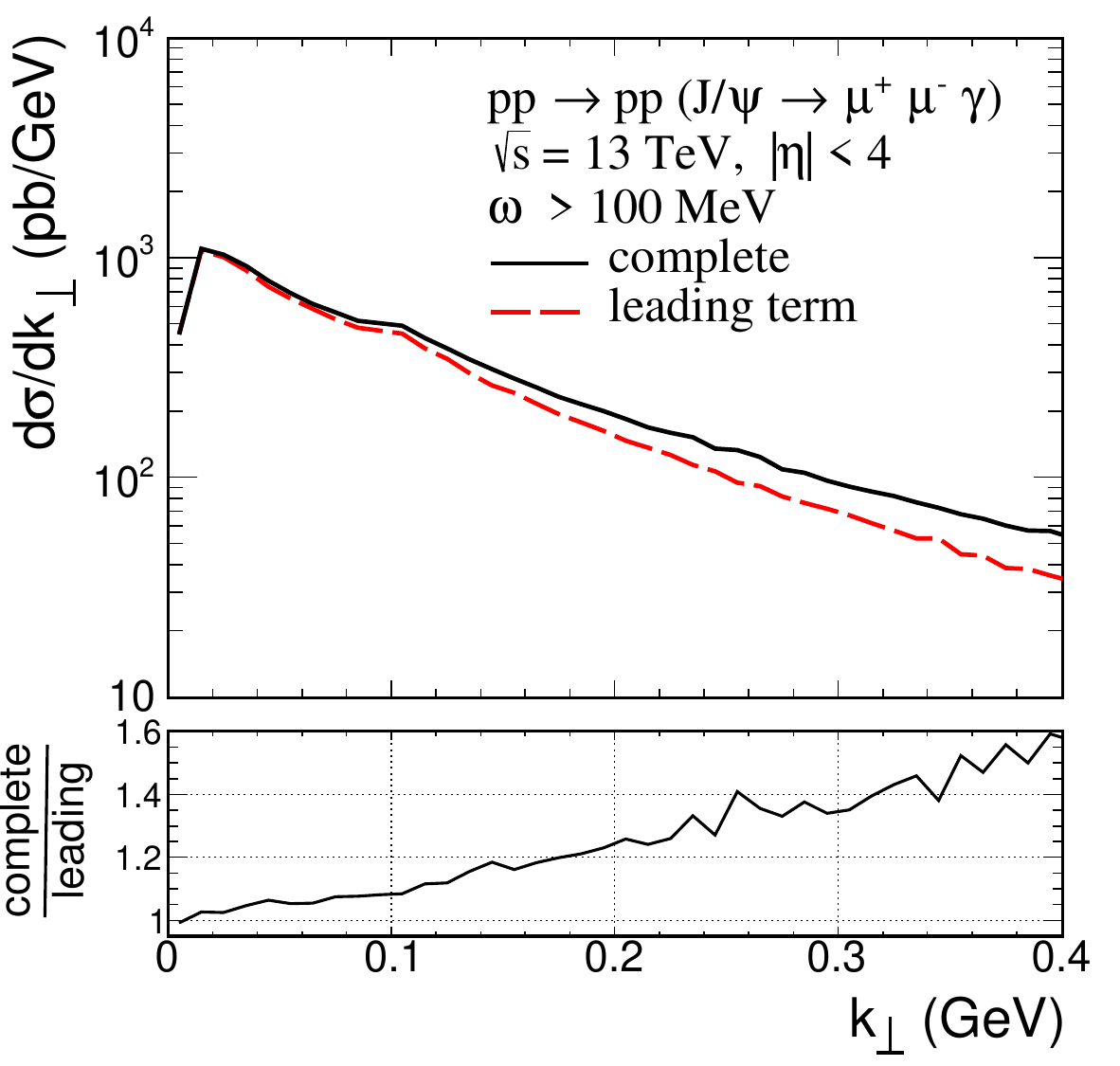}
\caption{\label{fig:pp_ppJpsi_mumugamma} The differential distributions
  in $\omega = E_\gamma$, the energy of the photon, and in $k_{\perp} = k_{\text{T}}$, the
  transverse momentum of the photon, for the
  $\text{p} \text{p} \to \text{p} \text{p} (\text{J}/\psi \to \mu^{+} \mu^{-}
  \gamma)$ reaction. The calculations were done for
  $\sqrt{s} = 13 \, \text{TeV}$, $3.5 < \eta < 5$,
  $1 \, \text{MeV} < k_{\text{T}} < 100 \, \text{MeV}$ (the top panels) and
  $|\eta| < 4$, $E_\gamma > 100 \, \text{MeV}$ (the bottom panels). The black solid
  line corresponds to the complete decay contribution
  (\ref{approach_1}), the red dashed line corresponds to SPA (leading
  term, see Eq.~(\ref{approach_2}). The ratios of the two contributions
  are shown in the lower panels.}
\end{figure}

In Tab.~\ref{Table_PL} we have collected integrated cross sections for
the exclusive reactions
$\text{p} \text{p} \to \text{p} \text{p} \mu^{+} \mu^{-}$ and
$\text{p} \text{p} \to \text{p} \text{p} \mu^{+} \mu^{-} \gamma$
calculated for $\sqrt{s} = 13$~TeV with different acceptance cuts.
Both photons and muons cover the same pseudorapidity range,
$k_{\text{T}}$ and $E_\gamma$ are the absolute value of the transverse momentum
and the energy of the photon in the overall c.m. system, respectively.
The absorptive corrections were not included in the calculations.
We show the results for the SPA (Eq.~\eqref{approach_2})
and complete decay (Eq.~\eqref{approach_1}).
Note that the cut on $k_{\text{T}}$ significantly reduces
the cross section compared to that of the cut on $E_\gamma$,
especially at forward photon rapidities.
In Fig.~\ref{fig:pp_ppJpsi_mumugamma}, we show the distributions
for the $\text{p} \text{p} \to \text{p} \text{p} \mu^{+} \mu^{-} \gamma$ reaction
for the complete result and SPA (leading term).
One can see from the top panels that both results are very close to each other.
The deviation of the SPA from the complete result are up to
around 5\% in the kinematic range considered ($3.5 < \eta < 5$,
$k_{\text{T}} < 0.06$~GeV, $E_\gamma < 2$~GeV).
For soft-photon emission, the leading term dominates.
The other terms have no singularity for $k \to 0$.
Our findings are that these are indeed important in a much
larger range of $E_\gamma$ and $k_{\text{T}}$.
From the bottom panels, we see that the deviation increase rapidly
with growing $E_\gamma$ and $k_{\text{T}}$ considering $|\eta| < 4$.
Here, the accuracy 5\% occurs up to $E_\gamma \simeq 0.1$~GeV
and $k_{\text{T}} \simeq 0.05$~GeV, and is 10\%
at $E_\gamma \simeq 0.2$~GeV and $k_{\text{T}} \simeq 0.1$~GeV.

\begin{table}
\begin{tabular}{ l| r | r }
\hline
$\text{p} \text{p} \to \text{p} \text{p} (\text{J}/\psi \to \mu^{+} \mu^{-})$ & \multicolumn{2}{r}{$\sigma$ (pb) \phantom{111111111111111111}} \\

no cuts & \multicolumn{2}{r}{3458.2 \phantom{111111111111111111}} \\
$|\eta| < 1$ & \multicolumn{2}{r}{304.5 \phantom{111111111111111111}} \\
$|\eta| < 4$ & \multicolumn{2}{r}{2671.5 \phantom{111111111111111111}} \\
$4 < \eta < 5$ & \multicolumn{2}{r}{68.8 \phantom{111111111111111111}} \\
\hline\\
\hline
$\text{p} \text{p} \to \text{p} \text{p} (\text{J}/\psi \to \mu^{+} \mu^{-} \gamma)$
& $\sigma$ (pb), SPA, Eq.~(\ref{approach_2})
& $\sigma$ (pb), full decay, Eq.~(\ref{approach_1})\\
\hline
$E_\gamma > 100$~MeV,\;
$|\eta| < 1$
&  8.8
& 12.3 \\
$E_\gamma > 100$~MeV,\;
$|\eta| < 4$
& 147.8
& 188.3\\
$E_\gamma > 100$~MeV,\;
$4 < \eta < 5$
& 5.9
& 6.6\\
$10~{\rm MeV} < E_\gamma < 1~{\rm GeV}$,\;
$|\eta| < 4$
& 269.4
& 290.2\\
\hline
$k_{\perp} > 1$~MeV,\;
$3.5 < \eta < 5$
& 18.3
& 19.8\\
$1~{\rm MeV} < k_{\text{T}} < 10~{\rm MeV}$,\;
$3.5 < \eta < 5$
& 7.4
& 7.4\\
$10~{\rm MeV} < k_{\text{T}} < 100~{\rm MeV}$,\;
$3.5 < \eta < 5$
& 7.0
& 7.2\\
$100~{\rm MeV} < k_{\text{T}} < 1~{\rm GeV}$,\;
$3.5 < \eta < 5$
& 3.9
& 5.2\\
$k_{\text{T}} > 100$~MeV,\;
$|\eta| < 1$
&  8.1
& 11.6\\
$k_{\text{T}} > 100$~MeV,\;
$|\eta| < 4$
&  75.8
& 113.1\\
\hline
\end{tabular}
\caption{The integrated cross sections in pb for the exclusive reactions
  $\text{p} \text{p} \to \text{p} \text{p} \mu^{+} \mu^{-}$ and
  $\text{p} \text{p} \to \text{p} \text{p} \mu^{+} \mu^{-} \gamma$ via the
  photoproduction mechanism. The results have been calculated for
  $\sqrt{s} = 13 \, \text{TeV}$ and for some acceptance cuts for ALICE 3.
  No absorption ($\text{p} \text{p}$ rescattering) effects are included here.}
\label{Table_PL}
\end{table}

\section{Electrical conductivity via soft photons and dielectrons from heavy-ion collisions}
\label{sec_hics}
The soft-photon and dilepton radiation emitted from hot QCD matter
close to thermal equilibrium is of high interest for various reasons.

First, it is
governed by universal physics describable in terms of relativistic fluid
dynamics. More specifically, the fluctuations in electromagnetic
currents that produce such soft radiation are in a close-to-equilibrium situation. These fluctuations are related to electromagnetic response functions via the
fluctuation-dissipation relation. The response functions are in turn
governed by the physical laws for the space-time evolution of
electromagnetic currents, and at small frequencies and momenta, these
can be formulated in terms of relativistic fluid dynamics. Interesting
fluid properties such as charge susceptibilities, electric
conductivity and an associated relaxation time govern then the shape of
electromagnetic response functions.

One can also aim to constrain these transport properties
by studying how electric charges move during the quark-gluon plasma
phase. This would need some knowledge about their initial distribution
and a corresponding extension of fluid-dynamical descriptions to
propagate electromagnetic currents, as well as of the freeze-out
prescription. One could then compare the constraints obtained in this way
to the information about fluctuating currents obtained via the produced
electromagnetic radiation, and thus test the fluctuation-dissipation
relation and the degree of local thermalization of the quark-gluon
plasma.
Alternatively, if the fluctuation-dissipation relation is assumed, one
can constrain the electromagnetic transport properties from the
electromagnetic radiation and the dilepton spectra in the soft limit of
small frequencies and momenta.

From another viewpoint, electromagnetic (EM) radiation from the locally thermalized medium
produced in high-energy heavy-ion collisions provides unique access to
an in-medium spectral function of strongly interacting matter that
encodes, \eg, information about its degrees of freedom and its strongly-coupled properties. This is apparent from the differential thermal production rate of
lepton pairs, $l^+l^-$ which is given by
\begin{equation}
\frac{dN_{ll}}{d^4xd^4q} = \frac{\alpha_\text{EM}}{12 \pi^4 M^2} n_\text{B}(\omega;T) \rho(\omega,M;T,\mu_\text{B}) \left( 1+\frac{2m^2}{M^2} \right) \sqrt{1-\frac{4m^2}{M^2}} \theta(M^2 - 4 m^2),
\end{equation}
where $q^\mu = p_1^\mu + p_2^\mu$ is the momentum of the lepton pair,
$M$ the corresponding invariant mass, $m$ is the lepton mass,
$\alpha_\text{EM}=e^2/(4\pi)$ is the fine-structure constant, $T$ is the
temperature and $\mu_\text{B}$ the baryon chemical potential \cite{Feinberg:1976ua,McLerran:1984ay,Gale:1990pn}. We also
use the frequency in the fluid rest frame $\omega$. The function
$n_\text{B}(\omega;T)=1/(e^{\omega/T}-1)$ is the Bose-Einstein function,
and $\rho(\omega,M;T,\mu_\text{B})$ is the spectral function associated
to the trace of the electromagnetic current-current correlation
function.\footnote{In this paper the electromagnetic-current operators are defined
  including the electromagnetic coupling constant, $e=\sqrt{4 \pi \alpha_{\text{EM}}}$, i.e., the spectral function $\rho(\omega,M;T,\mu_{\text{B}})$ contains a factor $e^2=4 \pi \alpha_{\text{EM}}$.}

In the vacuum, this spectral function is well known from the inverse
process of electron-positron annihilation into hadrons. It can be
characterized by three regions which in turn set the stage for its
applications to heavy-ion collisions: In the low-mass region (LMR),
$M\lsim 1 \,\text{GeV}$, its strength is essentially saturated by the light
vector mesons $\rho$, $\omega$ and $\phi$, highlighting the formation of
massive and confined degrees of freedom as the basic excitations of the
chirally broken QCD vacuum. In the intermediate-mass region (IMR),
$1.5\, \text{GeV} \lsim M \lsim 3\, \text{GeV}$, it is essentially a continuum with a
strength given by perturbative $q\bar q$ production, \ie, a
short-distance production process which factorizes from the subsequent
hadronization.  A ``dip structure'' exists in the transition mass region (TMR)
from LMR to IMR, 1\,GeV\,$\lsim M \lsim$\,1.5\,GeV.

In experiment, thermal radiation will be observed as an integral over
the entire space-time evolution of the expanding fireball. However,
there are means to characterize the different emission times. For one,
there is a competition between the emitting three-volume and the Bose
occupation factor of the emitted (virtual) photon. The former favors
later emission, typically increasing with a power law $\sim 1/T^m$ with
$m\simeq 5$~\cite{Rapp:2011is}, while the latter exponentially favors
early emission where temperatures are higher. Since the exponential
dependence on temperature strongly increases with mass, the IMR is
dominated by early (QGP) emission while the LMR receives large
contributions from later (hadronic) emission. The mass variable in the
emission rate can further be used in connection with the collective
properties of the fireball: early emission is expected to carry less
transverse and elliptic flow than later emission, which can be detected
through the $q_T$ spectra in different mass bins~\cite{NA60:2008ctj}.

In the following, we focus on the prospects of dielectron measurements
in the very low-mass region, where the spectral function relates to the
electric conductivity of the created medium ($\sigma_{\text{EM}}$). After
motivating the need for experimental constraints on this transport
coefficient (Sec.~\ref{motivation}), we elaborate on the relation
between the mass spectra of ${\rm e^{+}e^{-}}$ pairs produced thermally
in the fireball and $\sigma_{\text{EM}}$ (Sec.~\ref{predictions}). For this
purpose, predictions from the model in \cite{Rapp:2013nxa,Atchison:2022yxm} are
used. Additional physical background sources of dielectrons are studied
in Sec.~\ref{physicsbackground} in the case of a measurement with the
ALICE 3 experiment~\cite{ALICE:2022wwr} planned at the LHC for
2035. These physical background sources are compared to the thermal
signal in Sec.~\ref{results}, where the main challenges of such
measurements are summarized.

\subsection{Electric conductivity}
\label{motivation}

In the low-frequency and -momentum regime, the spectral function
$\rho(\omega, M;T, \mu_\text{B})$ in a medium is dominated by the
so-called transport peak and specifically by the electric conductivity
$\sigma_{\text{EM}}$. In the fluid rest frame and for vanishing absolute value of the three-momentum one has
\begin{equation}
\sigma_{\text{EM}} = \frac{1}{3} \lim_{\omega\to 0} \frac{
\rho(\omega, M;T, \mu_\text{B})
}{\omega} \ .
\end{equation}
For a more detailed discussion, also about the form of the spectral
function in the regime of small frequencies and momenta, where it can be
constrained by fluid dynamic considerations, see Ref.\
\cite{Floerchinger:2021xhb}.

The electric conductivity is one of the key bulk properties of the
medium created in heavy-ion collisions, as fundamental as the shear
viscosity or a heavy-flavor diffusion coefficient. Close to thermal
equilibrium, and in the fluid-rest frame, it relates the electric
current density with the electric field,
\begin{equation}
\boldsymbol{J} = \sigma_{\text{EM}} \boldsymbol{E}.
\end{equation}
In other words, it determines the diffusion of electric charges in the
medium and their response to electric fields. As a consequence, it
influences for example the time evolution of electromagnetic fields
generated by spectator protons in non-central heavy-ion
collisions~\cite{Huang:2015oca, Tuchin:2015oka}. The understanding of
phenomena related to the presence of these strong magnetic fields, like
the Chiral Magnetic Effect, requires precise knowledge of the electric
conductivity at the early stage of the collision.

\begin{figure} [ht!]
    \centering
    \includegraphics[width=0.63\textwidth]{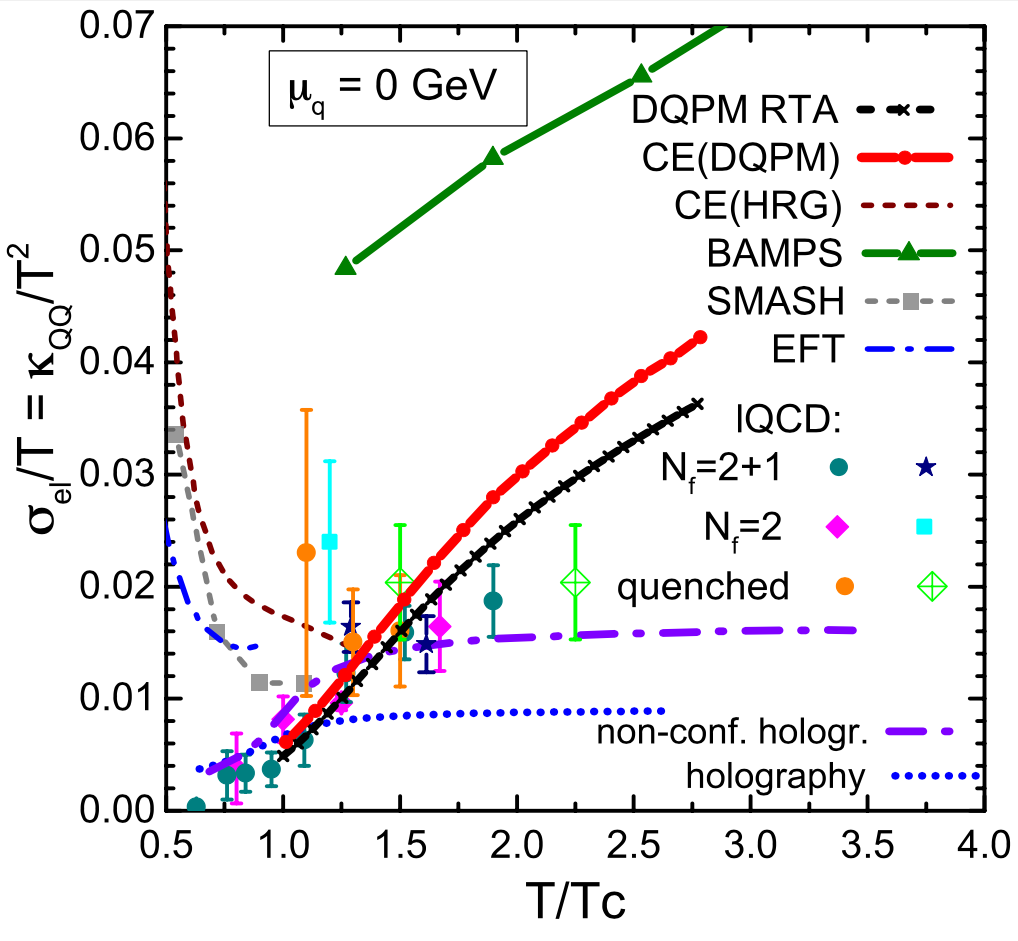}
    \caption{Different theoretical predictions for the electric
      conductivity ($\sigma_{\rm el} = \sigma_{\text{EM}}$) of the partonic
      and hadronic matter~\cite{Fotakis:2021diq}.}
    \label{fig:Elcontheo}
\end{figure}

The predicted temperature dependence of the ratio $\sigma_{\text{EM}}/T$ is
shown in Fig.~\ref{fig:Elcontheo} for different models employing kinetic
approaches~\cite{Fotakis:2021diq,Greif:2014oia,Rose:2020sjv,Hammelmann:2018ath,SMASH:2016zqf,Greif:2017byw,Fotakis:2019nbq,Greif:2016skc},
lattice QCD~\cite{Ding:2016hua,PhysRevLett.99.022002,Brandt:2012jc,Brandt:2015aqk,Aarts:2014nba,PhysRevD.102.054516},
holographic models~\cite{Rougemont:2015ona,Finazzo:2013efa} as well as
effective field theory~\cite{Torres-Rincon:2012sda}. While the
theoretical calculations show a large spread of the expected temperature
dependence below and above $T_{\rm c}= 158 \,\text{MeV}$, there is no constraint by measurements until now.

The experimental assessment of $\sigma_{\text{EM}}$ will likely require
dielectron measurements at masses and momenta below $\sim 100\, \text{MeV}$, which
we will refer to as very-low-mass region (vLMR).  Different electric
conductivities correspond to different heights and widths of the
spectral function near the zero-energy limit~\cite{Moore:2006qn,
  Floerchinger:2021xhb} and therefore different predicted yields of
thermal ${\rm e^{+}e^{-}}$ in the vLMR.  A decisive advantage over other
transport coefficients is that it directly connects to the microscopic
physics of low-mass measurements that have progressed substantially over
the last two decades~\cite{Rapp:2009yu,Rapp:2016xzw}.

\subsection{Thermal radiation signal for two different scenarios}
\label{predictions}
\begin{figure}[ht!]
    \centering
    \includegraphics[width=0.49\textwidth]{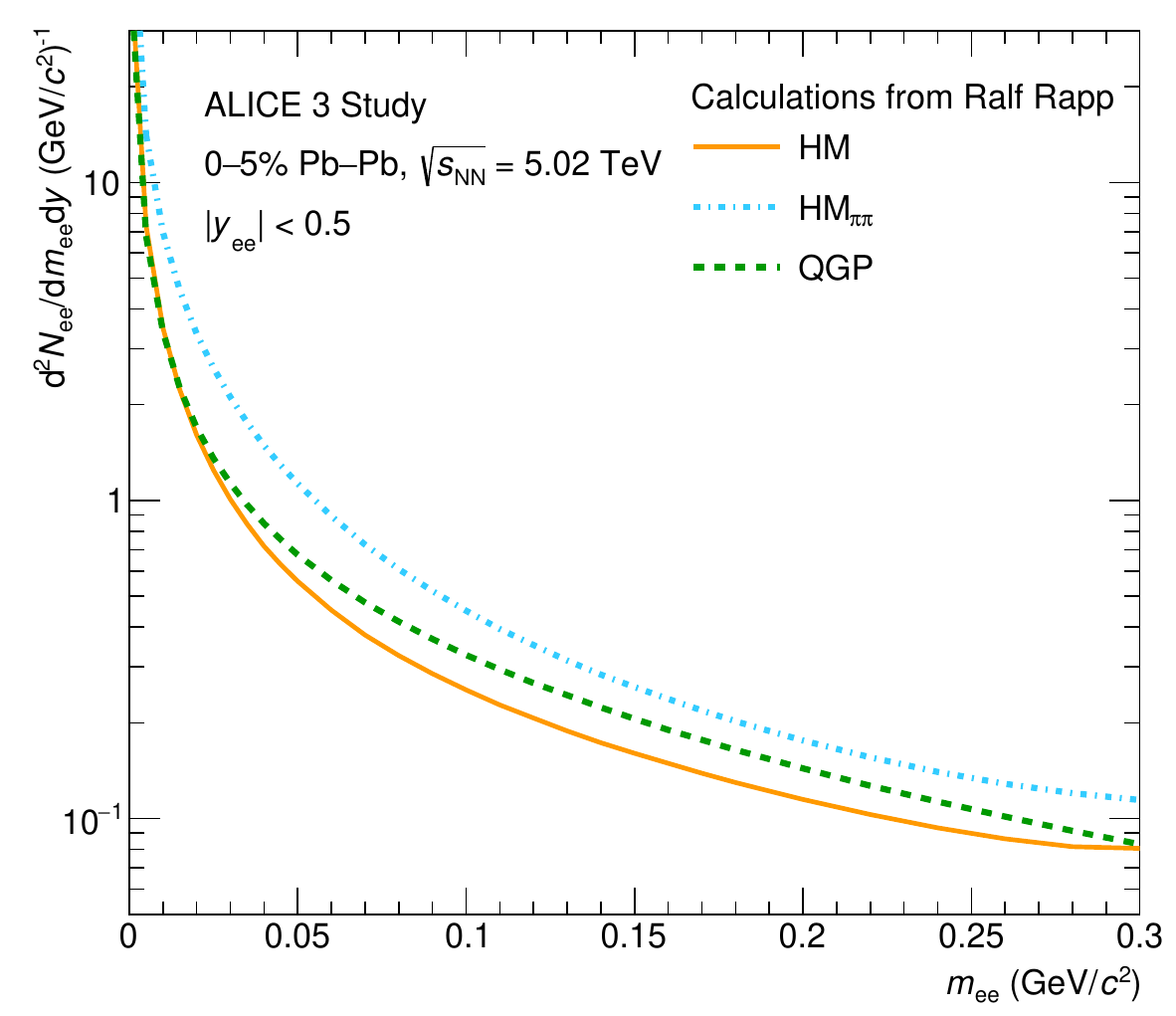}
    \includegraphics[width=0.49\textwidth]{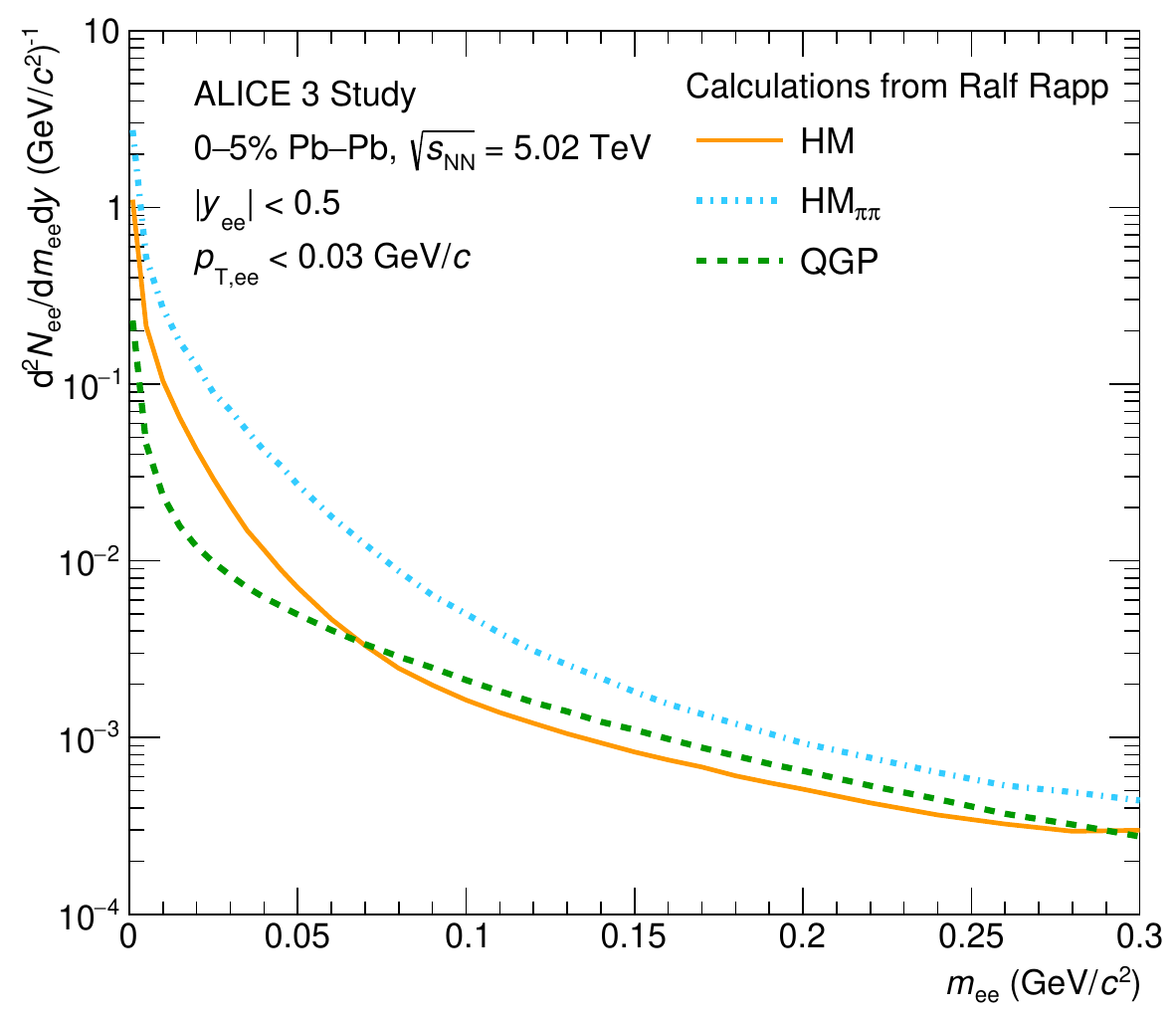}
    \caption{Predicted thermal ${\rm e^{+}e^{-}}$ yields from the QGP
      (green) and from the hadronic matter with ($\rm{HM}_{\pi\pi}$,
      blue) and without (HM, yellow) $\pi\pi$-bremsstrahlung at
      midrapidity ($|y_{\rm{ee}}|=0$) in the 5\% most central Pb--Pb
      collision at $\sqrt{s_{\rm{NN}}}=5.02\:\rm{TeV}$~\cite{Rapp:2013nxa,   Atchison:2022yxm}. The yields are integrated over $p_{\rm T,ee}$
      in the left panel, and for $p_{\rm T,ee} < 0.03$\,GeV/$c$ in the
      right panel.}
    \label{fig:Rapp}
\end{figure}

The thermal radiation signal used in the present study is calculated with in-medium spectral functions that are convoluted over an expanding fireball model~\cite{Rapp:2011is,Rapp:2013nxa,Rapp:2016xzw,Atchison:2022yxm}. This model is successful in describing dilepton measurements at SPS
energies~\cite{CERESNA45:2002gnc, CERES:2006wcq, NA60:2008dcb,
  NA60:2006ymb, NA60:2008ctj} as well as at RHIC
energies~\cite{PHENIX:2015vek, STAR:2013pwb}. It is also compatible with first
${\rm e^{+}e^{-}}$ results at the LHC~\cite{ALICE:2018ael}, although the
uncertainties of the data are still large. In the QGP,
it employs a perturbative dilepton rate based on $q\bar q$ annihilation that is combined with a transport peak which approximately reproduces the conductivity computed in current lattice-QCD computations, albeit with very large uncertainties. In the hot hadronic phase, the thermal emission rate of
dielectrons is calculated within hadronic many-body theory, including an
in-medium modified vector-meson spectral function. The production of
dileptons includes processes like annihilation, resonance Dalitz decays,
meson-exchange scattering, and bremsstrahlung. The temperature evolution is determined with an equation of state based on lattice-QCD results at vanishing net-baryon density coupled to a hadron-resonance gas with partial chemical equilibrium. The expected yields of
thermal radiation from the QGP and from the hadronic phase as a function
of $m_{\rm ee}$, i.e., the invariant mass of the ${\rm e^{+} e^{-}}$
pairs, are shown in Fig.~\Ref{fig:Rapp} for the 5\% most central Pb--Pb
collisions at $\sqrt{s_{\text{NN}}} = 5.02 \, \text{TeV}$ at midrapidity.
Two different versions of the radiation from the hadronic phase are shown,
i.e., one including $\pi\pi$-bremsstrahlung and one in which it is
switched off manually. With $\pi\pi$-bremsstrahlung the electric
conductivity in the hadronic medium is effectively reduced which then
leads to a 50\% enhancement of the thermal ${\rm e^{+}e^{-}}$ yield at
$m_{\text{ee}} \approx 0.1 \, \text{GeV}/c^{2}$ integrated over the pair
transverse momentum ($p_{\rm T,ee}$), as shown in Fig.~\ref{fig:Rapp}
(left). By selecting $p_{\rm T,ee} < 0.03 \,\text{GeV}/c$, the enhancement of
the predicted thermal ${\rm e^{+}e^{-}}$ yield from the hadronic matter
is even more pronounced as can be seen on the right panel of
Fig.~\ref{fig:Rapp}.

\begin{figure}[ht!]
    \centering
    \includegraphics[width=0.55\textwidth]{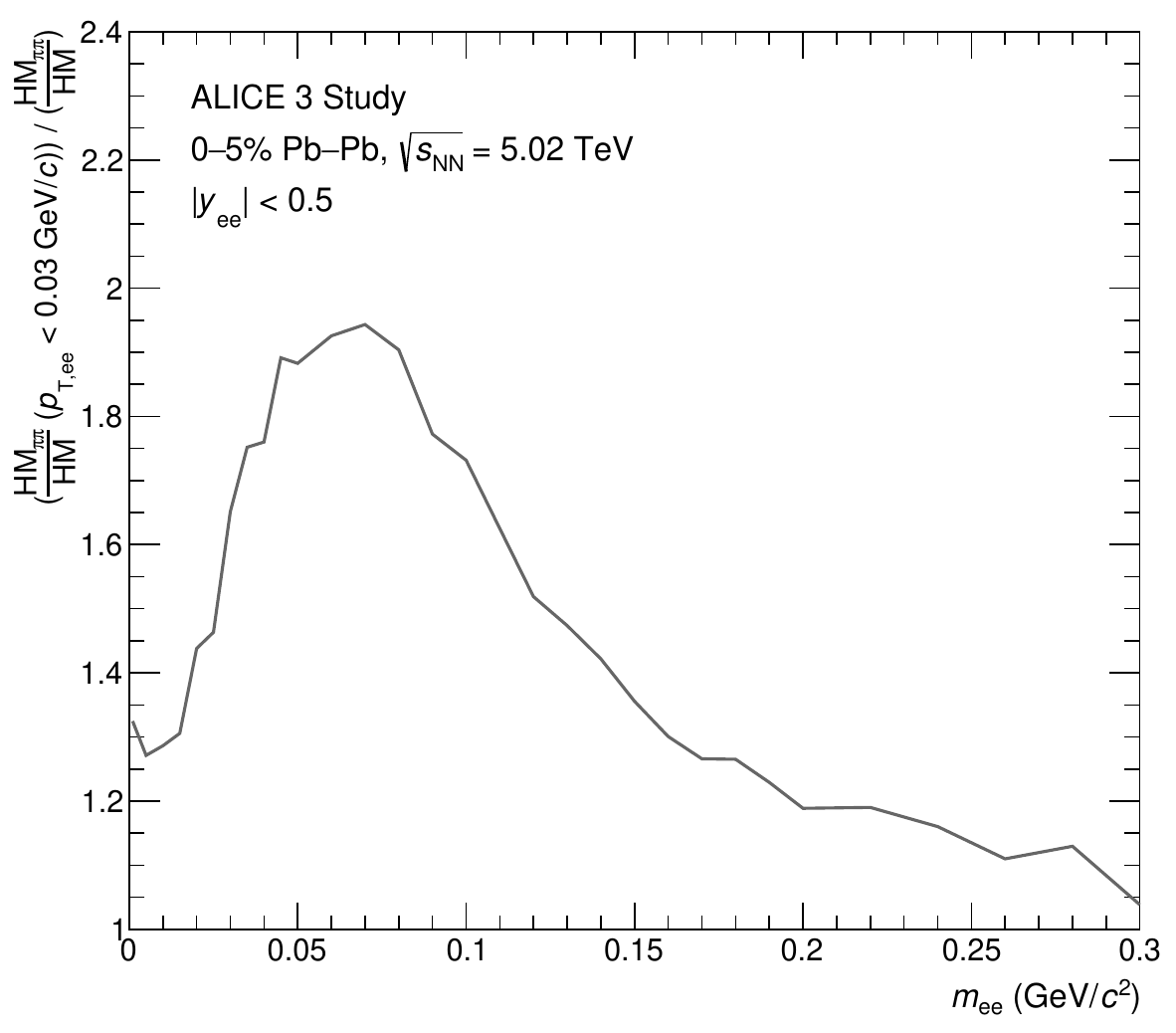}
    \caption{Double ratio of thermal radiation from the hadronic matter
      with and without $\pi\pi$-bremsstrahlung and a $p_{\rm{T,ee}}$ cut
      ($p_{\rm{T,ee}}<0.03\:\rm{GeV}/$$c$) for the 5\% most central
      Pb--Pb collisions at
      $\sqrt{s_{\rm{NN}}}=5.02\:\rm{TeV}$~\cite{Rapp:2013nxa,
        Atchison:2022yxm}.}
    \label{fig:Doubleratio}
\end{figure}

The double ratio of the predicted thermal radiation yield from the
hadronic matter with and without
$\pi\pi$-bremsstrahlung is shown in Fig.~\ref{fig:Doubleratio} for
$p_{\rm T,ee} < 0.03 \,\text{GeV}/c$ and integrated over $p_{\rm
  T,ee}$. It reaches a factor of about two at $m_{\rm ee} \approx
0.07\,\text{GeV}/c^{2}$, whereas it is still of the order of 1.2 at
$m_{\rm ee} \approx 0.15\,
\text{GeV}/c^{2}$.  This suggests measuring dielectrons not only at low
$m_{\rm ee}$ but also at low $p_{\rm
  T,ee}$. Note that no requirement on the transverse momentum of the
electron or positron ($p_{\rm
  T,e}$) or on their pseudorapidity ($\eta_{\rm
  e}$) is applied, although experimentally the detectors used to track
and identify the particles have a finite acceptance and efficiency. The latter is ignored in this study.

\subsection{Physical backgrounds}
\label{physicsbackground}

Thermal radiation from the QGP and the hadronic phase are not the only
source of ${\rm e^{+}e^{-}}$ pairs at low $m_{\rm ee}$ and
$p_{\rm T,ee}$. Hadronic decays, in particular the decays of the
$\pi^{0}$ and $\eta$ mesons to $\gamma {\rm e^{+}e^{-}}$, are expected
to contribute to the dielectron yield. In addition, the highly
Lorentz-contracted lead nuclei produce extremely strong electromagnetic
fields which can be treated as a flux of quasi-real photons generated
coherently, i.e., the charges of the $Z$ protons in the nucleus act
coherently leading to a $Z^{2}$ dependence of the quasi-real photon
flux. Such photons from the two colliding nuclei are expected to
interact via the Breit–Wheeler process~\cite{PhysRev.46.1087} to produce
dileptons characterized by a small transverse pair momentum
($p_{\rm T,ee} \leq 0.3 \, \text{GeV}/c$). In the following, these background
sources will be studied.

\subsubsection{Hadronic decay background}

The expected dielectron yield from the decays of known hadrons produced
in hadronic Pb--Pb collisions is calculated with a fast Monte Carlo
program taking parametrizations of the $p_{\rm T}$ spectra of the
hadrons as input. For this purpose, the $p_{\rm T}$-differential yield
of $\pi^{\pm}$ mesons, measured down to a $p_{\rm T}$ of $0.1\, \text{GeV}/c$ in
the 5\% most central Pb--Pb collisions at $\sqrt{s_{\text{NN}}} =5.02 \,\text{TeV}$
by ALICE\,\cite{ALICE:2019hno}, is parameterised and extrapolated to
$p_{\rm T} = 0 \,\text{GeV}/c$ with a two-component
function\,\cite{Bylinkin:2012bz,Bylinkin:2015xya}. A correction to the
input $\pi^{0}$ parametrization is applied to take into account
isospin-violating decays leading to differences between $\pi^{0}$ and
$\pi^{\pm}$~\cite{ALICE:2020mfy}, mainly originating from decays of the $\eta$
meson. The effect is assumed to be similar in pp and in central Pb--Pb
collisions at the same collision energy.
The $p_{\rm T}$ spectrum of $\eta$ is computed as the average of the
spectra obtained using the parameterizations retrieved from the
$\eta/\pi^{0}$ ratio as a function of $p_{\rm T}$ in pp
collisions\,\cite{ALICE:2020umb} and from the $\text{K}^{\pm}/\pi^{\pm}$
ratio as a function of $p_{\rm T}$ measured down to
$p_{\rm T} = 0.3 \,\text{GeV}/c$ in Pb--Pb
collisions\,\cite{ALICE:2019hno}. The ratio of the resulting $p_{\rm T}$
spectrum of $\eta$ to the one of $\pi^{0}$ at very low $p_{\rm T}$
\mbox{($\pt \leq 0.1\, \text{GeV}/c$)} is in agreement within
uncertainties with the $\eta/\pi^{0}$ ratio measured in hadronic
collisions by CERES/TAPS~\cite{Agakichiev:1998ign}.

Other hadrons, $\omega$, $\eta^{\prime}$, $\rho$, and $\phi$, are
generated assuming $m_{\rm T}$-scaling, implying that the spectra of all
light mesons as a function of
$m_{\rm T} = \sqrt{m^{2}+p_{\rm T}^{2}}$ follow a universal form
and differ only by a normalization factor.

\begin{figure}[ht!]
    \centering
    \includegraphics[width=0.55\textwidth]{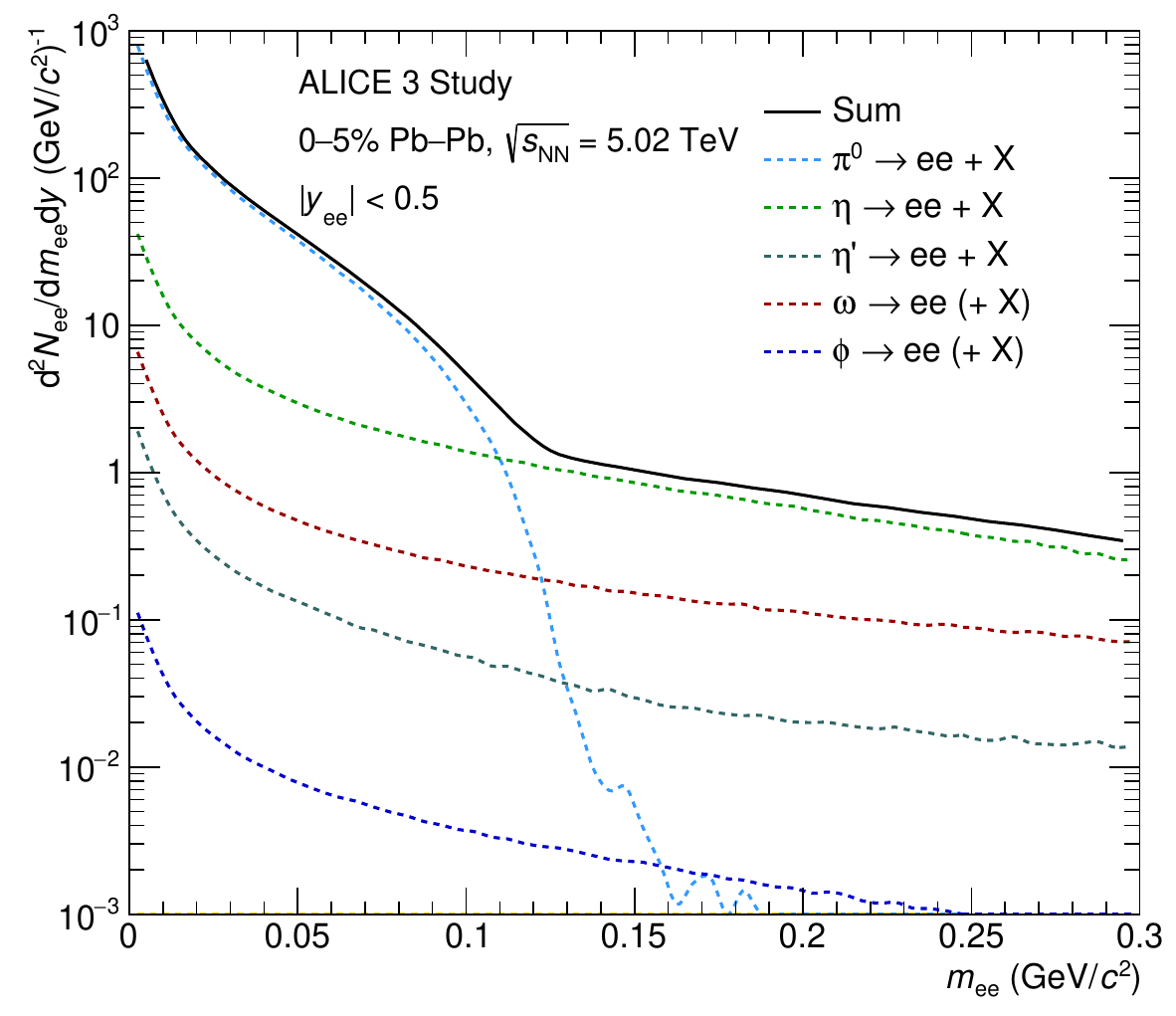}
    \caption{Expected dielectron yield from hadronic decays in the 5\%
      most central Pb--Pb collisions at
      $\sqrt{s_{\rm{NN}}}=5.02\,\text{TeV}$. The sum (black), as well as
      the different hadron decay contributions, are shown.}
    \label{fig:Cocktail}
\end{figure}

The resulting expected yields of dielectrons from hadronic decays in the
5\% most central Pb--Pb collisions at
$\sqrt{s_{\text{NN}}} = 5.02 \, \text{TeV}$
integrated over $p_{\rm T,ee}$
are shown as a function of
$m_{\rm ee}$ in Fig.~\ref{fig:Cocktail}. The Dalitz decay of $\pi^{0}$ mesons dominates up to
the pion mass ($m_{\rm \pi^{0}} \approx 0.135 \, \text{GeV}/c^{2}$), whereas above
the $\eta \to \gamma {\rm e^{+}e^{-}}$ decay channel is the main source
of ${\rm e^{+}e^{-}}$ pairs from hadronic decays.

\subsubsection{Electromagnetic background}

The highly Lorentz contracted electromagnetic fields generated by the
ultra-relativistic lead ions can manifest themselves in quasi-real
photons that in turn can interact via the Breit-Wheeler
process~\cite{Breit:1934zz} and produce a dilepton pair with small pair
transverse momentum.

The $\gamma\gamma \to \lele$ process has been first observed in
ultra-peripheral collisions (UPCs) characterized by an impact
parameter $b$ that is larger than twice the nuclear radius $R_{A}$,
i.e., with no hadronic overlap~\cite{STAR:2004bzo}. Nevertheless, in
recent years, the STAR~\cite{STAR:2018ldd},
ATLAS~\cite{ATLAS:2018pfw,ATLAS:2022yad}, and ALICE~\cite{ALICE:2022hvk}
collaborations successfully measured this process in collisions with
hadronic overlap (HOCs), i.e., with $b < 2\times R_A$, down to the most
central Pb--Pb collisions for ATLAS. Medium effects, like
electromagnetic scatterings of the leptons in the QGP or the presence of
magnetic fields trapped in the emerging QGP, were initially suggested to
explain the broader dilepton transverse momentum $p_{\rm T,ll}$ or
acoplanarity distributions observed in such collisions compared to
UPCs. However, it turns out that with more recent calculations the room
for such interactions with the medium is significantly reduced. Nevertheless, the data can be used to map the electromagnetic fields of
the nuclei and to better understand QED phenomena. An extensive review can be found in Ref.~\cite{Brandenburg:2021lnj}.

\begin{figure}[ht!]
    \centering
    \includegraphics[width=0.49\textwidth]{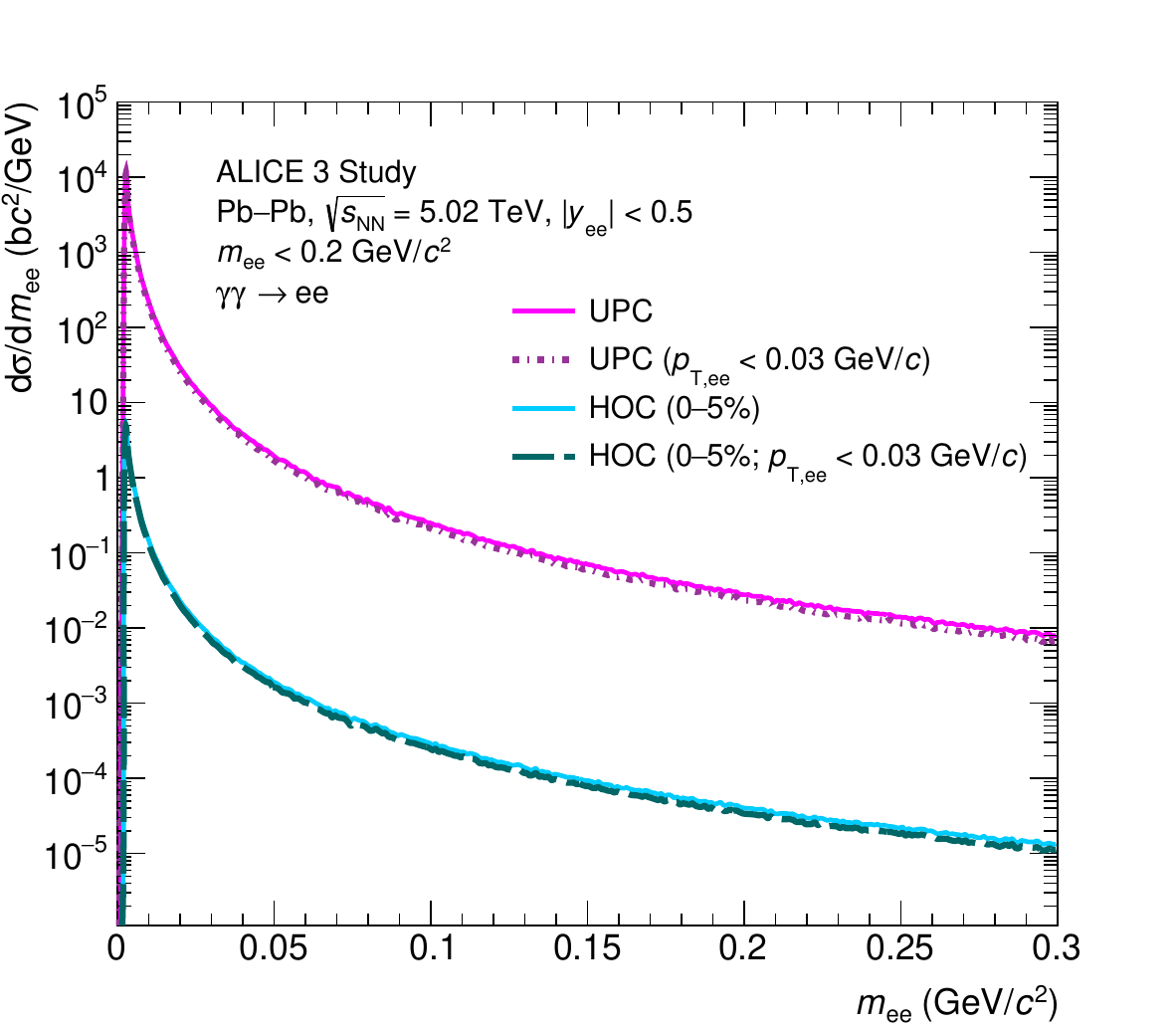}
    \includegraphics[width=0.49\textwidth]{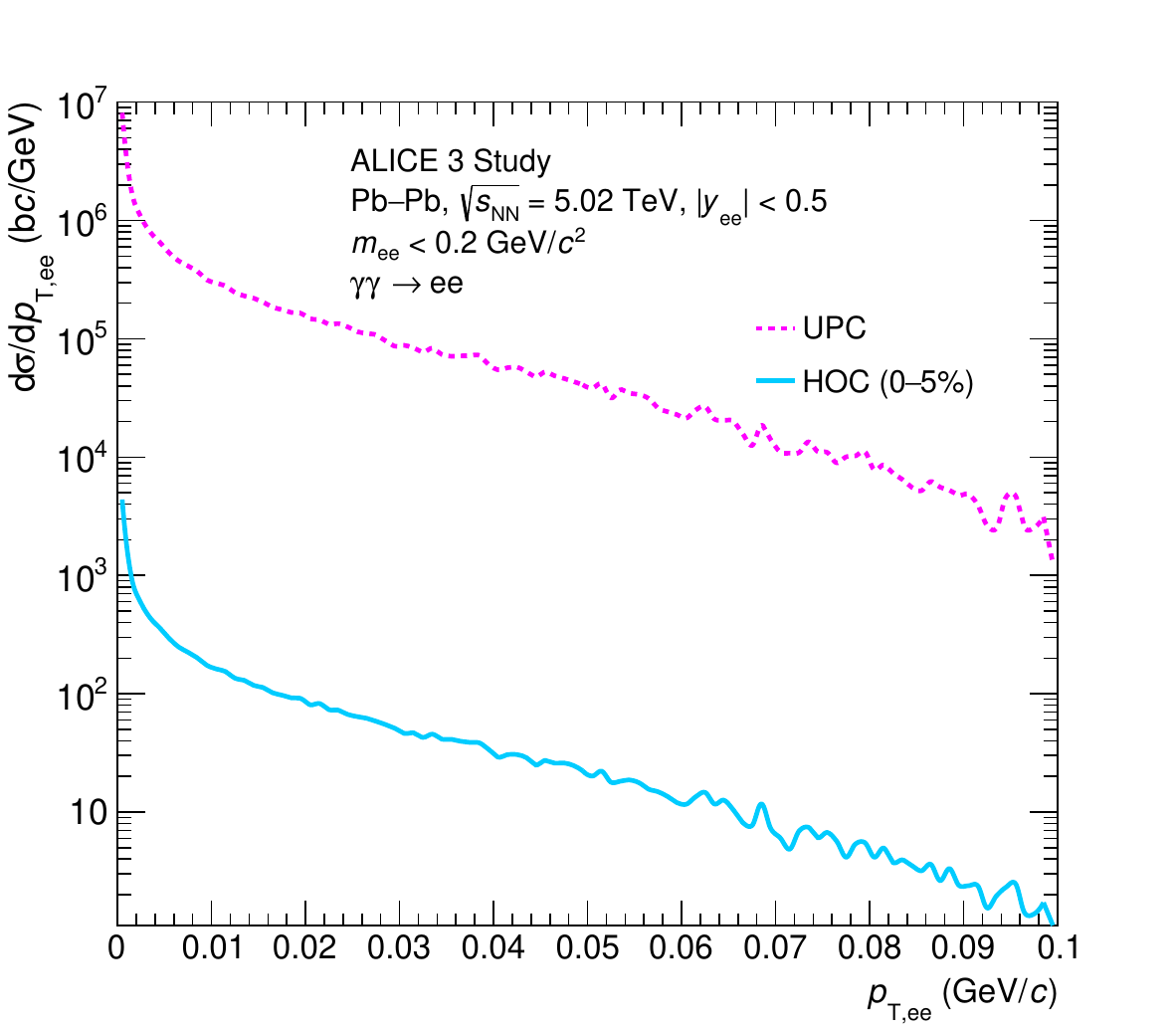 }
    \caption{Invariant mass $m_{\rm ee}$ (left) and pair transverse momentum
      $p_{\rm T,ee}$ (right) distributions of dielectrons originating from
      the $\gamma\gamma\to \text{e}^{+}\text{e}^{-}$ process in ultra-peripheral
      Pb--Pb collisions and in the 5\% most central collisions at
      midrapidity for $\sqrt{s_{\text{NN}}} = 5.02\,\text{TeV}$, as
      prediced by STARlight~\cite{Klein:2016yzr}. The effect of a
      selection of pairs with $p_{\rm T,ee} < 0.03 \, \text{GeV}/c$ is
      also shown.}
    \label{fig:SLsimulations}
\end{figure}

Different models are available to estimate the ${\rm e^{+}e^{-}}$ yields
from $\gamma\gamma$ interactions, i.e., the STARlight event
generator~\cite{Klein:2016yzr}, lowest-order QED
calculations~\cite{Hencken:1994my,Alscher:1996mja,Brandenburg:2020ozx,Zha:2018tlq}
(LO-QED) or approaches using the Wigner
formalism~\cite{Li:2019sin,Klein:2020jom,Klusek-Gawenda:2020eja,Wang:2021kxm}. All
of them neglect higher-order corrections in the QED predictions and
assume that the nuclei maintain their velocities to validate the
external and coherent field approximation. The latter may lead to
important corrections to the initial electromagnetic fields in central
collisions, where the photon flux is expected to be generated
predominantly by the participant nucleons, whereas the former could be
relevant for both UPCs and HOCs~\cite{Baur:2008hn, Zha:2021jhf}. For
this study, the STARlight generator was used. It has been found to reproduce
the measurements in UPCs reasonably well~\cite{ALICE:2013wjo,
  CMS:2018erd, ATLAS:2020epq}, while predicting too narrow
$p_{\rm T,ll}$ or acoplanarity distributions in HOCs~\cite{STAR:2018ldd,
  ATLAS:2018pfw, ATLAS:2022yad, ALICE:2022hvk}, mostly due to a partial
treatment of the impact-parameter dependence. In this model, the shape
of the transverse-momentum distribution of the quasi-real photons is
assumed to be impact-parameter independent, in contrast to the
broadening of the measured $p_{\rm T,ll}$ or acoplanarity spectra
observed by the PHENIX, STAR, and ALICE collaborations towards central
collisions. This broadening is however qualitatively reproduced by other
theoretical approaches including proper impact-parameter
dependences~\cite{Brandenburg:2020ozx, Zha:2018tlq,
  Klusek-Gawenda:2020eja, Wang:2021kxm}. The $p_{\rm T,ll}$-integrated
cross sections calculated with STARlight in HOCs are nevertheless in
fair agreement with the data.

Using the STARlight~\cite{Klein:2016yzr} event generator the cross sections for UPCs and HOCs were calculated. For the calculations of the HOC $\gamma\gamma\to{\rm e^{+}e^{-}}$ process the impact parameter in the event generation was restricted to $0 < b < 3.5$~fm to match the selection of 5\% most central Pb-Pb events. For UPCs the requirement was the criterion that no hadronic interaction took place by requiring $b > 2 R_{A}$.
The cross section for the $\gamma\gamma\to{\rm e^{+}e^{-}}$ process at midrapidity
($|y_{ee}| < 0.5$) is about three orders of magnitude larger in UPCs
(19572~b) than in the 5\% most central Pb--Pb collisions (10.7~b)
at $\sqrt{s_{\text{NN}}} = 5.02 \, \text{TeV}$. Both are shown as a function of
$m_{\rm ee}$ (left) and $p_{\rm T,ee}$ (right) in
Fig.~\ref{fig:SLsimulations}. The $p_{\rm T,ee}$ distributions are
similar in HOCs and in UPCs, as expected from the model, and peak at
very low $p_{\rm T,ee}$. The effect of requiring
$p_{\rm T,ee} < 0.03 \,\text{GeV}/c$ is also shown and reduces the
dielectron yields by less than 5\%.

\subsubsection{Normalization of the background sources}

Contrary to the hadronic decay background (HDB) and the
$\gamma\gamma \to {\rm e^{+}e^{-}}$ process in HOCs (GHOC), the
contribution from dielectrons from $\gamma\gamma$ interactions in UPCs
(GUPC) comes from a different Pb--Pb collision as the one producing the
thermal radiation signal, a so-called pile-up collision. Pile-up can be categorized into two different types: (I) out-of-bunch
pile-up where the UPC event occurs in a different bunch crossing than
the hadronic collision, and (II) in-bunch pile-up for which the UPC
happens in the same bunch crossing as the hadronic collision. Assuming a
timing resolution of the ALICE 3 Time-Of-Flight system in the order of
10~ps~\cite{ALICE:2022wwr}, i.e., three orders of magnitude smaller than
the bunch spacing at the LHC (10~ns), the type (I) of pile-up can be
neglected for measurements performed with this experiment. This means
that only UPCs happening in a bunch crossing where a central hadronic
collision occurs, have to be considered. This reduces the amount of GUPC
by a factor $0.05 \times \mu$, where $\mu$ is the probability to have
one hadronic collision per bunch crossing ($\mu =0.01$ for ALICE 3, see
Tab.~1 of the ALICE 3 letter of intent~\cite{ALICE:2022wwr}) and $0.05$ accounts for
the centrality selection.

The GUPC background can be further suppressed by requiring that the
electron and positron candidates point to the reconstructed primary
vertex of the hadronic collision along the beam axis
($z$-direction). The pointing resolution at low \pt is expected to be in
the order of \SI{100}{\micro\metre} for ALICE 3~\cite{ALICE:2022wwr}. Applying a
$5\sigma$ selection implies requiring an impact parameter in beam direction for the
reconstructed ${\rm e^{\pm}}$ trajectories and the collision vertex between \SI{-500}{\micro\metre} and \SI{500}{\micro\metre}. In the
following, the distributions of the hadronic collisions and UPCs along the beam direction are
assumed to be described by a Gaussian distribution with a width
of 5~cm, as it is currently the case in ALICE. The requirement on the
impact parameter of the electron and positron candidates is expected
to reduce the GUPC background by at least a factor
$\epsilon_{\rm DCA_{z}}$ with:
\begin{equation}
    \epsilon_{\rm DCA_z} = \frac{\int_{\SI{-500}{\micro\metre}}^{\SI{500}{\micro\metre}}G(z)dz}{\int_{\SI{-10}{\centi\metre}}^{\SI{10}{\centi\metre}}G(z)dz} = 0.0084,
\end{equation}
corresponding to the worst case where the hadronic collision and the UPC happen at the same $z$ value. Combining the above selection of bunch crossing with a central hadronic
collision and the $\epsilon_{\rm DCA_z}$ suppression factor, the
``visible'' cross section for the UPC $\gamma\gamma$ process can be
calculated as
$19572\,{\rm b} \times \mu \times 0.05 \times \epsilon_{\rm DCA_z} =
0.082\,{\rm b}$. The background contribution from dielectrons from the
$\gamma\gamma \to {\rm e^{+}e^{-}}$ process in UPCs is then expected to be
smaller than from $\gamma\gamma$ interactions in HOCs.

\subsection{Total dielectron yield}
\label{results}

\begin{figure}[ht!]
    \centering
    \includegraphics[width=0.49\textwidth]{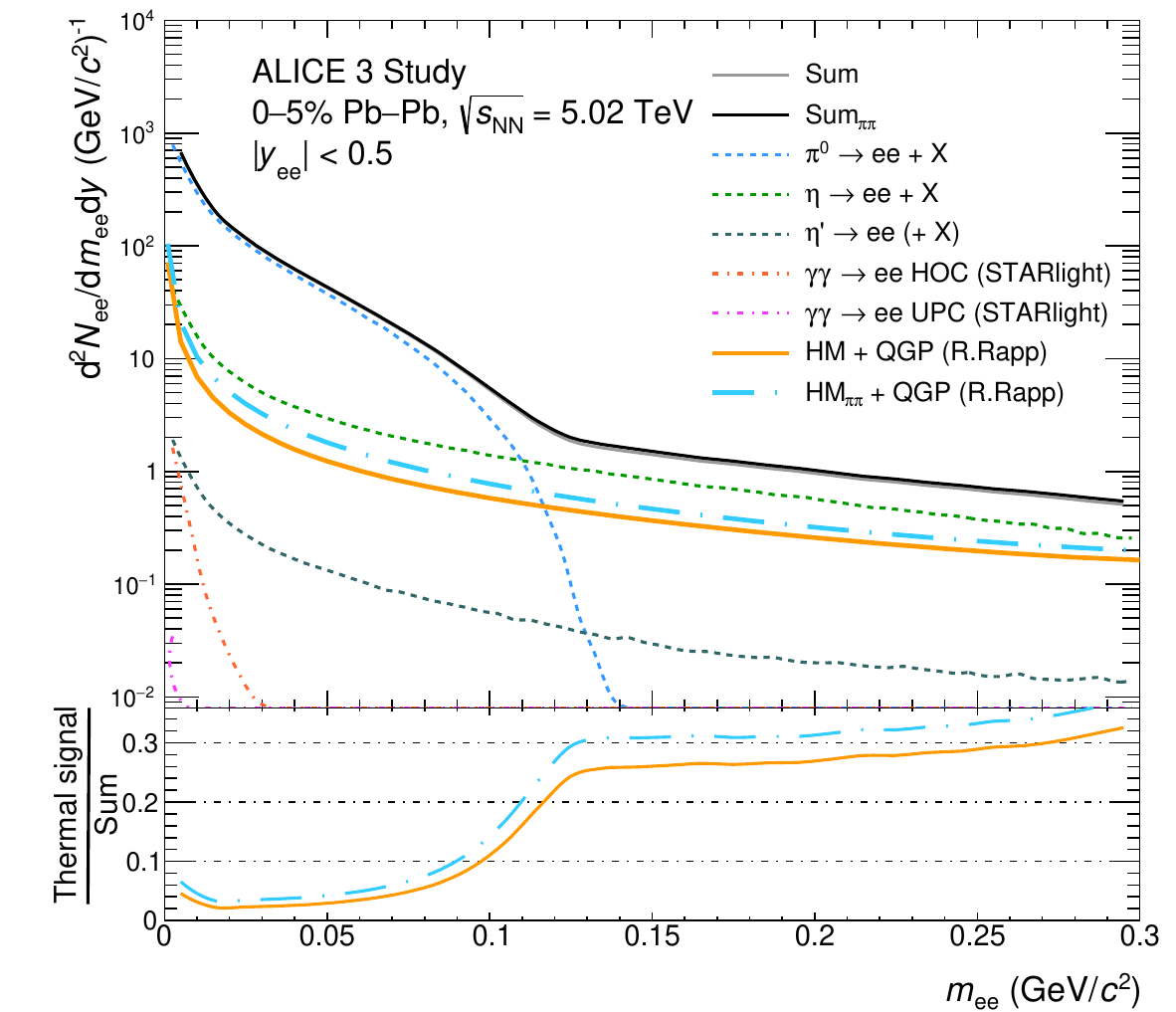}
    \includegraphics[width=0.49\textwidth]{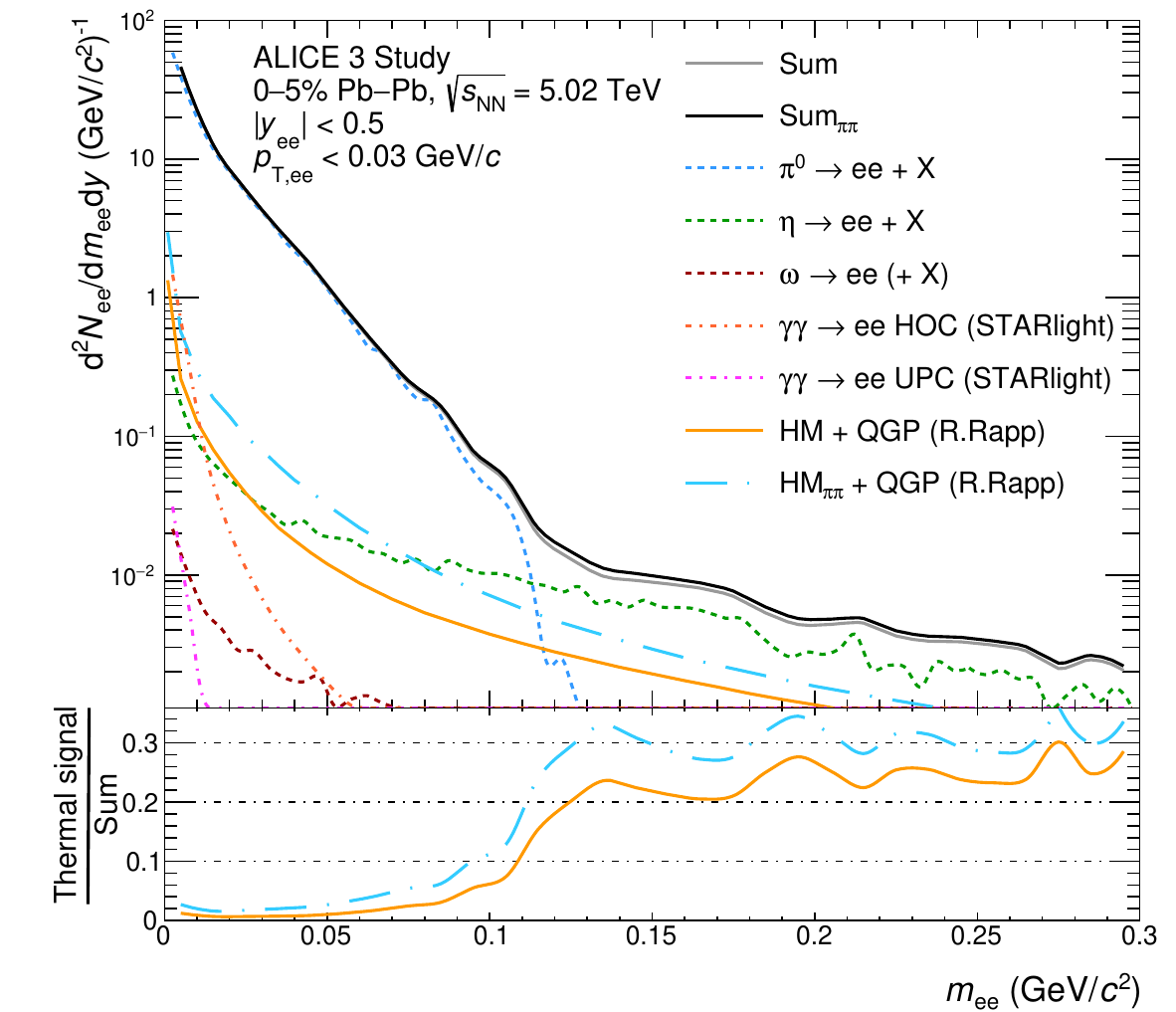}
    \caption{The expected total dielectron spectrum with its different contributions in the most 0--5\% central Pb--Pb collisions at $\sNN = 5.02\, \text{TeV}$ as a function of \mee integrated over \ptee (left) and for \mbox{$\ptee < 0.03 \, \text{GeV}/c$}. Two different calculations for thermal radiation, including ($\text{HM}_{\rm \pi\pi}$ + QGP) or not (HM +QGP) $\pi\pi$-bremsstrahlung in the hadronic phase, are shown~\cite{Atchison:2022yxm}. The relative contribution of thermal radiation from the QGP and the hadronic phase is shown on the bottom panels for the two different predictions.}
    \label{fig:CocktailComparison}
\end{figure}

Figure~\ref{fig:CocktailComparison} shows the expected total dielectron
yield, black with and grey without $\pi\pi$-brems\-strah\-lung in the hot
hadronic phase, as a function of \mee in the 5\% most central Pb--Pb
collisions at $\sqrt{s_{\text{NN}}} = 5.02 \, \text{TeV}$.
In the left panel the
yields are integrated over \ptee, whereas in the right panel they are
computed for $p_{\rm T,ee} < 0.03 \, \text{GeV}/c$. The different contributions
to the ${\rm e^{+}e^{-}}$ yield are also reported separately. Up to
around $\mee = 100 \, \MeVcc$ the dielectron yield is dominated by Dalitz
decays of \piZ mesons whereas above the $\eta$ Dalitz decay is the main
source of ${\rm e^{+}e^{-}}$ pairs. According to the STARlight event
generator~\cite{Klein:2016yzr} and the approach explained above, the
backgrounds from the $\gamma\gamma \to {\rm e^{+}e^{-}}$ process in HOCs
and UPCs are negligible, particularly for $\mee > 100 \, \MeVcc$. On the
bottom panels of Fig.~\ref{fig:CocktailComparison}, the relative
contribution from thermal radiation from the QGP and from the hadronic
matter~\cite{Atchison:2022yxm} is shown. Above the pole mass of the
\piZ, it amounts to 25--30\% and 20--30\% of the total ${\rm e^{+}e^{-}}$
yield, integrated over $p_{\rm T,ee}$ and for
$p_{\rm T,ee} < 0.03 \, \text{GeV}/c$, respectively, and depends on the
inclusion or not of $\pi\pi$-bremsstrahlung in the hot hadronic phase.

\begin{figure}[ht!]
    \centering
    \includegraphics[width=0.55\textwidth]{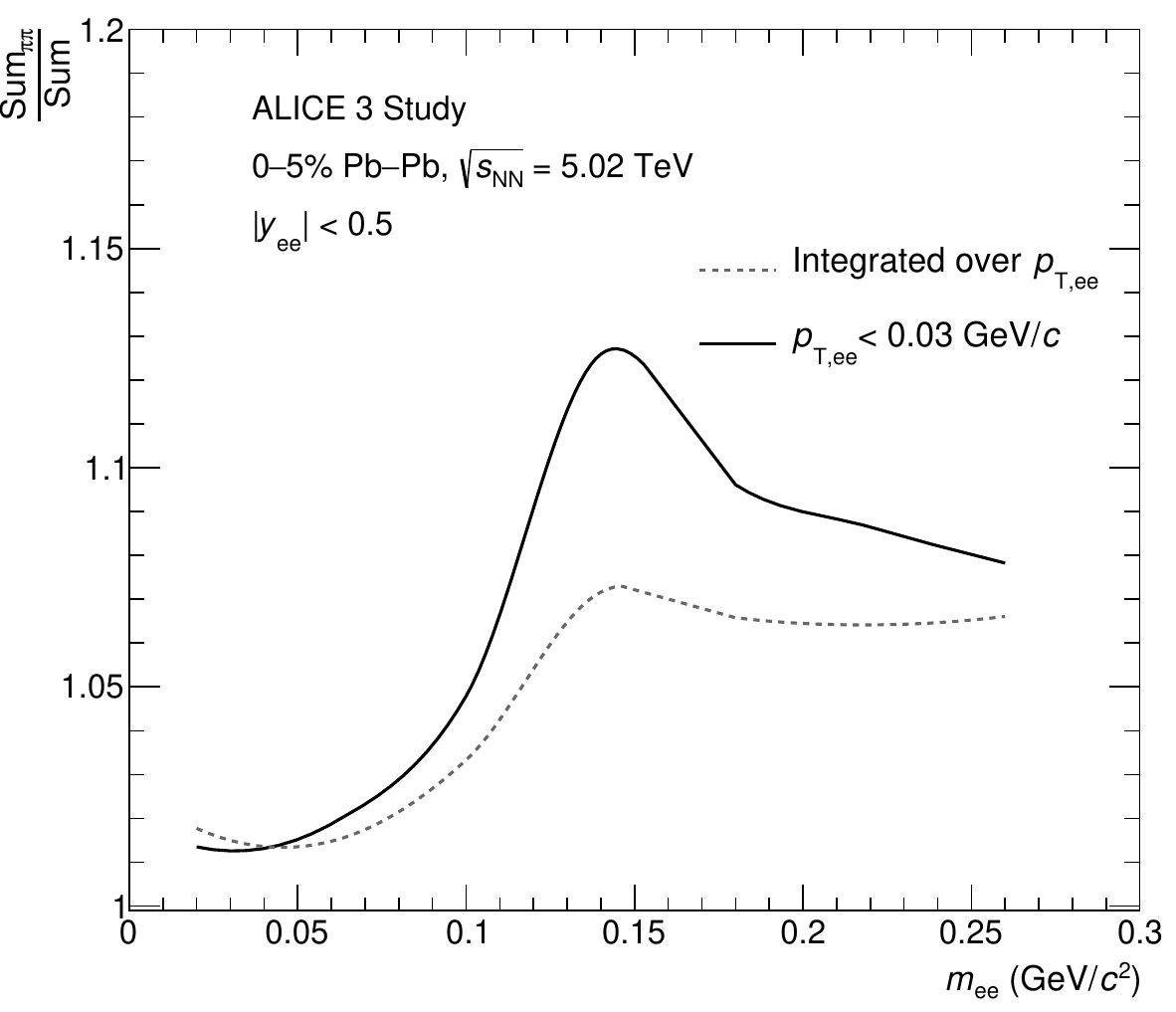}
    \caption{Difference between the two predictions for thermal
      radiation in hot hadronic matter~\cite{Rapp:2013nxa,Atchison:2022yxm}
      normalized to the total dielectron yield for
      $p_{\rm T,ee} < 0.03 \, \text{GeV}/c$, and integrated over $p_{\rm T,ee}$
      in the 5\% most central Pb--Pb collisions at $\sqrt{s_{\text{NN}}} =5.02 \, \text{TeV}$.}
    \label{fig:rat}
\end{figure}

To quantify the precision a measurement would need to differentiate
between the two different implementations for thermal radiation produced
in the hot hadronic matter, the ratio of the total expected dielectron
yield with and without $\pi\pi$-bremsstrahlung is calculated. The result is shown in Fig.~\ref{fig:rat} for the two
different \ptee selections. Integrated over \ptee, the relative
difference of the total dielectron yield is in the order 7\% for \mee
larger than $m_{\rm \pi^{0}}$. This means that a measurement would need
a total uncertainty below 1.5\% to reach a $5 \sigma$ significance in the
separation of both theoretical scenarios. For $\ptee < 0.03 \, \text{GeV}/c$,
the relative difference is larger by up to a factor 1.7 at
$\mee = m_{\rm \pi^{0}}$. Total uncertainties of about 2.5\% would then be
necessary.

According to this study, precise measurements of the dielectron
production at low \mee would be needed in order to constrain the
electric conductivity in the hot medium. Going to low \ptee increases the sensitivity to different $\sigma_{\text{EM}}$. Excellent knowledge of the hadronic background in particular from the Dalitz
decays of $\eta$ mesons is mandatory. No detector acceptance and
efficiencies were considered so far. They are nevertheless crucial,
together with an estimation of the combinatorial background, in order to
quantify the expected uncertainties of a dielectron measurement in this
phase space with ALICE 3.

\section{Conclusions and Outlook}

In this review we have summarized the current status and
future opportunities for the experimental investigation of the validity
of soft-photon theorems in the ALICE experiment at the LHC.
The emission of soft bremsstrahlung photons is related to fundamental
properties of local relativistic quantum field theories and has been investigated in a variety of approaches over the years. Already in the early developments of quantum electrodynamics (QED), the presence of
 infrared divergences of bremsstrahlung cross sections was an immediate issue which has
been resolved in the work by Bloch and Nordsieck by considering the
impossibility to resolve processes with the emission of one or many
``soft photons'' from those without these emissions. Taking into account
the finite energy resolution of detectors one obtains a finite effective
cross section \cite{Bloch:1937pw,Jauch1954,Jauch1955}. Another
resolution of the same IR problems is the analysis of the ``true
asymptotically free states'' of charged particles; due to the
long-ranged nature of the electromagnetic interaction, i.e., the
 vanishing mass of the photon, these asymptotically free states are rather
represented by ``infra-particle states'' which can be interpreted
intuitively as the ``naive bare-particle state'' surrounded by their
coherent electromagnetic field which can be pictured as a ``cloud of
soft photons''
\cite{Dollard:1964,PhysRev.140.B1110,Kibble1968,Kibble1968a,Kibble1968b,Kibble1968c,Kulish:1970ut}. Another
aspect is the close connection with the fundamental symmetry of
relativistic spacetime and the pertinent theory of unitary
representations of the proper orthochronous Poincar{\'e} group in terms
of local quantum fields \cite{wigner39,Weinberg1995}, which together
with the locality/microcausality assumption inevitably leads to the
realization of massless particles as gauge bosons with their
interactions via couplings to conserved charges
\cite{Weinberg:1964kqu}. Similar conclusions can be drawn from
$S$-matrix theory \cite{Weinberg:1964ew}. More recently the soft-photon
theorem has been related to asymptotic symmetries due to invariance
under ``large gauge transformations'' \cite{Strominger:2017zoo,He:2014cra,Arkani-Hamed:2020gyp,Campiglia:2015qka}.

Given these truly fundamental arguments for the validity of the related
soft-photon theorems, i.e., the factorization of scattering amplitudes in a universal
soft-photon factor and the amplitude for the corresponding process
without the emission of soft photons at leading power $1/\omega$ in the photon
energy, the soft-photon
puzzle, i.e., the excess of observed soft bremsstrahlung photons in
hadronic processes, is of great interest. From a theoretical perspective,
progress has been recently made regarding Low's theorem. In particular, recent papers addressed the problem of how to implement the subleading terms in the theorem (which arise from an expansion around $\omega=0$) in photon spectra which unavoidably require a non-vanishing soft momentum. In this regard, several proposals have been put forward on how to deal with the functional dependence of the factorized non-radiative amplitude and the corresponding ambiguities arising from momentum conservation constraints \cite{Lebiedowicz:2021byo, Lebiedowicz:2023ell, Balsach:2023ema, Fadin:2024tar}.

In spite of this recent progress, much remains to be investigated. A precise numerical estimation of next-to-leading power effects is only in its infancy since it has been computed only for a series of simple cases. For processes with generic hadronic final states and kinematical cuts in analogy to those that were implemented for the measurements that observed an excess of photons, further studies are necessary. Another aspect to be carefully examined is the question of non-perturbative QCD mechanisms of the
production processes of hadrons accompanied by soft bremsstrahlung,
particularly the hadronization processes and the involved space-time scales
which might be relevant for an estimation of the energy-momentum scale
defining the validity range of the leading-order soft-photon limit which must hold if the very fundamental properties of local relativistic quantum field theory are valid and the impact of subleading corrections.
It is also worth stressing that the leading term in Low’s theorem does not receive corrections as long as the photon energy is much smaller than the masses of the charged particles. When this condition is violated, e.g., for parametrically small masses and in perturbative QCD with massless partons, Low’s leading theorem receives loop corrections (as recently computed in \cite{Ma:2023gir}). However, the phenomenological implications of these results for processes where a perturbative description breaks down remain not clear.

Further, we have demonstrated that the ALICE 3 apparatus including the Forward Conversion Tracker will enable precision measurements of soft photon production in the region of the Low divergence in inclusive and exclusive processes with pp collisions at top LHC energy. This will settle the issue of the presence of the anomalous soft photon production and provide an experimental test of the infrared limit of quantum field theories such as QED and QCD.

Comparing with other LHC experiments,
the LHCb apparatus at the Large Hadron Collider covers the forward pseudorapidity range of the planned Forward Conversion Tracker and has already demonstrated photon conversion capabilities \cite{LHCb:2013ofo}. However, the relatively large material budget of the inner vertex tracker of 21.3\% in units of radiation lengths \cite{Bediaga:2013tje} and the photon energy threshold in the few GeV range will make it very difficult to provide a test of the Low theorem within the framework of LHCb.

With the recent advancement in silicon pixel detector technology that is largely employed in the ALICE~3 planning it would be intriguing to measure soft photon production at beam rapidity in coincidence with the deflected beam particles by use of Roman pots at LHC \cite{ATLAS:2008xda,TOTEM:2008lue}, a technique that is  well established in experimental particle physics.
A next-generation, extremely high-luminosity electron-positron collider may become available as early as during the next decade \cite{CEPCStudyGroup:2023quu,FCC:2018evy}. The clean environment of such collisions with abundantly produced high-energy jets would provide another precision test of the soft photon theorem in a different setting.

Furthermore, in heavy-ion collisions, the production of real and virtual soft photons from the locally thermalized medium is linked, at very low masses and momenta, to the electrical conductivity of this medium. We have summarized the status of the current theoretical predictions for $\sigma_{\text{EM}}$. Based on calculations from the model in \cite{Rapp:2013nxa,Atchison:2022yxm}, we have shown that measuring the ${\rm e^{+}e^{-}}$ yield not only at low masses ($m_{\rm ee}$ of about $0.15 \, \text{GeV}/c^{\rm 2}$) but also at low pair transverse momenta ($p_{\rm T,ee} < 0.03 \, \text{GeV}/c$) substantially increases the sensitivity to the electrical conductivity. The required experimental precision of such a dielectron measurement at midrapidity has been estimated to be of the order of 2\% in order to distinguish theoretical scenarios with significantly varying values for the electrical conductivity in the hadronic phase. Very good knowledge of the hadronic background, in particular from Dalitz decays of the $\eta$ meson, is mandatory while the background resulting from the $\gamma\gamma \to {\rm e^{+}e^{-}}$ process is expected to be small for the LHC and ALICE~3 setup. Further studies on the detector acceptance and efficiency at such low momenta will allow a better quantification of the expected uncertainties of a dielectron measurement in this phase space with ALICE~3.

\vspace{0.5cm}

\section*{Acknowledgement}
We thank Roger Balsach, Spencer Klein, Anna Kulesza, Otto Nachtmann, Joakim Nystrand, Rob Pisarski, George Sterman and Andrew Strominger for exciting discussions.
This work has been supported by the US National Science Foundation (NSF) under grant no.~PHY-2209335 (RR), by the DFG (German Research Foundation) -- Project-ID 273811115 -- SFB 1225 ISOQUANT and by the Excellence Cluster ORIGINS of the DFG under grant no.~EXC-2094-390783311.
We thank the ExtreMe Matter Institute EMMI at GSI, Darmstadt, for generous support in the framework of an EMMI Rapid Reaction Task Force meeting during which this work has been initiated.
\bibliography{ref.bib}

\end{document}